\documentclass[10pt]{article}

\usepackage[T1]{fontenc}
\usepackage[utf8]{inputenc}
\usepackage[margin=1.5in]{geometry}
\usepackage{graphicx}
\usepackage{amsmath,amssymb}
\usepackage{bm}
\usepackage{siunitx}
\usepackage{xcolor}
\usepackage{multicol}
\usepackage{lineno}

\usepackage{soul}
\usepackage{booktabs}
\usepackage{gensymb}
\usepackage{dcolumn}
\usepackage{epstopdf}
\usepackage{caption}
\usepackage[numbers,sort&compress]{natbib}
\usepackage{array,multirow,adjustbox,float,rotating,longtable,textcomp,pdflscape}
\usepackage[normalem]{ulem}
\usepackage[nomarkers]{endfloat}
\usepackage{authblk}

\usepackage[colorlinks=true, linkcolor=blue, citecolor=blue, urlcolor=blue]{hyperref}

\setstcolor{red}
\DeclareSIUnit\angstrom{\text{Å}}

\DeclareUnicodeCharacter{2212}{\ensuremath{-}}
\DeclareUnicodeCharacter{03B1}{\ensuremath{\alpha}}
\DeclareUnicodeCharacter{2264}{\ensuremath{\leq}}
\DeclareUnicodeCharacter{2265}{\ensuremath{\geq}}

\newcommand{\revadd}[1]{#1}

\newcommand{\cindexci}[2]{\ensuremath{#1\,{\pm}\,#2}}
\newcommand{\aucci}[3]{
  \ensuremath{#1\;(#2,\,#3)}
}
\newcommand{\auccib}[3]{
  \ensuremath{\mathbf{#1}\;(\mathbf{#2},\,\mathbf{#3})}
}

\newcommand{\localzscell}[3]{
  \ensuremath{#1/#2\;(#3)}
}
\newcommand{\dhcox}[2]{\ensuremath{#1\,/\,#2}}

\begin{document}
\addtocontents{toc}{\protect\setcounter{tocdepth}{-1}}


\title{TRACER, a time-to-event EHR foundation model for disease detection across health systems}

\author[1,2,3]{Mara Pleasure\textsuperscript{*}}
\author[1,2,4]{Ekaterina Redekop\textsuperscript{*}}
\author[1,2,3]{Vedrana Ivezic}
\author[1,2]{Rushikesh Kulkarni}
\author[5,6]{Aaron Chin}
\author[1,2]{William Speier}
\author[1,2,4,7]{Corey W. Arnold\textsuperscript{\dag}}

\affil[1]{Biomedical AI Research Lab, University of California, Los Angeles, Los Angeles, CA, USA}
\affil[2]{Department of Radiology, University of California, Los Angeles, Los Angeles, CA, USA}
\affil[3]{Bioinformatics, University of California, Los Angeles, Los Angeles, CA, USA}
\affil[4]{Department of Bioengineering, University of California, Los Angeles, Los Angeles, CA, USA}
\affil[5]{Department of Pediatrics, Division of Immunology, Allergy and Rheumatology, University of California, Los Angeles, Los Angeles, CA, USA}
\affil[6]{UCLA Health Department of Information Services and Solutions, University of California, Los Angeles, Los Angeles, CA, USA}
\affil[7]{Department of Computational Medicine, University of California, Los Angeles, Los Angeles, CA, USA}

\date{}

\maketitle
\begin{center}
\small
\textsuperscript{*}These authors contributed equally to this work. \\
\textsuperscript{\dag}Corresponding author: \texttt{cwarnold@mednet.ucla.edu}
\end{center}

\textbf{\small
Foundation models trained on longitudinal electronic health records (EHRs) can predict clinical diagnoses years in the future, making them a potential tool for disease screening. We developed TRACER (Temporal Risk Assessment with Clinical EHR Representations), a decoder-based foundation model pretrained on data from over 4.5 million patients from four health systems, and evaluated it on 16 incident diseases at 1-, 2-, and 5-year horizons in more than 2 million patients from two health systems excluded from model development. In contrast to typical evaluations that focus on global metrics (e.g., AUROC), we measured performance at the low false-positive rates that are realistic for screening.  We found that TRACER retained its discrimination in external health systems, and at 95\% specificity, the patients it flagged were up to eight times more likely to be diagnosed within a year than unselected patients. Evaluating at this operating point also revealed what global metrics alone did not: detection and false-alert burden varied across diseases and health systems, even when discrimination appeared stable. Our results illustrate the potential of foundation models for screening and underscore the need for evaluation in health systems excluded from development to assess disease detection and false-alert burden at levels appropriate for intended clinical use.}

\section*{\textbf{Introduction}}
Electronic health records (EHRs) capture longitudinal clinical histories containing signals of disease years before diagnosis \cite{placido2023deep, tang2024leveraging}. Foundation models pretrained on millions of structured clinical events learn patterns from these histories that can be used to estimate patients' risk of developing disease \cite{wornow2023ehrshot, li2020behrt, guo2023ehr, shmatko2025learning}. Common approaches include encoder-only transformers trained to reconstruct hidden clinical codes and decoder-only transformers trained to predict subsequent codes in a patient's sequence \cite{li2020behrt, rasmy2021med, pang2021cehr, steinberg2021language, kraljevic2024foresight, renc2024zero, redekop2025zero, zhang2026multimodal}. Epic's Curiosity illustrates the growing scale of these efforts, using decoder-only transformers pretrained on longitudinal, structured EHR data from Cosmos, a resource that aggregates records from 310 participating health systems \cite{waxler2025generative}. To support screening and early detection, these models must identify future cases while limiting false alerts to a level that clinicians and health systems can manage. In clinical decision support, more frequent and repeated reminders have been associated with alert fatigue, leading to lower clinical acceptance \cite{ancker2017effects}, while perceived EHR alert workload has also been associated with physical fatigue and cognitive weariness, two dimensions of clinical burnout \cite{gregory2017electronic}. 

Evaluations of EHR foundation models commonly emphasize global discrimination metrics, such as the area under the receiver operating characteristic curve (AUROC) and the concordance index (C-index), supplemented by measures of calibration or prediction error \cite{wornow2023ehrshot, guo2023ehr, redekop2025zero, gadd2025survivehr}. Neither measure quantifies how many future cases a model would identify while keeping false alerts manageable nor how often patients flagged as high risk would develop disease. This distinction is especially important for rare outcomes, for which very high specificity might not translate into a high proportion of true cases among patients flagged for review. For example, a retrospective evaluation of 42 case-finding algorithms for 34 rare diseases reported a median specificity of 99.9966\%, but a median positive predictive value (PPV) of 3.1\% after adjustment \cite{boardman2026evaluating}. Although some recent studies report sensitivity at high-specificity thresholds \cite{sun2025intercept}, these measures have rarely been jointly evaluated across multiple diseases, prediction horizons, and held-out health systems. 

Disease detection and false alerts should also be evaluated in health systems excluded from model development, as broad generalization in one clinical setting does not establish clinical usefulness in another  \cite{futoma2020myth}. Prognostic systems can lose calibration while maintaining discrimination \cite{justice1999assessing}, and differences in patient populations and care practices can cause over- or under-prediction \cite{finlayson2021clinician}. Recent benchmarks include subpopulation performance and testing across institutions \cite{jeanselme2025fomoh, guo2024multi, hegselmann2026large}, but comparisons at clinically relevant false-positive rates remain limited in systems excluded from both pretraining and downstream model development \cite{sun2025intercept, guo2023ehr, gadd2025survivehr}. Existing EHR foundation-model studies have evaluated prediction predominantly with the C-index, AUROC, ranking metrics, and Brier-type calibration scores \cite{steinberg2023motor, wornow2023ehrshot, shmatko2025learning, renc2024zero, kraljevic2024foresight, gadd2025survivehr, zhang2026multimodal, jeanselme2025fomoh, guo2024multi, hegselmann2026large}. These metrics summarize ranking across all thresholds and suit comparisons between representations, but they do describe how many future cases a model would identify, or how many patients it would flag, at the false-positive rates a clinical care setting could tolerate. Where operating points have been reported it has been for single diseases \cite{placido2023deep, yang2023transformehr, sun2025intercept}, and not in health systems excluded from both pretraining and downstream model development.  Performance over one year does not establish usefulness over longer horizons, particularly when distinguishing near-term diagnostic assessment from prevention. Comparisons must account for incomplete follow-up rather than assume that patients remain disease-free throughout that horizon, as excluding them  or labeling them as event-free can introduce bias \cite{hagar2014survival, kvamme2019time, lee2018deephit}. Time-to-event (TTE) modeling provides a framework to estimate risk across multiple horizons while accounting for right-censored follow-up \cite{lee2018deephit, uno2011c, kvamme2019time}. MOTOR and SurvivEHR have demonstrated the feasibility of this approach at the scale of EHR foundation models \cite{gadd2025survivehr, steinberg2023motor, zhang2026multimodal}. However, it remains unclear whether models favored by overall survival discrimination also identify future cases at low false-positive rates across prediction horizons.

To address these gaps, we developed TRACER (\textbf{T}emporal \textbf{R}isk \textbf{A}ssessment with \textbf{C}linical \textbf{E}HR \textbf{R}epresentations), a decoder-based foundation model for predicting disease risk over time from longitudinal structured EHRs. We pretrained TRACER on more than 4.5 million patients from four University of California health systems. External validation included more than 2 million patients from two health systems excluded from both pretraining and the training of disease-prediction models. Using a time-to-event approach, we evaluated 16 incident disease endpoints—12 chronic diseases and four rare cancers—at 1-, 2-, and 5-year horizons, with the cancers additionally evaluated at 6 months. We performed a detailed evaluation of TRACER, assessing disease detection, PPV, and case yield under specified false-positive constraints across prediction horizons and health systems. For heart failure we additionally quantified patient-selection rates alongside disease detection, using published cardiac screening to contextualize potential review requirements \cite{yao2021artificial}. Finally, to investigate the impact of different model architectures, we implemented an encoder-based version of the model and compared its performance with TRACER's under matched pretraining data and model scale.

\section{\textbf{Results}} 

\subsection{Study design and evaluation framework}

We evaluated future disease diagnoses using structured EHRs from six independently-operated University of California health systems (Fig. \ref{fig-main-framework}). TRACER was pretrained on more than 4.5 million patient sequences from four development systems (UC1-UC4). Patients held out from pretraining within these systems were used to train disease-prediction heads and for internal evaluation ("in-domain"). UC5 and UC6 were excluded from both pretraining and model development and were used only for external evaluation. Patients with less than one year of available EHR history were excluded. The pretrained model remained unchanged while DeepHit survival heads \cite{lee2018deephit} were trained on labeled in-domain patients.

We evaluated new diagnoses of 16 diseases, defined using SNOMED CT concepts (Methods). The 12 chronic diseases were evaluated at 1-, 2-, and 5-year horizons (52, 104, and 260 weeks), and the four rare cancers were additionally evaluated at 6 months (26 weeks). We maintained natural outcome frequencies rather than sampling to select fixed numbers of cases and controls. Outcome frequencies ranged from approximately 0.14\% to 15.5\% across diseases and evaluation cohorts (Supplementary Table \ref{tab:prevalence}).

Overall discrimination was assessed using cumulative/dynamic area under the receiver operating characteristic curve (AUROC) and Uno's inverse probability of censoring weighted (IPCW) concordance index (C-index) \cite{uno2011c, blanche2013estimating}. AUROC summarizes discrimination between patients diagnosed within a given horizon and those remaining disease-free across the full range of false-positive rates, whereas the C-index assesses whether higher-risk patients tend to be diagnosed sooner \cite{uno2011c, heagerty2005survival}. Standardized partial AUROC (pAUC) summarizes discrimination at false-positive rates up to 10\% \cite{mcclish1989analyzing, dodd2003partial}. 

At thresholds targeting 90\%, 95\%, and 99\% specificity, sensitivity quantified the proportion of future cases identified, PPV described the proportion of flagged patients diagnosed within the horizon, fold enrichment expressed PPV relative to the baseline incidence among all scored patients, and case yield estimated the number of future cases identified per 10,000 patients \cite{zheng2008time}. Estimates accounted for incomplete follow-up using IPCW. Thresholds were estimated separately within each cohort to compare detection at matched nominal false-positive rates. Between-model differences were assessed using paired patient-level bootstrap resampling (Methods).

\subsection{Overall discrimination was largely retained across health systems}

We evaluated TRACER in the UC5 and UC6 health systems without retraining the foundation model or disease-prediction heads. Across the 16 diseases, C-index estimates ranged from 0.624 to 0.879 in-domain, 0.617 to 0.870 at UC5, and 0.651 to 0.880 at UC6 (Supplementary Table \ref{tab:tracer-discrimination}). We examined stability within each disease across both external cohorts rather than treating similar overall ranges as evidence that disease-specific performance was preserved.

The three chronic diseases with the highest in-domain C-indices were Alzheimer's disease, heart failure, and chronic kidney disease. For Alzheimer's disease, the in-domain estimate was 0.879 and the external estimates ranged from 0.870 to 0.880; for chronic kidney disease, the corresponding values were 0.869 and 0.860 to 0.867. Heart-failure estimates spanned 0.836 to 0.869 across the in-domain and two external cohorts. In contrast, celiac disease had the lowest estimated C-index in each cohort (0.617-0.651 across cohorts).

\begin{figure*}[t]
\centering
\includegraphics[width=0.95\linewidth]{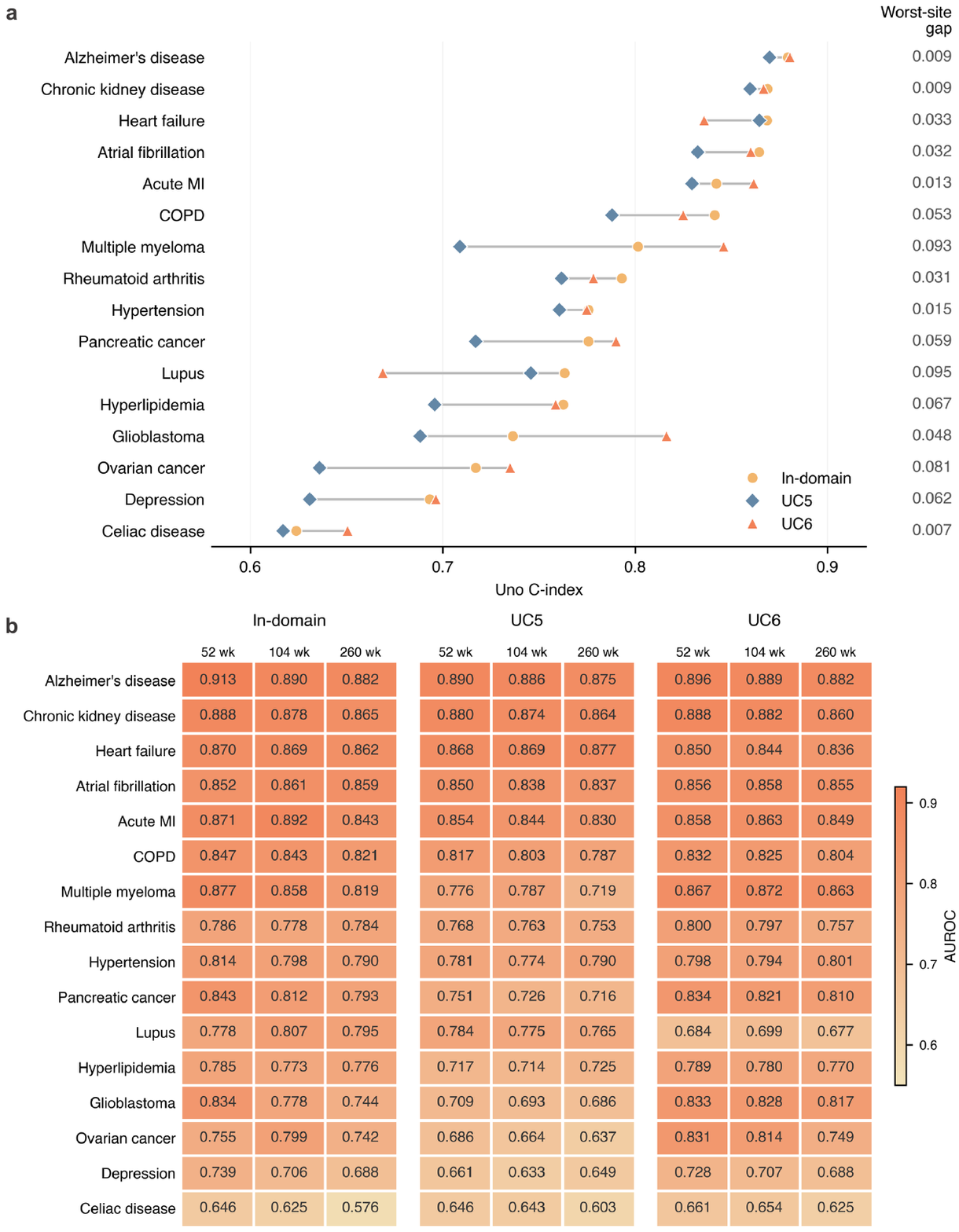}
\caption{\textbf{TRACER Discrimination across health systems and prediction horizons} \textbf{(a)} Uno's IPCW C-index for TRACER in the in-domain test cohort (yellow circles) and the held-out UC5 (blue diamonds) and UC6 (red-orange triangles) cohorts, with grey connectors spanning the lowest to the highest value for each endpoint. The right-hand column lists the worst-site gap, calculated as the in-domain C-index minus the lower of the two external values. \textbf{(b)} TRACER cumulative/dynamic AUROC at the 52-, 104-, and 260-week horizons for each cohort, with point estimates printed in each cell.}
\label{fig-tracer-transport}
\end{figure*}

To assess how closely external performance matched in-domain performance for each disease, we compared the in-domain C-index with the lower of its two external values. We observed comparable model performance on the in-domain and external systems with a median reduction of 0.041 (range 0.007-0.095) (Fig. \ref{fig-tracer-transport}a, Supplementary Table \ref{tab:tracer-discrimination}). The five smallest reductions occurred in chronic diseases, whereas the eight largest occurred in lupus, hyperlipidemia, depression, COPD, and the four cancers. The remaining three chronic diseases had intermediate reductions of 0.031-0.033. The largest reported losses were for lupus (0.095), multiple myeloma (0.093), and ovarian cancer (0.081). 

Stability across prediction horizons was also disease-dependent. In-domain AUROC at five years remained within 0.025 of its one-year value for seven of the 12 chronic diseases and for ovarian cancer. Heart failure provided an example of little change across all three horizons (0.870, 0.869, and 0.862, at 1, 2, and 5 years). The largest chronic-disease decline occurred for depression (0.739 at 1 year to 0.688 at 5 years) and celiac disease (0.646 to 0.576). Among cancers, pancreatic cancer, multiple myeloma, and glioblastoma showed AUROC reductions of 0.050 to 0.091. Similar 1- and 5-year values did not always imply stability throughout the interval; for example, ovarian cancer AUROC increased from 0.755 at one year to 0.799 at two years before decreasing to 0.742 at five years (Fig. \ref{fig-tracer-transport}b, Supplementary Table \ref{tab:tracer-horizons}).

\subsection{Disease detection at low false-positive rates}

\begin{figure*}[t]
\centering
\includegraphics[width=0.92\linewidth]{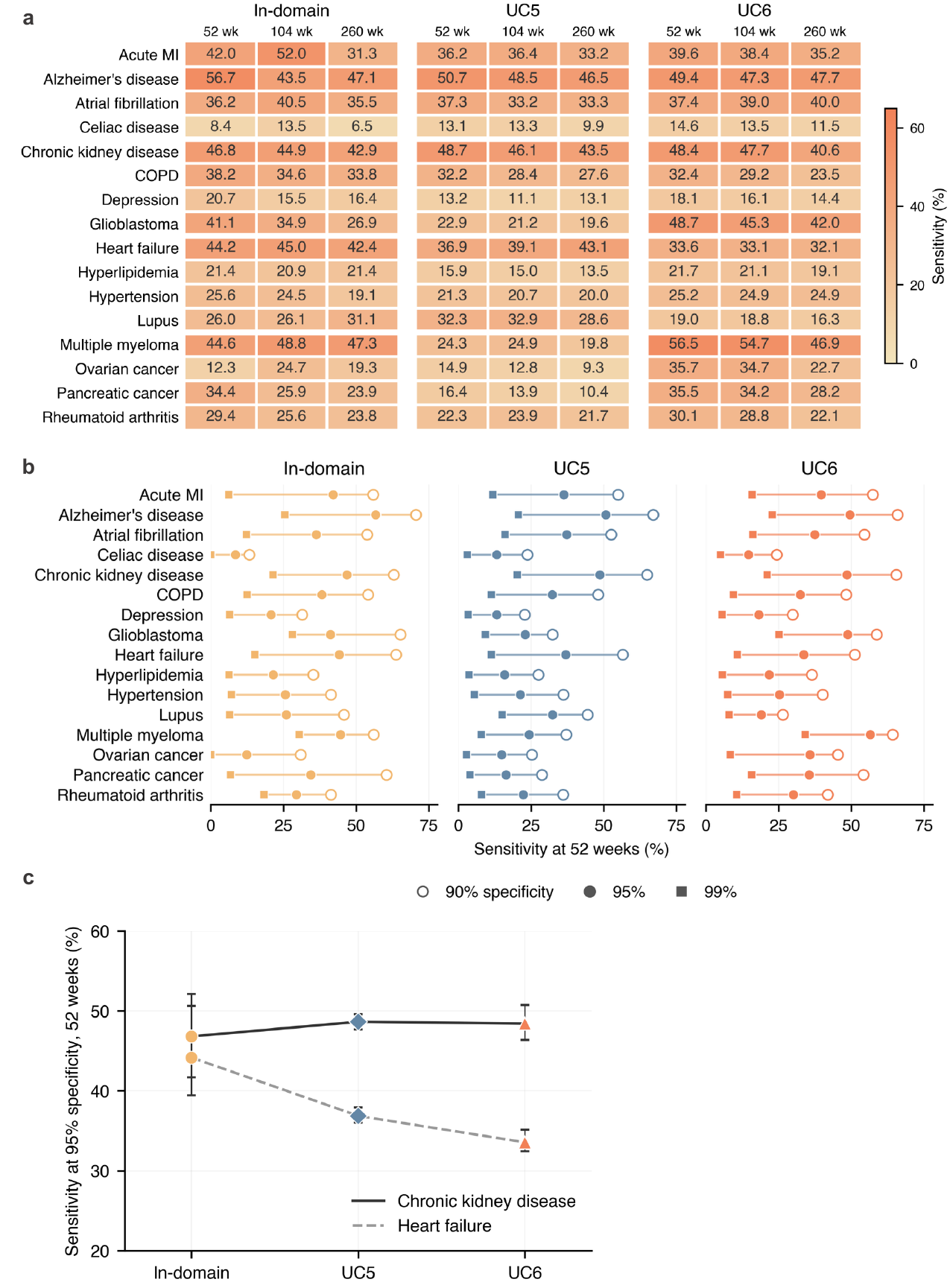}
\caption{\textbf{Disease detection at low false-positive rates.} Sensitivity was calculated for TRACER at cohort-specific thresholds selected to target the stated specificity, using the same censoring-aware case and control definitions as AUROC with IPCW. \textbf{(a)} Sensitivity (\%) at 95\% specificity for each endpoint at the 52-, 104-, and 260-week horizons in the in-domain, UC5, and UC6 cohorts with values printed in each cell. \textbf{(b)} Sensitivity at the 52-week horizon at threshold target 90\% (open circles), 95\% (filled circles), and 99\% specificity (squares), one panel per cohort. \textbf{(c)} Sensitivity at 95\% specificity and 52 weeks for chronic kidney disease (solid line) and heart failure (dashed line) across the three cohorts. Error bars are 95\% confidence intervals from 1,000 patient-level bootstrap replicates. Endpoints in (a) and (b) are listed alphabetically. Marker colors and shapes denote the evaluation cohort (yellow circle = in-domain, blue diamond = UC5, red-orange triangle = UC6).}
\label{fig-auroc-sens}
\end{figure*}

Next, we examined whether strong overall discrimination was accompanied by high sensitivity when false positives were limited. We evaluated sensitivity at cohort-specific thresholds corresponding to 90\%, 95\%, and 99\% specificity (Fig. \ref{fig-auroc-sens}). Sensitivity decreased as the specificity requirement increased, with the magnitude of this decrease varying across endpoints, prediction horizons, and health systems. For heart-failure, an in-domain one-year AUROC of 0.87 corresponded to a sensitivity of 44.2\% at the 95\% specificity threshold. Thus, more than half of future heart-failure cases were not identified at this threshold despite the model's overall high discrimination performance (Supplementary Table \ref{tab:tracer-sensitivity}).

The tradeoff between detection and false positives was evident across diseases. At one year and 90\% specificity in-domain, the four highest estimated sensitivities among the 12 chronic diseases were for Alzheimer’s disease (70.6\%), heart failure (63.7\%), chronic kidney disease (63.0\%), and acute myocardial infarction (55.8\%). Increasing  specificity to 99\% reduced sensitivity for these same diseases to 25.5\%, 15.1\%, 21.4\%, and 6.2\%, respectively. Reducing false positives, therefore, also reduced the proportion of future cases identified (Supplementary Table \ref{tab:tracer-sensitivity}).

To assess whether disease detection was maintained externally, we compared sensitivity at cohort-specific thresholds targeting 95\% specificity. Across 104 comparisons (52 disease-horizon combinations evaluated at each of the two external systems), estimated sensitivity was lower externally in 59 comparisons (56.7\%). The median change from in-domain to external evaluation was a decrease of 1.9 percentage points (Fig. \ref{fig-auroc-sens}a, Supplementary Table \ref{tab:sens-all-horizons-tracer}).

Chronic kidney disease and heart failure results highlight different external trajectories among diseases with high in-domain discrimination. At one year and 95\% specificity, chronic kidney disease sensitivity was 46.8\% in-domain, 48.7\% at UC5, and 48.4\% at UC6. For heart failure, however, sensitivity decreased from 44.2\% in-domain to 36.9\% and 33.6\%, respectively (Fig. \ref{fig-auroc-sens}c, Supplementary Table \ref{tab:tracer-sensitivity}). Detection was therefore close to or above the in-domain estimate at both external systems for one disease, but lower at both for the other.

The five-year heart failure results directly contrast changes in overall discrimination with changes in disease detection. TRACER's AUROC decreased from 0.862 in-domain to 0.836 at UC6, whereas sensitivity at 95\% specificity decreased from 42.4\% to 32.1\%, a loss of 10.3 percentage points (Supplementary Tables \ref{tab:auroc-all-cohorts} and \ref{tab:sens-all-horizons-tracer}). Because thresholds were re-estimated within each cohort, this difference reflects weaker detection at the same nominal false-positive rate rather than simply reflecting a threshold failure carried over from the in-domain systems.

The four rare cancers also showed substantial variation in detection. At one year and 95\% specificity, sensitivity ranged from 12.3\% to 44.6\% in-domain and 14.9\% to 56.5\% across external evaluations. Within these external evaluations, the ranges were 16.4-35.5\% for pancreatic cancer, 14.9-35.7\% for ovarian cancer, 22.9-48.7\% for glioblastoma, and 24.3-56.5\% for multiple myeloma. At 99\% specificity, the corresponding external ranges decreased to 4.0-15.7\%, 2.8-8.3\%, 9.3-25.1\%, and 7.9-34.1\% (Supplementary Table \ref{tab:tracer-sensitivity}). Even at 95\% specificity, these estimates indicate that substantial proportions of future cancer cases were not identified.

\subsection{PPV and case yield quantified potential review burden}

\begin{figure*}[t]
\centering
\includegraphics[width=1.0\linewidth]{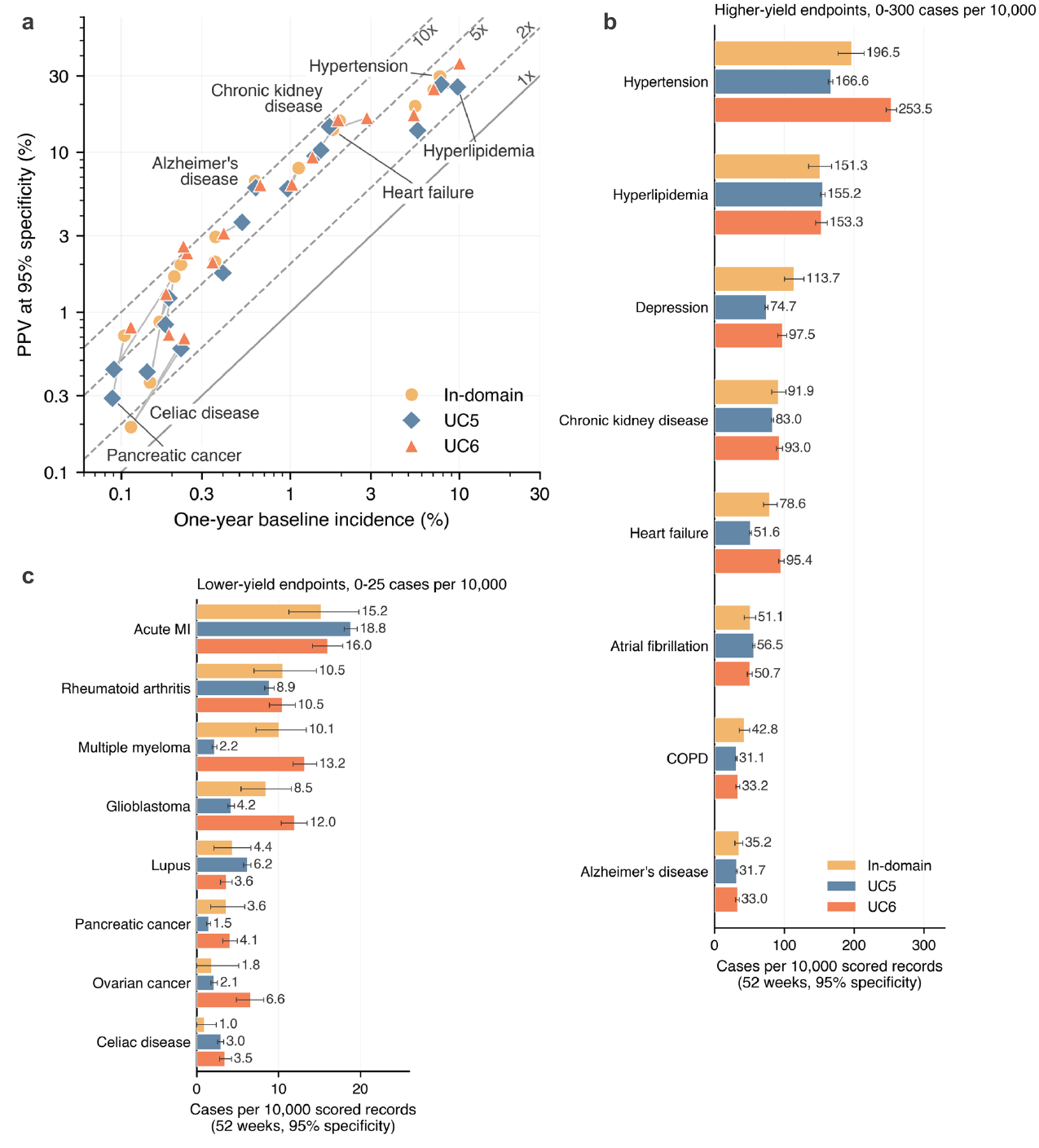}
\caption{\textbf{Positive predictive values and case yield at 95\% specificity over 52 weeks.} For each endpoint and cohort, the TRACER score threshold was selected within the evaluation cohort to attain nominal 95\% specificity, and PPV and case yield were censoring adjusted estimates. \textbf{(a)} PPV against one-year baseline incidence for each endpoint and cohort on logarithmic axes. Grey segments join each endpoint's in-domain point to its UC5 and UC6 points. The solid diagonal marks PPV equal to baseline incidence (no enrichment), and dashed lines mark 2-, 5-, and 10-fold enrichment. The seven endpoints discussed in the text are labeled. \textbf{(b,c)} Incident cases recovered per 10,000 score records at the 52-week horizon, calculated at 10,000 times the IPCW weighted number of true positives divided by the number of scored records. Endpoints with an in-domain yield of at least 20 cases per 10,000 are shown in (b) on a 0-300 scale, and the remaining endpoints in (c) on a 0-25 scale; within each panel, endpoints are ordered by in-domain yield. Error bars are 95\% confidence intervals. Colors denote the evaluation cohort (yellow = in-domain, blue = UC5, red-orange = UC6).}
\label{fig-case-yield}
\end{figure*}

Sensitivity describes the share of all actual future cases that a flag successfully captures, whereas PPV describes the share of flagged patients who go on to be diagnosed, which determines the review effort behind each identified case. Because PPV rises with disease frequency, we report it alongside baseline incidence, fold enrichment, and case yield per 10,000 patients. At one year and thresholds targeting 95\% specificity, hypertension had the highest estimated PPV among the 16 diseases in both external cohorts. PPV was 29.8\% in-domain and 26.5--36.0\% externally, against one-year base rates of 7.69\% in-domain and 7.82-10.06\% externally, a 3.4--3.9-fold enrichment over unselected patients. Case yield was 195.5 and 166.6-253.5 future cases per 10,000 scored records, and external sensitivity was 21.3-25.2\%. Hyperlipidemia had the second-highest external PPV, 24.8--25.6\%, against base rates of 7.06--9.78\%, with a 2.6--3.5-fold enrichment, yields of 153.3-155.2 cases per 10,000, and sensitivity of 15.9-21.7\%. Chronic kidney disease had the highest enrichment among the chronic diseases at 8.0--8.5-fold, with a PPV of 14.5--15.9\% against base rates of 1.71--1.96\%, and both were stable across the three cohorts (Fig. \ref{fig-case-yield}a, Supplementary Table \ref{tab:tracer-ppv-enrichment}).

For heart failure, TRACER selected 5.2-5.7\% of scored patients across the in-domain and two external cohorts. Sensitivity ranged from 33.6\% to 44.2\%, being highest in-domain and lower at both external systems. PPV was 13.80\% in-domain, 9.47\% at UC5, and 16.41\% at UC6, with corresponding case yields of 78.6, 51.6, and 95.4 future cases per 10,000 scored patients. Although sensitivity was lower at both external systems, PPV and case yield were lower at UC5 but higher at UC6 than the in-domain. The higher PPV at UC6 coincided with a higher one-year base rate there, 2.84\% against 1.78\% in-domain and 1.40\% at UC5. Enrichment, which adjusts for base rate, was 7.8-fold in-domain, 6.8-fold at UC5, and 5.8-fold at UC6. PPV and case yield therefore did not change in parallel with sensitivity across health systems, but enrichment did (ig. \ref{fig-case-yield}b, Supplementary Table \ref{tab:tracer-selection-52wk}, Supplementary Table \ref{tab:tracer-ppv-enrichment}).

The distinction between overall discrimination and yield among flagged patients was particularly apparent for pancreatic cancer. In-domain one-year AUROC was 0.843, but PPV at 95\% specificity was 0.71\% (Supplementary Table \ref{tab:tracer-selection-52wk}, Supplementary Table \ref{tab:tracer-ppv-enrichment}). Across the in-domain and two external cohorts, 5.4--5.6\% of scored patients were selected, a similar proportion to heart failure, and sensitivity ranged from 16.4\% to 35.5\%. PPV ranged from 0.29\% to 0.80\% and case yield from 1.45 to 4.05 future cases per 10,000 scored patients. Fewer than 1\% of patients selected for pancreatic cancer review were therefore estimated to receive the diagnosis within one year, against a one-year base rate of 0.09--0.11\%. Enrichment was 6.8-fold in-domain and 7.0-fold at UC6, comparable to heart failure, but only 3.3-fold at UC5, where sensitivity was also lowest. A similar initial review volume for the two diseases accompanied similar relative enrichment in two of the three cohorts, but an approximately twenty-fold difference in absolute PPV. 

Selection counts were direct, whereas sensitivity, PPV, and case yield were censoring-adjusted estimates. Patients censored before one year remained in the selection count and were not automatically classified as false positives. These results describe selection at one prediction visit, not repeated notifications or diagnoses established through an intervention. Patient-selection volume, the proportion of future cases identified, and the yield among selected patients therefore describe different aspects of potential clinical use.

\subsection{Risk signals are captured with different model architectures}

We compared TRACER, a decoder-only model pretrained to predict subsequent clinical codes, with an encoder-only counterpart (TRACER-Encoder) pretrained to reconstruct masked clinical codes. Both received the same ordered sequences of diagnoses, procedures, medications, and laboratory results, including visit boundaries and the time between visits. The pretraining population, vocabulary, tokenizer, sequence construction, model scale, and disease-prediction protocols were matched (Methods). Both pretrained models remained fixed while endpoint-specific DeepHit survival heads \cite{lee2018deephit} were trained on labeled in-domain patients and applied unchanged to the external cohorts.

\begin{figure*}[t]
\centering
\includegraphics[width=0.95\linewidth]{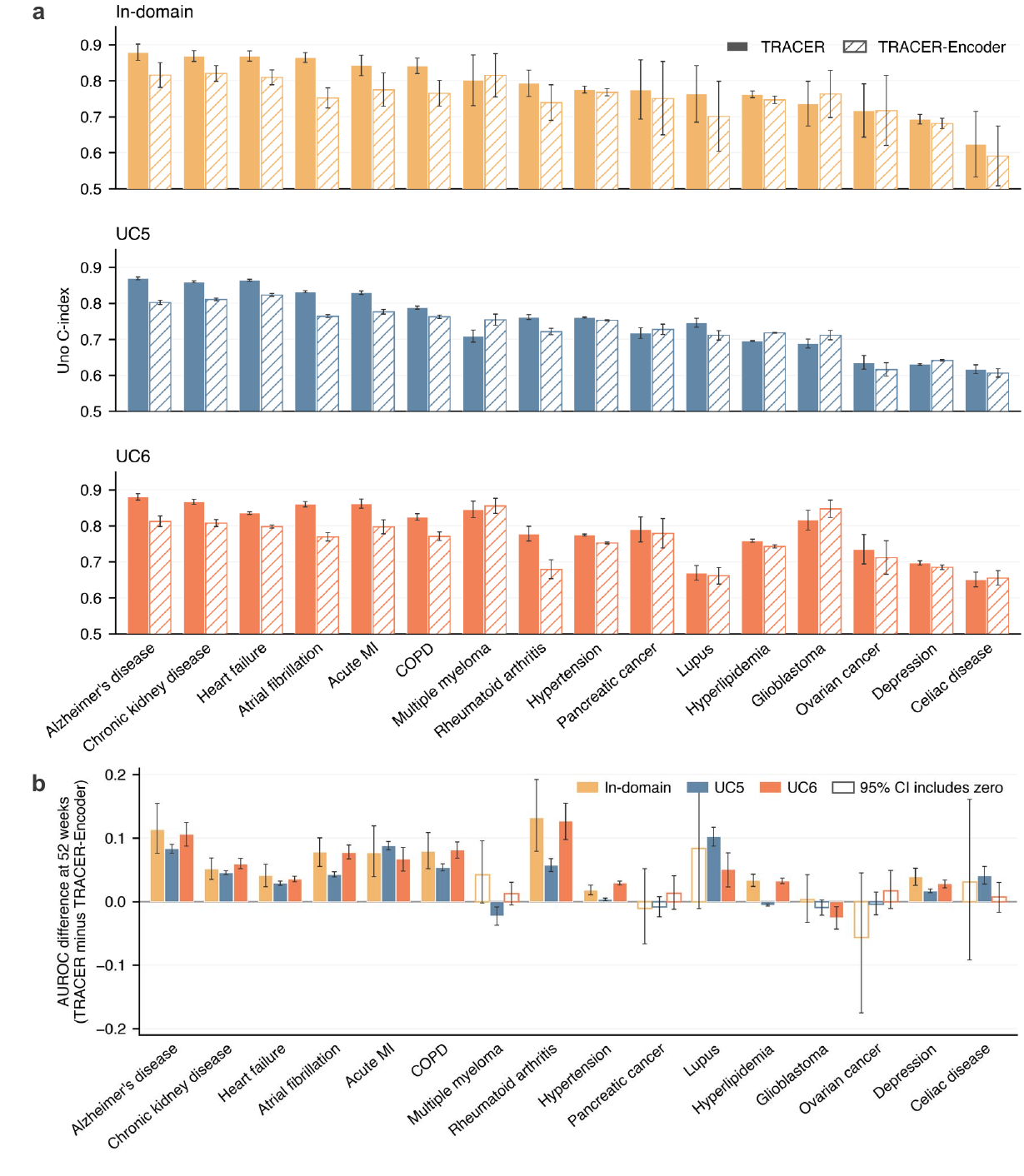}

\caption{\small{\textbf{Overall discrimination of TRACER and TRACER-Encoder across 16 disease endpoints and three health systems.}  Both pretrained models remained fixed while endpoint-specific DeepHit survival heads were trained on labeled in-domain patients and applied unchanged to the held-out UC5 and UC6 cohorts. Endpoints are ordered by TRACER in-domain C-index in both panels. \textbf{(a)} Uno's IPCW C-index for TRACER (solid bars) and TRACER-Encoder (hatched bars) in the in-domain test cohort, UC5, and UC6. Error bars are model-specific 95\% confidence intervals. \textbf{(b)} Paired difference in cumulative/dynamic AUROC at the 52-week horizon, calculated as TRACER minus TRACER-Encoder, for each endpoint and cohort. Positive values favor TRACER and negative values favor TRACER-Encoder. Error bars are 95\% confidence intervals from 1,000 paired patient-level bootstrap replicates in which the same resampled patients were applied to both models; bars who intervals includes zero are drawn in outline only. Colors denote the evaluation cohort (yellow = in-domain, blue = UC5, red-orange = UC6.}}
\label{fig-c-index}
\end{figure*}

Both models captured risk signals across the 16 diseases. TRACER generally achieved better overall discrimination for chronic diseases in the in-domain cohort and the two external health systems (Fig. \ref{fig-c-index}, Supplementary Table \ref{tab:cindex-by-site}). At one year, TRACER's AUROC advantage was statistically supported for 10 of 12 chronic diseases in-domain, with differences ranging from +0.018 for hypertension to +0.132 for rheumatoid arthritis (Fig. \ref{fig-c-index}, Supplementary Table \ref{tab:auroc-difference-52wk}). However, overall and low-false-positive performance did not always favor the same model. Across the 144 comparisons (16 diseases evaluated at 1-, 2-, and 5-year horizons in the in-domain cohort and two external cohorts), AUROC and partial AUROC favored opposite models with statistical support in 15 comparisons (10.4\%), all in the external cohorts (Supplementary Table \ref{tab:model-choice}). In 14 of these, AUROC favored TRACER, whereas partial AUROC favored TRACER-Encoder (Supplementary Results, Section \ref{sec:model-comparison-details}). Complete results, paired comparisons, and analyses of model-ranking disagreements are provided in the Supplementary Results.

\section*{\textbf{Discussion}} 
We developed TRACER to predict future disease diagnoses from longitudinal structured EHRs and evaluated its performance across 16 diseases and multiple prediction horizons, including external validation on two health systems held out from model development. The central question was not only how accurately a model ranked patients by risk, but how many future diagnoses it identified when false positives were restricted. Alongside AUROC and C-index, which summarize global discrimination, we therefore assessed sensitivity, PPV, enrichment over base rate, and case yield at matched specificity. Strong discrimination for several diseases coexisted with limited sensitivity or PPV at that operating point, and discrimination that was relatively preserved across external systems did not guarantee that sensitivity or PPV was preserved with it. TRACER's contribution is, therefore, a framework for risk prediction and a rigorous external validation of what its predictions identify, what they miss, and what review burden they might impose if deployed.

Absolute performance at a clinical operating point cannot be inferred from AUROC or C-index alone. TRACER's in-domain one-year heart-failure AUROC was 0.870. At 95\% specificity, selecting 5.2--5.7\% of scored patients identified 33.6--44.2\% of future cases across cohorts, with an in-domain PPV of 13.8\% against a one-year base rate of 1.78\%, a 7.8-fold enrichment. For pancreatic cancer, in-domain one-year AUROC was 0.843, and a comparable selection rate of 5.4--5.6\% identified 16.4--35.5\% of future cases, with in-domain PPV of 0.71\% against a base rate of 0.10\%, a 6.8-fold enrichment. The two diseases were flagged at nearly the same rate and enriched to a similar degree, yet their PPVs differed substantially due to their base rate difference. A pancreatic cancer flag therefore carried a relative enrichment similar to a heart failure flag but a far lower absolute probability of diagnosis.

These results do not establish that either use would be clinically worthwhile or unusable, but they highlight that their value depends on the assessment a flag prompts and on the consequences of testing or a missed case. They identify where improvement in EHR foundation models would matter most: sensitivity and yield at a fixed false-positive rate. By the metrics that EHR foundation-model studies usually report, TRACER transferred well, with a median C-index reduction of 0.041 across the two external systems. The operating-point evaluation showed that this stability did not extend to detection, PPV, or case yield, which varied by disease and health system in ways that the global discrimination metrics did not signal. A complete evaluation of EHR foundation models should therefore pair the C-index and AUROC with sensitivity, PPV, and yield at prespecified clinically relevant specificities in held-out systems.

External performance stability is separate from absolute performance, and a small change in one metric can conceal a large change in another. Five-year heart failure AUROC changed from 0.862 in-domain to 0.836 at UC6, while sensitivity at 95\% specificity decreased by 10.3 percentage points. At one year, heart failure PPV rose from 13.8\% in-domain to 16.4\% at UC6, even though sensitivity fell because the one-year base rate at UC6 was 2.84\% compared with 1.78\% in-domain. Enrichment, which removes the base-rate dependence, fell from 7.8-fold to 5.8-fold. For pancreatic cancer at UC5, by contrast, base rate was nearly unchanged while PPV fell from 0.71\% to 0.29\% and enrichment from 6.8-fold to 3.3-fold, a loss of operating-point performance rather than a shift in prevalence. A change in PPV should therefore be interpreted alongside base rate rather than read directly as a change in discrimination. Conversely, one-year chronic kidney disease sensitivity and enrichment, 8.0--8.5-fold, were maintained at both external systems. This highlights that each measure must be evaluated for the intended disease, horizon, and population. 

Performance across prediction horizons also varied by disease, reinforcing the need to evaluate the intended use rather than extrapolate from a single near-term result. Some cardiometabolic and renal diseases retained discrimination over several years, whereas other endpoints declined. Differences in how consistently prediagnostic information is recorded offer one explanation, but supplementary analyses of patient records, local survival-head training, and model attribution did not identify a clear cause for the external shifts (Supplementary Results).

Published screening studies offer context for selection rates, although populations and targets differ. TRACER's one-year heart failure selection rates of 5.2-5.7\% were similar to the approximately 6\% AI-ECG positivity rate in the EAGLE study, which assessed new low-ejection fraction diagnoses within 90 days among adults without prior heart failure undergoing routine electrocardiography \cite{yao2021artificial}. A selection rate of this magnitude has therefore been delivered prospectively in practice.

Selection rates can also be set against existing clinical workload. Tai-Seale et al. reported a mean of 114 system-generated inbox messages per physician per week, with above-average volume associated with a 40\% higher probability of burnout \cite{tai2019physicians}. To contextualize TRACER's selection rates against this existing workload, we estimated the additional weekly messages they might generate under an illustrative delivery scenario. Assuming 2,000 patients are assessed annually per physician \cite{bazemore2025physician} for one disease and one initial notification per selected patient, a representative 5.5\% selection rate would generate 110 messages annually, or approximately 2.1 per week if distributed across the year. This increment represents approximately 1.9\% of the 114 weekly system-generated messages reported \cite{tai2019physicians}. However, the 114-message mean is a descriptive statistic rather than a validated burnout threshold, and burnout is subjective and multifactorial \cite{ancker2017effects, gregory2017electronic}. This scenario also fixes panel size and delivery frequency and excludes follow-up and repeated or overlapping alerts.

Our study has several limitations. TRACER used structured EHR events and did not incorporate free-text notes, imaging, pathology, genomics, or other modalities that can contain important predictive information for selected diseases. The UC Data Warehouse does not include non-UC encounter data available through Care Everywhere. Because UC sites often serve as referral centers, patients can receive care across multiple health systems, potentially limiting the completeness of longitudinal clinical histories and outcome ascertainment within the warehouse. Disease onset was defined by the first recorded diagnosis concept and therefore represents documented rather than biological onset. We counteracted this drawback of the data by requiring at least one year of observed history within the EHR in order for a patient to be included in downstream evaluation. Prediction visits were sampled without an explicit washout period, so short-horizon performance may partly reflect the diagnostic work-up preceding diagnosis. 

Matching pretraining procedures reduced several major sources of confounding but did not isolate the effects of architecture, pretraining objective, pooling, optimization, or their interactions. Primary analyses used frozen representations with separately trained survival heads; end-to-end adaptation or different downstream heads could alter both global and operating-region rankings. Cohort-specific thresholds were estimated retrospectively, enabling matched-specificity comparison, but prospective evaluation is needed to assess threshold transport, calibration, costs, and patient benefit. Direct selection counts describe scored evaluation patients and do not establish complete coverage of all clinically eligible patients. We did not measure clinician-level message exposure, responses, or burnout. 

In summary, TRACER demonstrates disease-risk prediction from longitudinal EHRs in external health systems while making its potential screening trade-offs explicit. Strong overall discrimination did not guarantee high sensitivity or PPV at specified false-positive rates, and relatively preserved discrimination did not ensure stable disease detection or yield across settings. Joint assessment of sensitivity, PPV, enrichment, case yield, and potential review volume is necessary to understand how EHR foundation models can begin to be translated into clinical practice.

\section{\textbf{Methods}} 
\subsection{Study Setting and Data Sources}
Data for this study were obtained from the six University of California (UC) academic health centers, labeled UC1-UC6. All records were accessed through the University of California Data Discovery Platform (UCDDP), providing a common data structure across all six systems. Five of these centers are full health systems that include one or more hospitals and an outpatient network, while one is an outpatient-only system. These six centers were divided into two groups. Four centers (UC1-UC4) comprised an "in-domain" dataset used for foundation-model pretraining and downstream survival-head training. The in-domain set includes the smaller outpatient-only network. Two centers (UC5 and UC6) were held out entirely from pretraining and survival-head training and were used only for external validation. Although all six health systems fall under the University of California health network, each system covers a distinct patient population, has its own EHR, and operates independently.

Records from UC1-UC4 were partitioned into non-overlapping pretraining and downstream datasets. The larger partition contained 4,514,400 patient sequences used for foundation model pretraining. The remaining 501,307 patient sequences were reserved for downstream time-to-event (TTE) model development and in-domain evaluation. This downstream partition was further divided into training, validation, and test sets.

We evaluated incident disease-onset prediction for 16 diseases (see Disease endpoints and cohort construction). For each endpoint downstream survival heads were trained using the in-domain training data, selected using the validation data, and evaluated on the fixed in-domain test cohort and the complete held-out cohorts from UC5 and UC6.

Because this study used only retrospective, fully de-identified records, it was determined to be exempt from institutional review board review under 45 CFR \S46.104(d)(4), and informed consent was not required.

\subsection{Data Representation}
\label{sec:data_rep}
An individual’s EHR was represented as a temporally ordered sequence of clinical events structured into visits (see Fig.~\ref{fig-main-framework}). Clinical events were mapped to standard vocabularies, including diagnoses coded using the Systematized Nomenclature of Medicine Clinical Terms (SNOMED CT), procedures coded using Current Procedural Terminology (CPT), medications coded using Prescription Normalized Nomenclature (RxNorm), and laboratory tests coded using Logical Observation Identifiers Names and Codes (LOINC).  Encounters were separated by a \texttt{[SEP]} token, and each encounter was preceded by a discretized time-gap token representing the interval since the prior visit, following \cite{redekop2025zero}. Care-site tokens encoding the location within the health system at which an encounter occurred were also included in the input sequence. As a preprocessing step, the first two visits of each sequence were concatenated into a single visit group prior to sequence construction. Event ordering within each individual's trajectory was preserved using positional embeddings over the full token sequence. 

Numeric laboratory values were discretized into quantile-based tokens using boundaries estimated from the foundation-model pretraining corpus \cite{bellamy2023labrador}. The same boundaries were subsequently applied without re-estimation to the downstream in-domain and held-out health-system cohorts. The resulting representation therefore combined clinical concepts, discretized laboratory values, encounter structure, and elapsed time within a single longitudinal token sequence. 

The same tokenized sequences were provided to both foundation-model backbones. For downstream time-to-event modeling, each backbone remained frozen, and patient-level representations were computed from its final-layer hidden states.

Padding and model-control tokens without patient-specific clinical or temporal information were excluded from the pooling step. Unless otherwise specified, all remaining clinical and structural token representations were mean-pooled. Let $\mathcal{K}_i$ denote the set of token positions included for patient $i$, with $m_i = |\mathcal{K}_i|$. The patient representation was:

$$z_i = \frac{1}{m_i}\sum_{k \in \mathcal{K}_i} h_{ik}^{(L)},$$

where $h_{ik}^{(L)}$ is the final-layer representation of token $k$ for patient $i$. Pooling and structural-token ablations are described in the Supplementary Results (Supplementary Table \ref{tab:supp-pooling}).

\subsection{TRACER Foundation Model Implementations}
We implemented two approximately 125-million-parameter foundation models that differed in architecture, attention pattern, and pretraining objective. TRACER-Encoder used a bidirectional ModernBERT-style encoder trained with masked-token prediction \cite{warner2025smarter}, whereas TRACER used a LLaMA-style causal decoder trained with next-token prediction \cite{grattafiori2024llama}. The pretraining corpus, tokenizer, sequence construction, model scale, and downstream survival protocol were matched across implementations. Full architecture and pretraining configurations for both implementations are provided in Supplementary Methods.

\subsubsection{TRACER}
TRACER used a decoder-only LLaMA-style transformer architecture with causal self-attention \cite{grattafiori2024llama}. The causal attention mask restricts the hidden state at position $t$ to information from positions $1,...,t$, preventing access to subsequent tokens in the sequence (Fig. \ref{fig-main-framework}). For a sequence $x_{1:n}$, the model specifies the conditional probability of the next token $x_{t+1}:$
\begin{equation}
    p(x_{t+1}|x_{1}, ..., x_{t})
\end{equation}
This architecture mirrors large autoregressive language models and naturally aligns with sequence generation and forecasting tasks. The model was pretrained using a causal next-token prediction (NTP) objective:
\begin{equation}
\mathcal{L}_{\mathrm{NTP}}(\theta_{\text{D}})
= - \sum_{t=1}^{n-1} \log p\!\left( x_{t+1} \mid x_1, \ldots, x_t \right),
\end{equation}

\noindent where $\theta_{\text{D}}$ denotes the TRACER parameters. Unlike the encoder objective, which reconstructs selected tokens using bidirectional context, the decoder objective learns each conditional distribution using only the temporally preceding sequence context.

\subsubsection{TRACER-Encoder}
\begin{figure*}[t]
\centering
\includegraphics[width=0.95\linewidth]{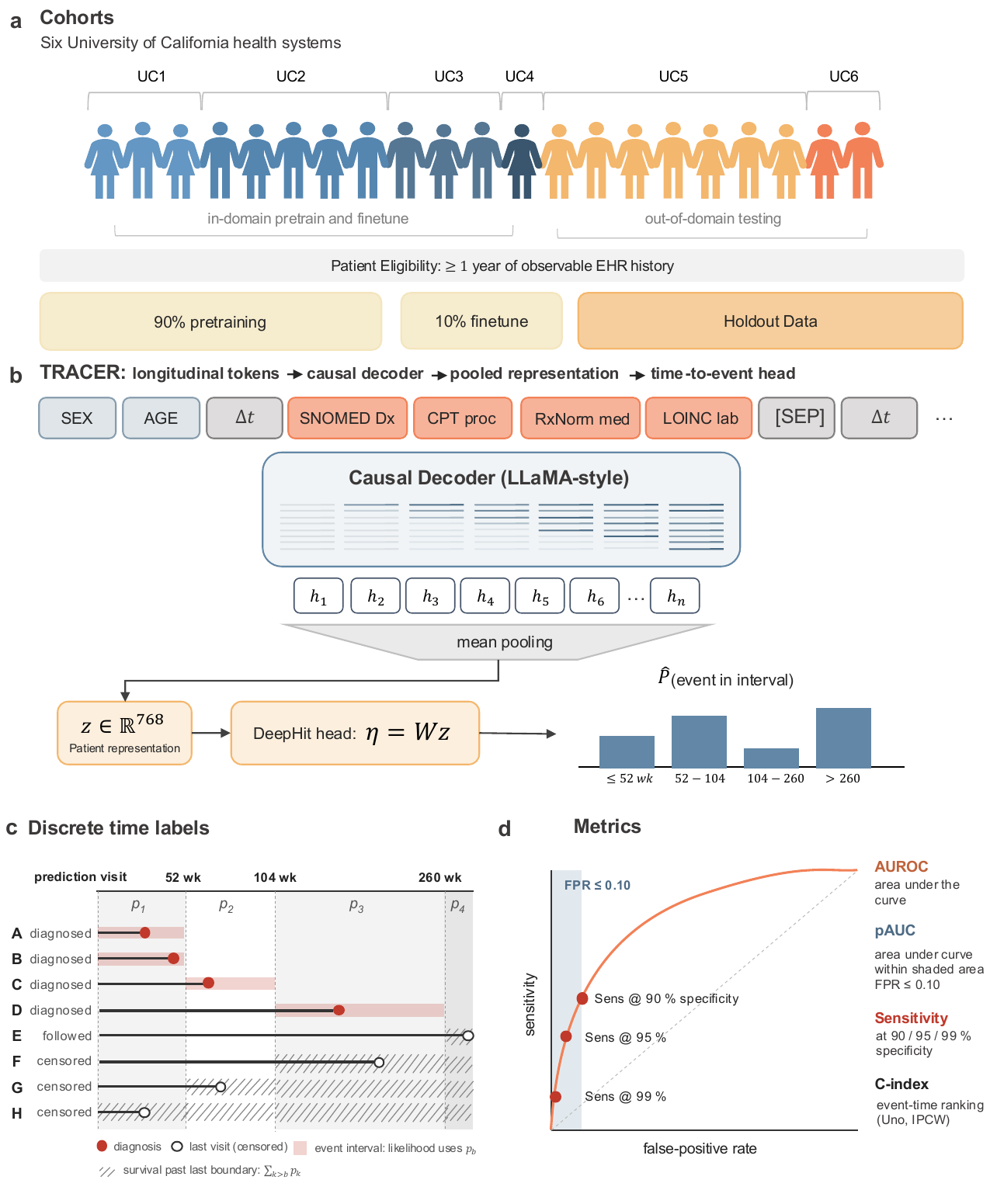}
\caption{\textbf{Multi-center EHR foundation modeling and time-to-event prediction framework.}
\textbf{(a)} Cohort allocation across six University of California (UC) health systems. Patients with less than one year of observable EHR history were excluded. UC1-UC4 provided the pretraining data and the held-out in-domain patients used to train and evaluate disease-prediction heads; UC5 and UC6 were excluded from both and used only for external evaluation.
\textbf{(b)} TRACER modeling pipeline. Each patient's history is a temporally ordered sequence of clinical tokens processed by a causal decoder-only transformer pretrained to predict the next token. \textbf{(c)} Discrete-time labels. Each patient contributes one interval index to the DeepHit likelihood. \textbf{(d)} Evaluation at each horizon. Cases are patients diagnosed by the horizon and controls are patients followed event-free beyond it with inverse probability of censoring weights (IPCW) for earlier censoring. Metrics include cumulative/dynamic AUROC, the standardized partial AUROC over false-positive rates 0-10|\% (shaded), and sensitivity, PPV, and case yield at cohort-specific thresholds targeting 90\%, 95\%, and 99\% specificity (points). Uno's IPCW C-index summarizes event-time ranking across all horizons.}
\label{fig-main-framework}
\end{figure*}

TRACER-Encoder used an encoder-only ModernBERT-style transformer architecture\cite{warner2025smarter}. Given a token sequence
$$x_{1:n} = (x_1,...,x_n)$$

\noindent we sampled a set of token positions

$$\mathcal{M} \subset \{1, ...,n\}$$

\noindent and constructed a corrupted sequence $\tilde{x}_{1:n}$ according to the masking procedure. The model was trained to reconstruct the original tokens at those positions:

$$\mathcal{L}_{\text{MLM}}(\theta_{\text{E}}) = -\frac{1}{|\mathcal{M}|}\sum_{t\in \mathcal{M}}\log p_{\theta_{\text{E}}}(x_t | \tilde{x}_{1:n}),
$$
where $\theta_{\text{E}}$ denotes the TRACER-Encoder parameters. Bidirectional self-attention allows the prediction at a masked position to use unmasked clinical events occurring before and after that position within the observed patient history (Fig. \ref{fig-main-framework}).

\subsection{Disease endpoints and cohort construction}
The analysis included 16 disease-onset endpoints: twelve chronic conditions (heart failure, acute myocardial infarction, lupus, hypertension, celiac disease, hyperlipidemia, depression, chronic kidney disease, atrial fibrillation, chronic obstructive pulmonary disease, Alzheimer's disease, and rheumatoid arthritis) and four rare cancers (pancreatic cancer, ovarian cancer, glioblastoma, and multiple myeloma). Ovarian cancer cohorts were restricted to female patients; no endpoint used case-control sampling, so natural outcome frequencies were retained.

Each endpoint was represented by a prespecified SNOMED CT concept set containing the index concept and including descendant or equivalent concepts. Disease onset was defined as the first visit containing any concept in the endpoint-specific set.  Patients without the endpoint code were treated as event-free through their final observed visit and were right-censored at that time. 

For each endpoint, patients were required to have at least one year between the start of the observed history and the event visit or final observed visit. Patients not satisfying this requirement were excluded. For each retained patient, we uniformly sampled one prediction visit from visits occurring strictly more than one year after the start of the record and strictly before the disease-onset or censoring visit. The date of this visit defined the prediction landmark, but the visit itself was excluded from model input. Each patient contributed one prediction example to each endpoint cohort. The observed follow-up time $T_i$ was defined as the number of weeks from the prediction visit to the first qualifying endpoint-coded visit for patients with an event or to the final observed visit for censored patients. 

\subsection{Downstream discrete-time survival modeling}
Foundation-model parameters remained frozen during all downstream survival analyses. For each foundation model and endpoint, a single-risk DeepHit head mapped the pooled patient representation $z_i \in \mathbb{R}^d$ ($d$ = 768) to a discrete probability distribution over event time. The head consisted of a single bias-free linear layer with no hidden layers. Let $K$ denote the number of finite event-time intervals and let 

$$\eta_i = Wz_i, \quad W\in \mathbb{R}^{K\times d}$$

\noindent denote the corresponding interval logits. Following the PMF parameterization implemented in \cite{kvamme2019time}, the $K$ logits were augmented with a fixed zero logit and normalized using the softmax function:

$$\widehat{p}_i=\text{softmax}([\eta_i^{T}, O]^{T}).$$

\noindent The resulting vector
$$\widehat{p}_i = (\widehat{p}_{i,1},...,\widehat{p}_{i,K},\widehat{p}_{i,K+1})$$

\noindent contains $K+1$ components: one per evaluated interval and a terminal component carrying the residual survival mass beyond the last interval boundary. For the 12 chronic-disease endpoints, continuous follow-up was discretized using boundaries at 52, 104, and 260 weeks, with follow-up beyond 260 weeks assigned to a terminal interval. The cancer endpoints additionally included a 26-week boundary. 

Accordingly, chronic disease metrics were evaluated at 52, 104, and 260 weeks, whereas cancer metrics were evaluated at 26, 52, 104, and 260 weeks. 

Let $\delta_i \in \{0, 1\}$ denote the event indicator and $b_i$ the discretized event or censoring index. For patients with an observed event, $b_i$ identifies the interval containing the event. For censored patients, $b_i$ identifies the last interval boundary through which the patient was observed to remain event-free. The discrete-time negative log-likelihood for a mini-batch of size $B$ is

$$\mathcal{L}_{\text{NLL}}=-\frac{1}{B}\sum_{i=1}^{B}\left[ \delta_i \log\widehat{p}_{i,b_i}+(1-\delta_i)\log \left( \sum_{k=b_i+1}^{K+1}\widehat{p}_{ik}\right) \right].$$

The event contribution is the predicted probability mass assigned to the observed event interval. The censoring contribution is the predicted probability of surviving beyond the discretized censoring time. Because the summation includes the terminal component, patients censored at the final evaluated boundary retain positive probability of survival beyond 260 weeks.

A pairwise ranking term encouraged earlier observed events to receive higher predicted cumulative risk. Following the DeepHit formulation as implemented in \texttt{pycox}, the ranking loss applied an exponential penalty based on cumulative incidence at the earlier event time. We defined the predicted cumulative incidence through interval $r$ as

$$\widehat{F}_i(\tau_r) = \sum_{k=1}^r\widehat{p}_{ik}$$

\noindent where $\tau_r$ is the upper boundary of interval $r$. Within each mini-batch, the comparability indicator was

$$A_{ij} = \mathbb{I}\{\delta_i = 1 \text{ and } (b_i < b_j \text{ or } (b_i = b_j \text{ and } \delta_i = 0))\}  $$

\noindent Thus, $A_{ij}$ equals 1 when patient $i$ has an observed event before patient $j$, including the discretized-time case in which patient $i$ had an event and patient $j$ was censored in the same interval. The ranking loss is

$$\mathcal{L}_{\text{rank}} = \frac{1}{B^2} \sum_{i=1}^B\sum_{j=1}^B A_{ij} \exp \left[ -\frac{\widehat{F}_{i}(\tau_{b_i}) - \widehat{F}_j(\tau_{b_i})} {\sigma} \right]$$

\noindent  where noncomparable pairs had $A_{ij} = 0$ and therefore contributed zero. We used $\sigma = 0.1$ under the \texttt{pycox} convention. The total DeepHit loss was

$$\mathcal{L}_{\text{total}} = \alpha \mathcal{L}_{\text{NLL}} + (1-\alpha)\mathcal{L}_{\text{rank}}$$

We used $\alpha = 0.1$. Survival heads were optimized using Adam (learning rate $10^{-3}$) with a batch size of 256 for up to 100 epochs with early stopping based on validation loss (patience, 10 epochs). 

For the primary analysis, survival heads trained on the in-domain data from UC1-UC4 were applied unchanged to the UC5 and UC6 cohorts. The foundation model backbones remained frozen.

\subsection{Evaluation Metrics}
Let $T_i$ denote the observed time for patient $i$ (the minimum of the event time and the censoring time), let $\delta_i \in \{0,1\}$ denote the event indicator, and let $\widehat{S}_i(t)$ and $\widehat{F}_i(t) = 1-\widehat{S}_i(t)$ denote the predicted survival probability and cumulative incidence, respectively, at horizon $t$. The censoring survival function was estimated using the Kaplan-Meier estimator and is denoted by $\widehat{G}(t)$.

\subsubsection{Global discrimination and probabilistic performance}
The primary global survival-discrimination metric was Uno's IPCW C-index \cite{uno2011c}. For each endpoint, the C-index was truncated at the longest evaluated horizon, ($\tau = 260$ weeks), and calculated from predicted cumulative incidence $\widehat{F}_i(\tau)$ as the scalar risk score. Higher values indicate better ordering of patients by event risk. Model-specific 95\% C-index confidence intervals were obtained using an analytic influence-function variance estimator rather than bootstrap resampling.

We additionally calculated the cumulative/dynamic AUROC at each prediction horizon \cite{heagerty2005survival, blanche2013estimating}. At horizon $t$ cumulative cases were patients with an observed event by $t$, $\delta_i =1$ and $T_i \leq t$, whereas dynamic controls were patients known to remain event-free beyond $t$, $T_i > t$. Patients censored before $t$ contributed to neither group. AUROC was calculated using the horizon-specific cumulative-risk score $\widehat{F}_i(t)$, with case contributions weighted by $1/\widehat{G}(T_i^-)$. Controls share the common weight $1/\widehat{G}(t)$, which cancels when the control distribution is normalized. This estimator quantifies the probability that a randomly selected cumulative case receives a higher predicted risk than a randomly selected dynamic control while accounting for right censoring.

Probabilistic performance was evaluated using the integrated Brier score (IBS) \cite{graf1999assessment} and the integrated negative binomial log-likelihood (IBLL), as implemented in \texttt{pycox} \cite{kvamme2019time}. Lower IBS and IBLL values indicate better probabilistic performance (Supplementary Results) .

\subsubsection{Low-false-positive discrimination and fixed operating points}
Global AUROC includes false-positive rates that can be irrelevant for screening or clinical alerting. We therefore calculated raw and standardized partial AUROC over FPR $\in [0, 0.10],$ corresponding to specificity from 90\% to 100\%. At each horizon $t$, the raw partial AUROC was calculated from the IPCW cumulative/dynamic ROC curve as

$$\text{pAUC}(t;a) = \int_{0}^a \widehat{TPR}_t(u)du,$$

\noindent where $u$ denotes the false positive rate. The standardized partial AUROC \cite{mcclish1989analyzing} was

$$\text{pAUC}_{\text{std}}(t;a) = \frac{1}{2}\left[ 1 + \frac{\text{pAUC(t;a)} - a^2/2}{a-a^2/2} \right],$$

\noindent where $a^2/2$ is the area under the chance diagonal and $a$ is the maximum attainable area over the interval $[0,a]$. A standardized value of 0.5 corresponds to chance discrimination within the restricted region, and 1.0 corresponds to perfect discrimination.

Sensitivity was also evaluated at cohort-specific thresholds selected to achieve nominal specificity of 90\%, 95\%, and 99\%. At each horizon, the threshold for target specificity $s$ was selected from the control-score distribution, and sensitivity was then calculated among horizon-specific cases using the same censoring-aware case and control definitions as AUROC.

\subsubsection{Positive predictive value, case yield, and fold enrichment}
Time-dependent PPV quantified the proportion of records above a risk threshold that were associated with a diagnosis by prediction horizon $t$\cite{zheng2008time}.  For each model, disease, horizon, and evaluation cohort, PPV was evaluated at the same cohort-specific thresholds used for sensitivity, targeting 90\%, 95\%, and 99\% specificity. Cases and controls used the same horizon-specific definitions and IPCW scheme as the discrimination analyses. Let $P_w(t)$ denote the total IPCW-weighted case count, and let $\text{TP}_w(t,c)$ and $\text{FP}_w(t,c)$ denote the weighted case and control counts, respectively, among records with predicted risk $\widehat{F}_i(t)\geq c$. Then
$$\widehat{\text{PPV}}(t,c) = \frac{\text{TP}_w(t,c)}{\text{TP}_w(t,c)+\text{FP}_w(t,c)}, \quad \text{TP}_w(t,c)=\widehat{\text{Sensitivity}}(t,c)P_w(t)$$

For an evaluation cohort containing $N$ scored patient records, baseline incidence and case yield were defined as
$$\widehat{\pi}(t)=\frac{P_w(t)}{N}, \quad \widehat{\text{Case Yield}}(t,c)=10,000\frac{\text{TP}_w(t,c)}{N}$$

The denominator included all scored records in the disease-specific evaluation cohort, including those with incomplete follow-up. Weighted false-positive counts were normalized using the denominator. Case yield, therefore, represented the estimated number of future cases identified per 10,000 scored records, not the number selected for review. Fold enrichment was defined as $\widehat{\text{PPV}}(t,c)/\widehat{\pi}(t)$, the ratio of PPV to baseline incidence.

For heart failure and pancreatic cancer, we counted patients whose predicted risk was at or above each saved cohort-specific threshold targeting 90\%, 95\%, and 99\% specificity. Selection rates used the full scored evaluation cohort as the denominator, including patients censored before the prediction horizon. 

\subsection{Statistical analysis}
Except where an analytic variance estimator was available, 95\% confidence intervals were estimated using $B$ = 1,000 patient-level bootstrap replicates. Each endpoint cohort contained one prediction visit per patient; patients were, therefore, sampled with replacement as the resampling unit, with each resample containing the same number of patients as the original evaluation cohort. Censoring weights were estimated once per evaluation cell with the Kaplan-Meier censoring estimator and carried through resampling as fixed patient-level weights. Fixed-specificity thresholds were re-estimated within every bootstrap replicate from the resampled control-score distribution, so the reported intervals reflect threshold-selection variability. 

For the 52-week fixed-specificity analyses, sensitivity and PPV confidence intervals were the 2.5th and 97.5th percentiles of the bootstrap estimates. Case yield interval endpoints were calculated by multiplying the corresponding sensitivity bounds, expressed as proportions, by 10,000$P_w(t)/N$, with the weighted case pool and scored-record denominator held fixed. These intervals did not include re-estimation of incidence or the censoring distribution. Baseline incidence and fold enrichment were reported as point estimates without confidence intervals.

For head-to-head comparisons, we used a paired bootstrap. The same resampled patients were applied to TRACER and TRACER-Encoder. Differences were defined as TRACER minus TRACER-Encoder. Positive values therefore favored TRACER, whereas negative values favored TRACER-Encoder. A nominal AUROC-pAUC disagreement occurred when the two point differences had opposite signs. A significant result required the two metrics to favor different models and the paired 95\% confidence interval for both differences to exclude zero.

The primary comparative analysis used survival heads trained on 100\% of the in-domain downstream training cohort across 16 endpoints, one in-domain cohort, two external cohorts, and three horizons (144 endpoint-site-horizon evaluations), with the four cancers contributing 12 additional 26-week evaluations. 

For fixed-specificity analyses, thresholds were weighted quantiles of the control-score distribution, and patients whose risk score equaled the threshold were classified as alerts; therefore, tied scores crossed the threshold jointly, and the achieved specificity could differ slightly from the nominal target when ties occurred.

\subsection{Secondary analyses and ablations}
Analyses reported only in the Supplementary Results are described in the Supplementary Methods: downstream labeled-data availability, local survival-head training at UC5 and UC6, the Cox proportional-hazards survival-head ablation, follow-up analyses of AUROC-pAUC ranking reversals, token-level attribution, pooling ablations, and characterization of health-system record differences.

\section*{Acknowledgments}
Data that support the findings of this study were obtained from the University of California Data Discovery Platform (UCDDP), a HIPAA limited data set which contains patient data from all six UC Health academic medical centers (Davis, Irvine, Los Angeles, Riverside, San Diego, and San Francisco). Due to privacy and confidentiality restrictions, the data are not publicly available. Minimal de-identified aggregate data in the form of tables are available from the corresponding author on reasonable request and subject to corresponding author’s institutional approval.

The authors thank the Center for Data-driven Insights and Innovation at UC Health (\href{https://www.ucop.edu/uc-health/departments/center-for-data-driven-insights-and-innovations-cdi2.html}{CDI2}), for its analytical and technical support related to use of the UC Health Data Warehouse and related data assets.

\section*{Competing interests}
The authors declare no competing interests.

\bibliographystyle{unsrtnat}
\bibliography{ref}
\clearpage

\clearpage

\processdelayedfloats
\clearpage
\makeatletter
\efloat@restorefloats
\makeatother
\appendix
\newgeometry{margin=1in,headheight=14pt}
\setcounter{table}{0}
\setcounter{figure}{0}
\setcounter{section}{0}
\renewcommand{\thetable}{S\arabic{table}}
\renewcommand{\thefigure}{S\arabic{figure}}
\renewcommand{\thesection}{S\arabic{section}}
\renewcommand{\theHtable}{supp.\arabic{table}}
\renewcommand{\theHfigure}{supp.\arabic{figure}}
\renewcommand{\theHsection}{supp.\arabic{section}}
\setcounter{secnumdepth}{3}
\setlength{\emergencystretch}{2em}
\setlength{\parskip}{3pt}
\widowpenalty=10000
\clubpenalty=10000
\captionsetup{font=small,labelfont=bf,skip=7pt,list=false}
\captionsetup[figure]{font=small,labelfont=bf}

\nolinenumbers
\begin{center}{\Large\bfseries Supplementary Material}\par\medskip{\large TRACER, a time-to-event EHR foundation model for disease detection across health systems}\end{center}

\renewcommand{\contentsname}{Supplementary contents}
\addtocontents{toc}{\protect\setcounter{tocdepth}{3}}
\tableofcontents

\clearpage\section{\revadd{Tables supporting the main text}}\label{sec:main-support}

\begin{table}[H]
\centering
\caption{\small
\textbf{Observed endpoint-event frequency across development and held-out cohorts.}
Values are the percentage of eligible patients with an observed incident endpoint event during available follow-up in the development-domain downstream training, validation, and test cohorts and the held-out UC5 and UC6 cohorts. Rows are ordered by event frequency in the in-domain training cohort within each endpoint group. Cohort sizes varied by endpoint-specific eligibility criteria; ovarian cancer cohorts were restricted to female patients.
}
\label{tab:prevalence}
\small
\setlength{\tabcolsep}{8pt}
\renewcommand{\arraystretch}{1.08}
\begin{tabular}{@{}lrrrrr@{}}
\toprule
& \multicolumn{3}{c}{Development domain}
& \multicolumn{2}{c}{Held-out health systems} \\
\cmidrule(lr){2-4}
\cmidrule(lr){5-6}
Disease endpoint
& Training & Validation & Test
& UC5 & UC6 \\
\midrule
\multicolumn{6}{@{}l}{\textit{Chronic disease endpoints}} \\
Hypertension
& 10.00 & 10.00 & 10.00 & 11.70 & 15.20 \\
Hyperlipidemia
& 9.23 & 9.23 & 9.25 & 15.47 & 9.42 \\
Depression
& 6.81 & 6.82 & 6.81 & 8.78 & 6.88 \\
Chronic kidney disease
& 2.53 & 2.57 & 2.54 & 2.51 & 2.53 \\
Heart failure
& 2.38 & 2.37 & 2.38 & 2.30 & 4.93 \\
Atrial fibrillation
& 2.04 & 2.06 & 2.06 & 2.34 & 2.06 \\
Chronic obstructive pulmonary disease
& 1.47 & 1.47 & 1.46 & 1.53 & 1.38 \\
Alzheimer's disease
& 0.97 & 0.98 & 0.99 & 1.08 & 0.98 \\
Acute myocardial infarction
& 0.65 & 0.64 & 0.65 & 0.86 & 0.65 \\
Rheumatoid arthritis
& 0.45 & 0.45 & 0.47 & 0.59 & 0.43 \\
Celiac disease
& 0.23 & 0.23 & 0.23 & 0.35 & 0.36 \\
Lupus
& 0.20 & 0.19 & 0.20 & 0.27 & 0.30 \\
\addlinespace
\multicolumn{6}{@{}l}{\textit{Rare-cancer endpoints}} \\
Glioblastoma
& 0.29 & 0.29 & 0.29 & 0.26 & 0.29 \\
Multiple myeloma
& 0.27 & 0.31 & 0.28 & 0.15 & 0.27 \\
Ovarian cancer
& 0.25 & 0.26 & 0.25 & 0.21 & 0.25 \\
Pancreatic cancer
& 0.15 & 0.17 & 0.15 & 0.14 & 0.15 \\
\bottomrule
\end{tabular}
\par\smallskip
\end{table}

\begin{table}[H]
\centering
\caption{\small
\textbf{Global discrimination across the in-domain and held-out health-system cohorts.}
Uno's IPCW C-index is reported for TRACER and TRACER-Encoder using survival heads trained on the full development-domain downstream training cohort and applied unchanged to the in-domain, UC5, and UC6 evaluation cohorts. Values are shown as the point estimate \(\pm\) the half-width of the model-specific analytic 95\% confidence interval.
}
\label{tab:cindex-by-site}

\scriptsize
\setlength{\tabcolsep}{3pt}
\renewcommand{\arraystretch}{1.12}

\begin{tabular*}{\textwidth}{
@{\extracolsep{\fill}}
lcccccc
@{}
}
\toprule
& \multicolumn{2}{c}{In-domain}
& \multicolumn{2}{c}{UC5}
& \multicolumn{2}{c}{UC6} \\
\cmidrule(lr){2-3}
\cmidrule(lr){4-5}
\cmidrule(lr){6-7}
Disease endpoint
& Decoder & Encoder
& Decoder & Encoder
& Decoder & Encoder \\
\midrule
\multicolumn{7}{@{}l}{\textit{Chronic disease endpoints}} \\
Heart failure
& \cindexci{0.869}{0.015}
& \cindexci{0.810}{0.021}
& \cindexci{0.865}{0.003}
& \cindexci{0.823}{0.004}
& \cindexci{0.836}{0.004}
& \cindexci{0.798}{0.005} \\
Acute myocardial infarction
& \cindexci{0.842}{0.029}
& \cindexci{0.775}{0.046}
& \cindexci{0.830}{0.005}
& \cindexci{0.776}{0.007}
& \cindexci{0.862}{0.013}
& \cindexci{0.798}{0.020} \\
Lupus
& \cindexci{0.763}{0.079}
& \cindexci{0.701}{0.098}
& \cindexci{0.746}{0.012}
& \cindexci{0.711}{0.012}
& \cindexci{0.669}{0.020}
& \cindexci{0.661}{0.023} \\
Hypertension
& \cindexci{0.776}{0.010}
& \cindexci{0.768}{0.010}
& \cindexci{0.761}{0.002}
& \cindexci{0.753}{0.002}
& \cindexci{0.775}{0.003}
& \cindexci{0.753}{0.003} \\
Celiac disease
& \cindexci{0.624}{0.091}
& \cindexci{0.591}{0.083}
& \cindexci{0.617}{0.012}
& \cindexci{0.606}{0.012}
& \cindexci{0.651}{0.021}
& \cindexci{0.655}{0.020} \\
Hyperlipidemia
& \cindexci{0.763}{0.010}
& \cindexci{0.747}{0.011}
& \cindexci{0.696}{0.002}
& \cindexci{0.718}{0.002}
& \cindexci{0.759}{0.005}
& \cindexci{0.743}{0.005} \\
Depression
& \cindexci{0.693}{0.014}
& \cindexci{0.681}{0.015}
& \cindexci{0.631}{0.002}
& \cindexci{0.641}{0.003}
& \cindexci{0.696}{0.006}
& \cindexci{0.685}{0.006} \\
Chronic kidney disease
& \cindexci{0.869}{0.015}
& \cindexci{0.820}{0.022}
& \cindexci{0.860}{0.003}
& \cindexci{0.811}{0.004}
& \cindexci{0.867}{0.007}
& \cindexci{0.808}{0.010} \\
Atrial fibrillation
& \cindexci{0.865}{0.014}
& \cindexci{0.753}{0.028}
& \cindexci{0.833}{0.003}
& \cindexci{0.764}{0.005}
& \cindexci{0.860}{0.007}
& \cindexci{0.769}{0.012} \\
Chronic obstructive pulmonary disease
& \cindexci{0.841}{0.022}
& \cindexci{0.765}{0.036}
& \cindexci{0.788}{0.004}
& \cindexci{0.762}{0.005}
& \cindexci{0.825}{0.009}
& \cindexci{0.771}{0.012} \\
Alzheimer's disease
& \cindexci{0.879}{0.023}
& \cindexci{0.816}{0.034}
& \cindexci{0.870}{0.004}
& \cindexci{0.803}{0.006}
& \cindexci{0.880}{0.010}
& \cindexci{0.813}{0.015} \\
Rheumatoid arthritis
& \cindexci{0.793}{0.036}
& \cindexci{0.739}{0.050}
& \cindexci{0.762}{0.007}
& \cindexci{0.721}{0.009}
& \cindexci{0.778}{0.021}
& \cindexci{0.679}{0.026} \\
\addlinespace
\multicolumn{7}{@{}l}{\textit{Rare-cancer endpoints}} \\
Glioblastoma
& \cindexci{0.736}{0.062}
& \cindexci{0.763}{0.066}
& \cindexci{0.688}{0.013}
& \cindexci{0.711}{0.013}
& \cindexci{0.816}{0.028}
& \cindexci{0.848}{0.024} \\
Multiple myeloma
& \cindexci{0.801}{0.071}
& \cindexci{0.815}{0.060}
& \cindexci{0.709}{0.017}
& \cindexci{0.754}{0.016}
& \cindexci{0.846}{0.023}
& \cindexci{0.856}{0.021} \\
Ovarian cancer
& \cindexci{0.717}{0.074}
& \cindexci{0.717}{0.097}
& \cindexci{0.636}{0.019}
& \cindexci{0.616}{0.019}
& \cindexci{0.735}{0.041}
& \cindexci{0.712}{0.046} \\
Pancreatic cancer
& \cindexci{0.776}{0.083}
& \cindexci{0.752}{0.102}
& \cindexci{0.717}{0.014}
& \cindexci{0.728}{0.015}
& \cindexci{0.790}{0.035}
& \cindexci{0.779}{0.041} \\
\bottomrule
\end{tabular*}
\end{table}

\begin{landscape}
\begin{table}[H]
\centering
\caption{\small
\textbf{Paired differences in 52-week time-dependent AUROC between TRACER and TRACER-Encoder.}
Values are paired differences in dynamic AUROC, calculated as TRACER minus TRACER-Encoder, using survival heads trained on the full development-domain downstream training cohort. Parentheses show 95\% percentile confidence intervals obtained using paired patient-level bootstrap resampling. Positive values favor TRACER, whereas negative values favor TRACER-Encoder. Boldface indicates comparisons for which the paired 95\% confidence interval excluded zero.
}
\label{tab:auroc-difference-52wk}
\small
\setlength{\tabcolsep}{8pt}
\renewcommand{\arraystretch}{1.16}
\begin{adjustbox}{max width=\linewidth}
\begin{tabular}{@{}lccc@{}}
\toprule
Disease endpoint & In-domain & UC5 & UC6 \\
\midrule
\multicolumn{4}{@{}l}{\textit{Chronic disease endpoints}} \\
Atrial fibrillation & \auccib{+0.078}{+0.055}{+0.100} & \auccib{+0.043}{+0.039}{+0.047} & \auccib{+0.077}{+0.067}{+0.089} \\
Acute myocardial infarction & \auccib{+0.077}{+0.039}{+0.119} & \auccib{+0.088}{+0.081}{+0.094} & \auccib{+0.067}{+0.048}{+0.085} \\
Alzheimer's disease & \auccib{+0.113}{+0.076}{+0.154} & \auccib{+0.083}{+0.076}{+0.090} & \auccib{+0.106}{+0.087}{+0.124} \\
Chronic kidney disease & \auccib{+0.051}{+0.035}{+0.068} & \auccib{+0.045}{+0.042}{+0.048} & \auccib{+0.059}{+0.052}{+0.068} \\
Chronic obstructive pulmonary disease & \auccib{+0.079}{+0.052}{+0.109} & \auccib{+0.054}{+0.048}{+0.059} & \auccib{+0.081}{+0.068}{+0.094} \\
Celiac disease & \aucci{+0.031}{-0.092}{+0.161} & \auccib{+0.041}{+0.027}{+0.055} & \aucci{+0.007}{-0.017}{+0.030} \\
Depression & \auccib{+0.039}{+0.025}{+0.052} & \auccib{+0.017}{+0.014}{+0.020} & \auccib{+0.028}{+0.022}{+0.034} \\
Heart failure & \auccib{+0.041}{+0.023}{+0.059} & \auccib{+0.029}{+0.025}{+0.032} & \auccib{+0.036}{+0.031}{+0.040} \\
Hyperlipidemia & \auccib{+0.034}{+0.024}{+0.043} & \auccib{-0.006}{-0.007}{-0.004} & \auccib{+0.032}{+0.028}{+0.037} \\
Hypertension & \auccib{+0.018}{+0.010}{+0.026} & \auccib{+0.004}{+0.002}{+0.005} & \auccib{+0.029}{+0.026}{+0.032} \\
Lupus & \aucci{+0.084}{-0.011}{+0.185} & \auccib{+0.102}{+0.087}{+0.117} & \auccib{+0.051}{+0.023}{+0.077} \\
Rheumatoid arthritis & \auccib{+0.132}{+0.079}{+0.192} & \auccib{+0.057}{+0.047}{+0.067} & \auccib{+0.127}{+0.098}{+0.155} \\
\addlinespace
\multicolumn{4}{@{}l}{\textit{Rare-cancer endpoints}} \\
Glioblastoma & \aucci{+0.003}{-0.033}{+0.042} & \aucci{-0.009}{-0.022}{+0.003} & \auccib{-0.025}{-0.043}{-0.008} \\
Multiple myeloma & \aucci{+0.042}{-0.002}{+0.096} & \auccib{-0.023}{-0.037}{-0.009} & \aucci{+0.012}{-0.006}{+0.030} \\
Ovarian cancer & \aucci{-0.056}{-0.175}{+0.045} & \aucci{-0.004}{-0.021}{+0.015} & \aucci{+0.017}{-0.011}{+0.049} \\
Pancreatic cancer & \aucci{-0.011}{-0.067}{+0.052} & \aucci{-0.008}{-0.024}{+0.007} & \aucci{+0.013}{-0.012}{+0.041} \\
\bottomrule
\end{tabular}
\end{adjustbox}
\end{table}
\end{landscape}

\begin{table}[H]
\centering
\caption{\textbf{Model favored within the low-false-positive region, by disease, evaluation cohort, and prediction horizon.} Each cell reports the model with the higher point-estimate standardized partial AUROC over the prespecified low-false-positive region ($\mathrm{FPR}\leq 0.10$): D, TRACER; E, TRACER-Encoder. Plain cells: global AUROC favors the same model. Italic cells: global AUROC favors the other model at the point estimate (nominal disagreement). Bold cells: significant reversal, with both paired 95\% confidence intervals excluding zero. The table identifies where metric choice changes the preferred model.}
\label{tab:model-choice}
\small
\setlength{\tabcolsep}{4pt}
\renewcommand{\arraystretch}{1.1}
\begin{adjustbox}{max width=\linewidth}
\begin{tabular}{@{}lccc@{\hspace{1.5em}}ccc@{\hspace{1.5em}}ccc@{}}
\toprule
& \multicolumn{3}{c}{In-domain} & \multicolumn{3}{c}{UC5} & \multicolumn{3}{c}{UC6} \\
\cmidrule(lr){2-4} \cmidrule(lr){5-7} \cmidrule(lr){8-10}
Disease & 52 wk & 104 wk & 260 wk & 52 wk & 104 wk & 260 wk & 52 wk & 104 wk & 260 wk \\
\midrule
Heart failure & D & D & D & \textbf{E} & \textbf{E} & \textit{E} & \textbf{E} & \textbf{E} & \textbf{E} \\
Rheumatoid arthritis & D & D & \textit{E} & \textbf{E} & \textbf{E} & \textbf{E} & D & D & D \\
Atrial fibrillation & D & D & D & \textbf{E} & \textbf{E} & \textit{E} & D & D & D \\
Depression & D & D & D & \textbf{E} & \textbf{E} & D & D & D & D \\
Hyperlipidemia & D & D & D & \textbf{D}$^{*}$ & \textit{E} & \textbf{E} & D & D & D \\
Hypertension & D & D & \textit{E} & \textbf{E} & D & D & D & D & D \\
\midrule
COPD & D & D & \textit{E} & \textit{E} & \textit{E} & D & D & D & D \\
Celiac disease & \textit{E} & \textit{E} & \textit{D} & D & D & D & D & D & \textit{E} \\
\midrule
Acute myocardial infarction & D & D & D & D & D & D & D & D & D \\
Alzheimer's disease & D & D & D & D & D & D & D & D & D \\
Chronic kidney disease & D & D & D & D & D & D & D & D & D \\
Lupus & D & D & D & D & D & D & D & D & D \\
\midrule
Pancreatic cancer & E & \textit{E} & \textit{E} & E & E & E & \textit{E} & \textit{E} & \textit{E} \\
Ovarian cancer & E & \textit{D} & D & E & D & D & D & D & D \\
Glioblastoma & \textit{E} & E & E & E & E & E & E & E & E \\
Multiple myeloma & D & D & D & E & E & E & \textit{E} & \textit{E} & \textit{E} \\
\bottomrule
\end{tabular}
\end{adjustbox}
\par\smallskip
\footnotesize
$^{*}$Opposite-direction strict reversal (TRACER-Encoder favored by AUROC, TRACER by pAUC). The additional 26-week cancer evaluations (not shown) contained five nominal disagreements and no strict reversals.
\end{table}

\begin{table}[H]
\centering
\caption{\small \textbf{TRACER overall discrimination across health systems.} Uno's IPCW C-index for the development-trained TRACER model, applied unchanged to the in-domain and two external evaluation cohorts. Values are point estimates $\pm$ the half-width of the model-specific analytic 95\% confidence interval. The reported worst-site gap is the in-domain C-index minus the lower of the UC5 and UC6 values.}
\label{tab:tracer-discrimination}
\begingroup
\small
\setlength{\tabcolsep}{4pt}
\renewcommand{\arraystretch}{1.13}
\begin{tabular*}{\textwidth}{@{\extracolsep{\fill}}lcccc@{}}
\toprule
Disease & In-domain & UC5 & UC6 & \shortstack{Worst-site\\gap} \\
\midrule
\multicolumn{5}{@{}l}{\textit{Chronic disease endpoints}} \\
Heart failure & \(0.869\,\pm\,0.015\) & \(0.865\,\pm\,0.003\) & \(0.836\,\pm\,0.004\) & \(+0.033\) \\
Acute myocardial infarction & \(0.842\,\pm\,0.029\) & \(0.830\,\pm\,0.005\) & \(0.862\,\pm\,0.013\) & \(+0.013\) \\
Lupus & \(0.763\,\pm\,0.079\) & \(0.746\,\pm\,0.012\) & \(0.669\,\pm\,0.020\) & \(+0.095\) \\
Hypertension & \(0.776\,\pm\,0.010\) & \(0.761\,\pm\,0.002\) & \(0.775\,\pm\,0.003\) & \(+0.015\) \\
Celiac disease & \(0.624\,\pm\,0.091\) & \(0.617\,\pm\,0.012\) & \(0.651\,\pm\,0.021\) & \(+0.007\) \\
Hyperlipidemia & \(0.763\,\pm\,0.010\) & \(0.696\,\pm\,0.002\) & \(0.759\,\pm\,0.005\) & \(+0.067\) \\
Depression & \(0.693\,\pm\,0.014\) & \(0.631\,\pm\,0.002\) & \(0.696\,\pm\,0.006\) & \(+0.062\) \\
Chronic kidney disease & \(0.869\,\pm\,0.015\) & \(0.860\,\pm\,0.003\) & \(0.867\,\pm\,0.007\) & \(+0.009\) \\
Atrial fibrillation & \(0.865\,\pm\,0.014\) & \(0.833\,\pm\,0.003\) & \(0.860\,\pm\,0.007\) & \(+0.032\) \\
Chronic obstructive pulmonary disease & \(0.841\,\pm\,0.022\) & \(0.788\,\pm\,0.004\) & \(0.825\,\pm\,0.009\) & \(+0.053\) \\
Alzheimer's disease & \(0.879\,\pm\,0.023\) & \(0.870\,\pm\,0.004\) & \(0.880\,\pm\,0.010\) & \(+0.009\) \\
Rheumatoid arthritis & \(0.793\,\pm\,0.036\) & \(0.762\,\pm\,0.007\) & \(0.778\,\pm\,0.021\) & \(+0.031\) \\
\multicolumn{5}{@{}l}{\textit{Rare-cancer endpoints}} \\
Glioblastoma & \(0.736\,\pm\,0.062\) & \(0.688\,\pm\,0.013\) & \(0.816\,\pm\,0.028\) & \(+0.048\) \\
Multiple myeloma & \(0.801\,\pm\,0.071\) & \(0.709\,\pm\,0.017\) & \(0.846\,\pm\,0.023\) & \(+0.093\) \\
Ovarian cancer & \(0.717\,\pm\,0.074\) & \(0.636\,\pm\,0.019\) & \(0.735\,\pm\,0.041\) & \(+0.081\) \\
Pancreatic cancer & \(0.776\,\pm\,0.083\) & \(0.717\,\pm\,0.014\) & \(0.790\,\pm\,0.035\) & \(+0.059\) \\

\bottomrule
\end{tabular*}
\par\smallskip
\endgroup
\end{table}

\begin{table}[H]
\centering
\caption{\small \textbf{TRACER AUROC across prediction horizons in-domain.} Time-dependent AUROC at the 52-, 104- and 260-week horizons, using disease-prediction heads trained on the full in-domain training cohort. The final column reproduces the reported change from 52 to 260 weeks.}
\label{tab:tracer-horizons}
\begingroup
\small
\setlength{\tabcolsep}{4pt}
\renewcommand{\arraystretch}{1.13}
\begin{tabular*}{\textwidth}{@{\extracolsep{\fill}}lrrrr@{}}
\toprule
Disease & 52 weeks & 104 weeks & 260 weeks & $\Delta_{52\rightarrow260}$ \\
\midrule
\multicolumn{5}{@{}l}{
\textit{Small 52-to-260-week change under TRACER
(\(\lvert\Delta\rvert\leq0.025\))}
} \\
Lupus & 0.778 & 0.807 & 0.795 & \(+0.017\) \\
Atrial fibrillation & 0.852 & 0.861 & 0.859 & \(+0.008\) \\
Rheumatoid arthritis & 0.786 & 0.778 & 0.784 & \(-0.002\) \\
Heart failure & 0.870 & 0.869 & 0.862 & \(-0.008\) \\
Hyperlipidemia & 0.785 & 0.773 & 0.776 & \(-0.009\) \\
Ovarian cancer & 0.755 & 0.799 & 0.742 & \(-0.013\) \\
Chronic kidney disease & 0.888 & 0.878 & 0.865 & \(-0.023\) \\
Hypertension & 0.814 & 0.798 & 0.790 & \(-0.024\) \\
\multicolumn{5}{@{}l}{
\textit{Intermediate 52-to-260-week decline under TRACER
(\(0.025<\lvert\Delta\rvert\leq0.040\))}
} \\
Chronic obstructive pulmonary disease & 0.847 & 0.843 & 0.821 & \(-0.026\) \\
Acute myocardial infarction & 0.871 & 0.892 & 0.843 & \(-0.028\) \\
Alzheimer's disease & 0.913 & 0.890 & 0.882 & \(-0.030\) \\
\multicolumn{5}{@{}l}{
\textit{Large 52-to-260-week decline under TRACER
(\(\lvert\Delta\rvert>0.040\))}
} \\
Pancreatic cancer & 0.843 & 0.812 & 0.793 & \(-0.050\) \\
Depression & 0.739 & 0.706 & 0.688 & \(-0.050\) \\
Multiple myeloma & 0.877 & 0.858 & 0.819 & \(-0.058\) \\
Celiac disease & 0.646 & 0.625 & 0.576 & \(-0.070\) \\
Glioblastoma & 0.834 & 0.778 & 0.744 & \(-0.091\) \\

\bottomrule
\end{tabular*}
\par\smallskip
\endgroup
\end{table}
\clearpage

\begin{landscape}
\begingroup
\scriptsize
\setlength{\tabcolsep}{3pt}
\renewcommand{\arraystretch}{1.10}
\setlength{\LTcapwidth}{\linewidth}
\setlength{\LTleft}{\fill}
\setlength{\LTright}{\fill}
\begin{longtable}{@{}llrrrrrrrrrrrr@{}}
\caption{\small \textbf{Time-dependent AUROC by prediction horizon in the in-domain, UC5, and UC6 evaluation cohorts.} Point estimates of cumulative/dynamic IPCW AUROC for TRACER (D) and TRACER-Encoder (E), survival heads trained on the full in-domain training cohort and applied unchanged externally. Chronic endpoints have no 26-week evaluation. 95\% bootstrap confidence intervals are in the accompanying data file. } \label{tab:auroc-all-cohorts}\\
\toprule
& & \multicolumn{4}{c}{In-domain} & \multicolumn{4}{c}{UC5} & \multicolumn{4}{c}{UC6} \\
\cmidrule(lr){3-6}\cmidrule(lr){7-10}\cmidrule(lr){11-14}
Disease & Model & 26 wk & 52 wk & 104 wk & 260 wk & 26 wk & 52 wk & 104 wk & 260 wk & 26 wk & 52 wk & 104 wk & 260 wk \\
\midrule
\endfirsthead
\multicolumn{14}{@{}l}{\textit{(continued)}}\\
\toprule
& & \multicolumn{4}{c}{In-domain} & \multicolumn{4}{c}{UC5} & \multicolumn{4}{c}{UC6} \\
\cmidrule(lr){3-6}\cmidrule(lr){7-10}\cmidrule(lr){11-14}
Disease & Model & 26 wk & 52 wk & 104 wk & 260 wk & 26 wk & 52 wk & 104 wk & 260 wk & 26 wk & 52 wk & 104 wk & 260 wk \\
\midrule
\endhead
\midrule
\multicolumn{14}{r@{}}{\footnotesize Continued on next page}\\
\endfoot
\bottomrule
\endlastfoot
\multicolumn{14}{@{}l}{\textit{Chronic disease endpoints}} \\
Heart failure & D & --- & 0.870 & 0.869 & 0.862 & --- & 0.868 & 0.869 & 0.877 & --- & 0.850 & 0.844 & 0.836\\
 & E & --- & 0.829 & 0.821 & 0.822 & --- & 0.840 & 0.838 & 0.839 & --- & 0.815 & 0.820 & 0.814\\
\addlinespace
Acute myocardial infarction & D & --- & 0.871 & 0.892 & 0.843 & --- & 0.854 & 0.844 & 0.830 & --- & 0.858 & 0.863 & 0.849\\
 & E & --- & 0.794 & 0.812 & 0.771 & --- & 0.766 & 0.773 & 0.776 & --- & 0.791 & 0.811 & 0.801\\
\addlinespace
Lupus & D & --- & 0.778 & 0.807 & 0.795 & --- & 0.784 & 0.775 & 0.765 & --- & 0.684 & 0.699 & 0.677\\
 & E & --- & 0.694 & 0.637 & 0.675 & --- & 0.682 & 0.648 & 0.682 & --- & 0.633 & 0.629 & 0.641\\
\addlinespace
Hypertension & D & --- & 0.814 & 0.798 & 0.790 & --- & 0.781 & 0.774 & 0.790 & --- & 0.798 & 0.794 & 0.801\\
 & E & --- & 0.796 & 0.771 & 0.783 & --- & 0.777 & 0.767 & 0.765 & --- & 0.769 & 0.760 & 0.782\\
\addlinespace
Celiac disease & D & --- & 0.646 & 0.625 & 0.576 & --- & 0.646 & 0.643 & 0.603 & --- & 0.661 & 0.654 & 0.625\\
 & E & --- & 0.615 & 0.607 & 0.587 & --- & 0.605 & 0.592 & 0.575 & --- & 0.654 & 0.648 & 0.612\\
\addlinespace
Hyperlipidemia & D & --- & 0.785 & 0.773 & 0.776 & --- & 0.717 & 0.714 & 0.725 & --- & 0.789 & 0.780 & 0.770\\
 & E & --- & 0.751 & 0.739 & 0.749 & --- & 0.722 & 0.713 & 0.720 & --- & 0.756 & 0.745 & 0.739\\
\addlinespace
Depression & D & --- & 0.739 & 0.706 & 0.688 & --- & 0.661 & 0.633 & 0.649 & --- & 0.728 & 0.707 & 0.688\\
 & E & --- & 0.700 & 0.660 & 0.646 & --- & 0.645 & 0.624 & 0.604 & --- & 0.700 & 0.675 & 0.653\\
\addlinespace
Chronic kidney disease & D & --- & 0.888 & 0.878 & 0.865 & --- & 0.880 & 0.874 & 0.864 & --- & 0.888 & 0.882 & 0.860\\
 & E & --- & 0.837 & 0.826 & 0.832 & --- & 0.835 & 0.831 & 0.830 & --- & 0.828 & 0.827 & 0.819\\
\addlinespace
Atrial fibrillation & D & --- & 0.852 & 0.861 & 0.859 & --- & 0.850 & 0.838 & 0.837 & --- & 0.856 & 0.858 & 0.855\\
 & E & --- & 0.774 & 0.793 & 0.778 & --- & 0.807 & 0.800 & 0.785 & --- & 0.779 & 0.789 & 0.787\\
\addlinespace
Chronic obstructive pulmonary disease & D & --- & 0.847 & 0.843 & 0.821 & --- & 0.817 & 0.803 & 0.787 & --- & 0.832 & 0.825 & 0.804\\
 & E & --- & 0.768 & 0.782 & 0.757 & --- & 0.764 & 0.771 & 0.756 & --- & 0.751 & 0.753 & 0.756\\
\addlinespace
Alzheimer's disease & D & --- & 0.913 & 0.890 & 0.882 & --- & 0.890 & 0.886 & 0.875 & --- & 0.896 & 0.889 & 0.882\\
 & E & --- & 0.800 & 0.785 & 0.797 & --- & 0.807 & 0.802 & 0.794 & --- & 0.790 & 0.799 & 0.798\\
\addlinespace
Rheumatoid arthritis & D & --- & 0.786 & 0.778 & 0.784 & --- & 0.768 & 0.763 & 0.753 & --- & 0.800 & 0.797 & 0.757\\
 & E & --- & 0.654 & 0.703 & 0.754 & --- & 0.710 & 0.745 & 0.731 & --- & 0.674 & 0.706 & 0.671\\
\addlinespace
\multicolumn{14}{@{}l}{\textit{Rare-cancer endpoints}} \\
Pancreatic cancer & D & 0.807 & 0.843 & 0.812 & 0.793 & 0.754 & 0.751 & 0.726 & 0.716 & 0.859 & 0.834 & 0.821 & 0.810\\
 & E & 0.847 & 0.854 & 0.796 & 0.770 & 0.786 & 0.759 & 0.745 & 0.731 & 0.855 & 0.821 & 0.801 & 0.789\\
\addlinespace
Ovarian cancer & D & 0.684 & 0.755 & 0.799 & 0.742 & 0.698 & 0.686 & 0.664 & 0.637 & 0.829 & 0.831 & 0.814 & 0.749\\
 & E & 0.735 & 0.811 & 0.817 & 0.717 & 0.704 & 0.690 & 0.651 & 0.621 & 0.792 & 0.813 & 0.781 & 0.714\\
\addlinespace
Glioblastoma & D & 0.856 & 0.834 & 0.778 & 0.744 & 0.710 & 0.709 & 0.693 & 0.686 & 0.842 & 0.833 & 0.828 & 0.817\\
 & E & 0.827 & 0.831 & 0.808 & 0.760 & 0.707 & 0.718 & 0.708 & 0.702 & 0.858 & 0.858 & 0.859 & 0.839\\
\addlinespace
Multiple myeloma & D & 0.889 & 0.877 & 0.858 & 0.819 & 0.779 & 0.776 & 0.787 & 0.719 & 0.872 & 0.867 & 0.872 & 0.863\\
 & E & 0.810 & 0.835 & 0.835 & 0.815 & 0.791 & 0.799 & 0.804 & 0.761 & 0.844 & 0.855 & 0.862 & 0.857\\
\addlinespace
\end{longtable}
\endgroup
\end{landscape}

\begin{table}[H]
\centering
\caption{\small \textbf{TRACER patient selection and disease detection at 52 weeks.} Results at thresholds selected retrospectively within each evaluation cohort to target 95\% specificity. The full development-trained disease-prediction heads were applied unchanged; only the cohort-specific evaluation thresholds differed. ``Not identified'' is the estimated percentage of future cases below the threshold (100\% minus sensitivity), not the percentage of all scored patients.}
\label{tab:tracer-selection-52wk}
\begingroup
\footnotesize
\setlength{\tabcolsep}{3pt}
\renewcommand{\arraystretch}{1.13}
\begin{tabular*}{\textwidth}{@{\extracolsep{\fill}}lrrrrrrr@{}}
\toprule
& \multicolumn{3}{c}{Direct patient selection} & \multicolumn{4}{c}{Censoring-adjusted outcomes} \\
\cmidrule(lr){2-4}\cmidrule(lr){5-8}
Cohort & \shortstack{Patients\\scored} & \shortstack{Patients\\selected} & \shortstack{Selected\\(\%)} & \shortstack{Sensitivity\\(\%)} & \shortstack{Not identified\\(\%)} & \shortstack{PPV\\(\%)} & \shortstack{Case yield\\per 10,000} \\
\midrule
\multicolumn{8}{@{}l}{\textit{Heart failure}} \\
In-domain & 28,800 & 1,505 & 5.23 & 44.18 & 55.82 & 13.80 & 78.59 \\
UC5 & 919,344 & 51,998 & 5.66 & 36.88 & 63.12 & 9.47 & 51.59 \\
UC6 & 283,392 & 15,279 & 5.39 & 33.57 & 66.43 & 16.41 & 95.39 \\
\multicolumn{8}{@{}l}{\textit{Pancreatic cancer}} \\
In-domain & 29,176 & 1,624 & 5.57 & 34.42 & 65.58 & 0.71 & 3.59 \\
UC5 & 930,896 & 50,163 & 5.39 & 16.37 & 83.63 & 0.29 & 1.45 \\
UC6 & 145,880 & 8,075 & 5.54 & 35.48 & 64.52 & 0.80 & 4.05 \\

\bottomrule
\end{tabular*}
\par\smallskip
\endgroup
\end{table}

\begin{table}[H]
\centering
\caption{\small \textbf{TRACER sensitivity at fixed specificity at 52 weeks.} Entries are estimated sensitivity (\%) at cohort-specific thresholds targeting the specificity shown in the column headings. The same development-trained model was applied without local adaptation. Outcome estimates account for incomplete follow-up.}
\label{tab:tracer-sensitivity}
\begingroup
\scriptsize
\setlength{\tabcolsep}{4pt}
\renewcommand{\arraystretch}{1.13}
\begin{tabular*}{\textwidth}{@{\extracolsep{\fill}}lrrrrrrrrr@{}}
\toprule
& \multicolumn{3}{c}{In-domain} & \multicolumn{3}{c}{UC5} & \multicolumn{3}{c}{UC6} \\
\cmidrule(lr){2-4}\cmidrule(lr){5-7}\cmidrule(lr){8-10}
Disease & 90\% & 95\% & 99\% & 90\% & 95\% & 99\% & 90\% & 95\% & 99\% \\
\midrule
\multicolumn{10}{@{}l}{\textit{Chronic disease endpoints}} \\
Heart failure & 63.7 & 44.2 & 15.1 & 56.5 & 36.9 & 11.3 & 51.2 & 33.6 & 10.8 \\
Acute myocardial infarction & 55.8 & 42.0 & 6.2 & 54.9 & 36.2 & 11.8 & 57.3 & 39.6 & 15.9 \\
Lupus & 45.8 & 26.0 & 6.4 & 44.5 & 32.3 & 15.0 & 26.4 & 19.0 & 7.9 \\
Hypertension & 41.4 & 25.6 & 7.1 & 36.0 & 21.3 & 5.5 & 40.1 & 25.2 & 7.4 \\
Celiac disease & 13.2 & 8.4 & 0.0 & 23.7 & 13.1 & 3.0 & 24.3 & 14.6 & 4.9 \\
Hyperlipidemia & 35.3 & 21.4 & 6.2 & 27.5 & 15.9 & 3.6 & 36.4 & 21.7 & 5.6 \\
Depression & 31.4 & 20.7 & 6.4 & 22.8 & 13.2 & 3.4 & 29.8 & 18.1 & 5.5 \\
Chronic kidney disease & 63.0 & 46.8 & 21.4 & 64.9 & 48.7 & 20.2 & 65.4 & 48.4 & 21.0 \\
Atrial fibrillation & 53.7 & 36.2 & 12.2 & 52.5 & 37.3 & 16.0 & 54.5 & 37.4 & 16.1 \\
Chronic obstructive pulmonary disease & 54.1 & 38.2 & 12.5 & 48.0 & 32.2 & 11.3 & 48.2 & 32.4 & 9.4 \\
Alzheimer's disease & 70.6 & 56.7 & 25.5 & 67.0 & 50.7 & 20.6 & 65.9 & 49.4 & 22.8 \\
Rheumatoid arthritis & 41.3 & 29.4 & 18.3 & 36.0 & 22.3 & 8.0 & 41.9 & 30.1 & 10.6 \\
\multicolumn{10}{@{}l}{\textit{Rare-cancer endpoints}} \\
Pancreatic cancer & 60.4 & 34.4 & 6.8 & 28.7 & 16.4 & 4.0 & 54.2 & 35.5 & 15.7 \\
Ovarian cancer & 30.9 & 12.3 & 0.0 & 25.3 & 14.9 & 2.8 & 45.3 & 35.7 & 8.3 \\
Glioblastoma & 65.2 & 41.1 & 28.1 & 32.4 & 22.9 & 9.3 & 58.7 & 48.7 & 25.1 \\
Multiple myeloma & 56.0 & 44.6 & 30.4 & 37.1 & 24.3 & 7.9 & 64.2 & 56.5 & 34.1 \\

\bottomrule
\end{tabular*}
\par\smallskip
\endgroup
\end{table}

\begingroup
\scriptsize
\setlength{\tabcolsep}{3pt}
\renewcommand{\arraystretch}{1.10}
\setlength{\LTcapwidth}{\linewidth}
\setlength{\LTleft}{\fill}
\setlength{\LTright}{\fill}
\begin{longtable}{@{}lrrrrrrrrrr@{}}
\caption{\small \textbf{TRACER sensitivity at fixed specificity across all prediction horizons.} Estimated sensitivity (\%) at cohort-specific thresholds targeting 90\%, 95\%, and 99\% specificity, for survival heads trained on the full in-domain training cohort and applied unchanged to UC5 and UC6. Chronic endpoints were evaluated at 52, 104, and 260 weeks; cancers additionally at 26 weeks. Estimates account for incomplete follow-up (IPCW); 95\% bootstrap confidence intervals are provided in the accompanying data file. } \label{tab:sens-all-horizons-tracer}\\
\toprule
& & \multicolumn{3}{c}{In-domain} & \multicolumn{3}{c}{UC5} & \multicolumn{3}{c}{UC6} \\
\cmidrule(lr){3-5}\cmidrule(lr){6-8}\cmidrule(lr){9-11}
Disease & Horizon (wk) & 90\% & 95\% & 99\% & 90\% & 95\% & 99\% & 90\% & 95\% & 99\% \\
\midrule
\endfirsthead
\multicolumn{11}{@{}l}{\textit{(continued)}}\\
\toprule
& & \multicolumn{3}{c}{In-domain} & \multicolumn{3}{c}{UC5} & \multicolumn{3}{c}{UC6} \\
\cmidrule(lr){3-5}\cmidrule(lr){6-8}\cmidrule(lr){9-11}
Disease & Horizon (wk) & 90\% & 95\% & 99\% & 90\% & 95\% & 99\% & 90\% & 95\% & 99\% \\
\midrule
\endhead
\midrule
\multicolumn{11}{r@{}}{\footnotesize Continued on next page}\\
\endfoot
\bottomrule
\endlastfoot
\multicolumn{11}{@{}l}{\textit{Chronic disease endpoints}} \\
Heart failure & 52 & 63.7 & 44.2 & 15.1 & 56.5 & 36.9 & 11.3 & 51.2 & 33.6 & 10.8\\
 & 104 & 62.9 & 45.0 & 13.3 & 57.9 & 39.1 & 11.0 & 49.8 & 33.1 & 9.5\\
 & 260 & 60.4 & 42.4 & 14.5 & 60.3 & 43.1 & 15.4 & 48.4 & 32.1 & 9.9\\
\addlinespace
Acute myocardial infarction & 52 & 55.8 & 42.0 & 6.2 & 54.9 & 36.2 & 11.8 & 57.3 & 39.6 & 15.9\\
 & 104 & 63.2 & 52.0 & 13.4 & 53.8 & 36.4 & 11.7 & 57.9 & 38.4 & 13.8\\
 & 260 & 47.6 & 31.3 & 11.2 & 49.8 & 33.2 & 10.4 & 50.8 & 35.2 & 13.4\\
\addlinespace
Lupus & 52 & 45.8 & 26.0 & 6.4 & 44.5 & 32.3 & 15.0 & 26.4 & 19.0 & 7.9\\
 & 104 & 40.3 & 26.1 & 12.4 & 42.9 & 32.9 & 14.6 & 29.0 & 18.8 & 8.0\\
 & 260 & 46.5 & 31.1 & 12.3 & 40.4 & 28.6 & 10.6 & 26.9 & 16.3 & 7.1\\
\addlinespace
Hypertension & 52 & 41.4 & 25.6 & 7.1 & 36.0 & 21.3 & 5.5 & 40.1 & 25.2 & 7.4\\
 & 104 & 38.7 & 24.5 & 6.2 & 35.0 & 20.7 & 5.2 & 39.4 & 24.9 & 7.5\\
 & 260 & 35.9 & 19.1 & 3.5 & 36.3 & 20.0 & 4.8 & 40.1 & 24.9 & 7.7\\
\addlinespace
Celiac disease & 52 & 13.2 & 8.4 & 0.0 & 23.7 & 13.1 & 3.0 & 24.3 & 14.6 & 4.9\\
 & 104 & 16.2 & 13.5 & 0.0 & 22.3 & 13.3 & 2.7 & 24.1 & 13.5 & 4.8\\
 & 260 & 17.9 & 6.5 & 0.0 & 18.0 & 9.9 & 2.3 & 20.6 & 11.5 & 3.8\\
\addlinespace
Hyperlipidemia & 52 & 35.3 & 21.4 & 6.2 & 27.5 & 15.9 & 3.6 & 36.4 & 21.7 & 5.6\\
 & 104 & 34.0 & 20.9 & 5.3 & 26.3 & 15.0 & 3.5 & 35.1 & 21.1 & 5.1\\
 & 260 & 32.9 & 21.4 & 6.1 & 26.0 & 13.5 & 2.4 & 33.9 & 19.1 & 5.6\\
\addlinespace
Depression & 52 & 31.4 & 20.7 & 6.4 & 22.8 & 13.2 & 3.4 & 29.8 & 18.1 & 5.5\\
 & 104 & 26.1 & 15.5 & 4.5 & 19.9 & 11.1 & 2.5 & 27.6 & 16.1 & 5.1\\
 & 260 & 25.2 & 16.4 & 3.6 & 22.3 & 13.1 & 3.6 & 24.6 & 14.4 & 4.4\\
\addlinespace
Chronic kidney disease & 52 & 63.0 & 46.8 & 21.4 & 64.9 & 48.7 & 20.2 & 65.4 & 48.4 & 21.0\\
 & 104 & 61.4 & 44.9 & 17.7 & 62.8 & 46.1 & 19.2 & 63.8 & 47.7 & 20.1\\
 & 260 & 57.1 & 42.9 & 16.0 & 59.5 & 43.5 & 17.2 & 57.1 & 40.6 & 14.4\\
\addlinespace
Atrial fibrillation & 52 & 53.7 & 36.2 & 12.2 & 52.5 & 37.3 & 16.0 & 54.5 & 37.4 & 16.1\\
 & 104 & 55.8 & 40.5 & 18.2 & 48.6 & 33.2 & 13.6 & 54.4 & 39.0 & 17.8\\
 & 260 & 55.1 & 35.5 & 12.9 & 48.4 & 33.3 & 12.9 & 54.7 & 40.0 & 14.8\\
\addlinespace
Chronic obstructive pulmonary disease & 52 & 54.1 & 38.2 & 12.5 & 48.0 & 32.2 & 11.3 & 48.2 & 32.4 & 9.4\\
 & 104 & 50.7 & 34.6 & 10.8 & 44.3 & 28.4 & 8.9 & 44.5 & 29.2 & 9.2\\
 & 260 & 48.7 & 33.8 & 10.3 & 41.6 & 27.6 & 8.6 & 39.3 & 23.5 & 6.8\\
\addlinespace
Alzheimer's disease & 52 & 70.6 & 56.7 & 25.5 & 67.0 & 50.7 & 20.6 & 65.9 & 49.4 & 22.8\\
 & 104 & 62.1 & 43.5 & 18.7 & 64.4 & 48.5 & 18.9 & 63.8 & 47.3 & 18.8\\
 & 260 & 62.2 & 47.1 & 15.9 & 61.3 & 46.5 & 18.5 & 62.6 & 47.7 & 16.3\\
\addlinespace
Rheumatoid arthritis & 52 & 41.3 & 29.4 & 18.3 & 36.0 & 22.3 & 8.0 & 41.9 & 30.1 & 10.6\\
 & 104 & 40.4 & 25.6 & 12.5 & 37.5 & 23.9 & 8.2 & 40.7 & 28.8 & 11.4\\
 & 260 & 34.5 & 23.8 & 8.2 & 33.9 & 21.7 & 6.8 & 32.9 & 22.1 & 7.7\\
\addlinespace
\multicolumn{11}{@{}l}{\textit{Rare-cancer endpoints}} \\
Pancreatic cancer & 26 & 52.6 & 21.6 & 8.1 & 29.7 & 17.3 & 4.3 & 60.0 & 37.1 & 16.3\\
 & 52 & 60.4 & 34.4 & 6.8 & 28.7 & 16.4 & 4.0 & 54.2 & 35.5 & 15.7\\
 & 104 & 39.0 & 25.9 & 8.6 & 27.5 & 13.9 & 4.1 & 48.7 & 34.2 & 17.3\\
 & 260 & 32.2 & 23.9 & 8.4 & 21.7 & 10.4 & 1.9 & 41.3 & 28.2 & 12.6\\
\addlinespace
Ovarian cancer & 26 & 19.8 & 9.7 & 0.0 & 26.7 & 15.9 & 2.3 & 46.1 & 38.9 & 10.8\\
 & 52 & 30.9 & 12.3 & 0.0 & 25.3 & 14.9 & 2.8 & 45.3 & 35.7 & 8.3\\
 & 104 & 40.5 & 24.7 & 6.3 & 23.6 & 12.8 & 2.8 & 45.7 & 34.7 & 10.0\\
 & 260 & 31.0 & 19.3 & 3.6 & 17.5 & 9.3 & 2.2 & 33.0 & 22.7 & 7.6\\
\addlinespace
Glioblastoma & 26 & 70.6 & 57.2 & 42.5 & 31.4 & 23.1 & 8.2 & 60.5 & 50.0 & 30.5\\
 & 52 & 65.3 & 41.1 & 28.1 & 32.4 & 22.9 & 9.3 & 58.7 & 48.7 & 25.1\\
 & 104 & 51.0 & 34.9 & 25.1 & 31.1 & 21.2 & 8.1 & 55.4 & 45.3 & 20.7\\
 & 260 & 33.9 & 26.9 & 16.1 & 29.4 & 19.6 & 6.4 & 52.6 & 42.0 & 19.8\\
\addlinespace
Multiple myeloma & 26 & 55.4 & 45.8 & 27.9 & 39.2 & 25.0 & 10.0 & 64.5 & 58.6 & 37.5\\
 & 52 & 56.0 & 44.6 & 30.4 & 37.1 & 24.3 & 7.9 & 64.2 & 56.5 & 34.1\\
 & 104 & 61.5 & 48.8 & 22.0 & 37.3 & 24.9 & 7.9 & 64.4 & 54.7 & 28.6\\
 & 260 & 54.1 & 47.3 & 22.8 & 29.9 & 19.8 & 6.3 & 60.7 & 46.9 & 23.6\\
\addlinespace
\end{longtable}
\endgroup

\begin{landscape}
\begingroup
\footnotesize
\setlength{\tabcolsep}{4pt}
\renewcommand{\arraystretch}{1.12}
\setlength{\LTcapwidth}{\linewidth}
\setlength{\LTleft}{\fill}
\setlength{\LTright}{\fill}
\begin{longtable}{@{}p{0.70in}>{\raggedleft\arraybackslash}p{0.68in}>{\raggedleft\arraybackslash}p{0.75in}>{\raggedleft\arraybackslash}p{1.47in}>{\raggedleft\arraybackslash}p{1.47in}>{\raggedleft\arraybackslash}p{0.76in}>{\raggedleft\arraybackslash}p{1.65in}@{}}
\caption{{\color{black}\textbf{TRACER disease detection, PPV and fold enrichment at one year.}
Results at retrospectively selected, cohort-specific thresholds targeting 95\% specificity, using the frozen decoder-based TRACER representations and disease-prediction heads trained on 100\% of the in-domain downstream training cohort and applied unchanged externally. Sensitivity, PPV and case yield are shown with the existing 95\% confidence intervals in parentheses. Baseline incidence is the matched IPCW-weighted one-year case pool divided by the scored-record denominator. Fold enrichment equals PPV divided by that baseline incidence.}} \label{tab:tracer-ppv-enrichment}
\\
\toprule
{\color{black}Cohort} & {\color{black}\shortstack{Scored\\records}} & {\color{black}\shortstack{Baseline\\incidence (\%)}} & {\color{black}\shortstack{Sensitivity (\%)\\(95\% CI)}} & {\color{black}\shortstack{PPV (\%)\\(95\% CI)}} & {\color{black}\shortstack{Enrichment\\(fold)}} & {\color{black}\shortstack{Cases per 10,000\\(95\% CI)}} \\
\midrule
\endfirsthead
\multicolumn{7}{@{}l}{\textit{TRACER disease detection, PPV and fold enrichment at one year (continued)}}\\
\toprule
{\color{black}Cohort} & {\color{black}\shortstack{Scored\\records}} & {\color{black}\shortstack{Baseline\\incidence (\%)}} & {\color{black}\shortstack{Sensitivity (\%)\\(95\% CI)}} & {\color{black}\shortstack{PPV (\%)\\(95\% CI)}} & {\color{black}\shortstack{Enrichment\\(fold)}} & {\color{black}\shortstack{Cases per 10,000\\(95\% CI)}} \\
\midrule
\endhead
\midrule
\multicolumn{7}{r@{}}{\footnotesize Continued on next page}\\
\endfoot
\bottomrule
\endlastfoot
\multicolumn{7}{@{}l}{\textit{Heart failure}}\\*
In-domain & 28,800 & 1.779 & 44.18 (39.46--50.63) & 13.80 (11.86--15.95) & 7.76 & 78.59 (70.19--90.07)\\*
UC5 & 919,344 & 1.399 & 36.88 (36.02--37.94) & 9.47 (9.19--9.76) & 6.77 & 51.59 (50.40--53.07)\\*
UC6 & 283,392 & 2.842 & 33.57 (32.45--35.14) & 16.41 (15.85--17.14) & 5.78 & 95.39 (92.23--99.86)\\
\addlinespace
\multicolumn{7}{@{}l}{\textit{Acute myocardial infarction}}\\*
In-domain & 29,112 & 0.362 & 42.04 (31.15--54.81) & 2.96 (2.00--4.04) & 8.19 & 15.20 (11.26--19.82)\\*
UC5 & 927,720 & 0.520 & 36.21 (34.69--37.75) & 3.65 (3.47--3.84) & 7.02 & 18.83 (18.04--19.63)\\*
UC6 & 145,560 & 0.405 & 39.59 (34.99--44.16) & 3.12 (2.65--3.58) & 7.70 & 16.02 (14.16--17.87)\\
\addlinespace
\multicolumn{7}{@{}l}{\textit{Lupus}}\\*
In-domain & 29,160 & 0.168 & 25.98 (12.56--39.52) & 0.87 (0.39--1.38) & 5.16 & 4.38 (2.12--6.66)\\*
UC5 & 927,592 & 0.192 & 32.31 (29.91--34.76) & 1.23 (1.11--1.34) & 6.40 & 6.19 (5.73--6.66)\\*
UC6 & 290,896 & 0.192 & 18.95 (15.10--22.56) & 0.72 (0.56--0.88) & 3.77 & 3.63 (2.90--4.33)\\
\addlinespace
\multicolumn{7}{@{}l}{\textit{Hypertension}}\\*
In-domain & 24,920 & 7.685 & 25.57 (23.02--27.87) & 29.84 (27.52--32.07) & 3.88 & 196.49 (176.94--214.22)\\*
UC5 & 787,624 & 7.816 & 21.32 (20.88--21.70) & 26.55 (26.13--26.92) & 3.40 & 166.60 (163.17--169.60)\\*
UC6 & 227,352 & 10.056 & 25.20 (24.44--25.93) & 36.04 (35.18--36.78) & 3.58 & 253.45 (245.81--260.71)\\
\addlinespace
\multicolumn{7}{@{}l}{\textit{Hyperlipidemia}}\\*
In-domain & 26,360 & 7.057 & 21.43 (19.11--23.80) & 24.56 (22.13--26.80) & 3.48 & 151.25 (134.84--167.92)\\*
UC5 & 827,600 & 9.777 & 15.87 (15.55--16.21) & 25.60 (25.19--26.03) & 2.62 & 155.20 (152.00--158.44)\\*
UC6 & 131,952 & 7.068 & 21.69 (20.49--22.85) & 24.81 (23.65--25.89) & 3.51 & 153.31 (144.81--161.54)\\
\addlinespace
\multicolumn{7}{@{}l}{\textit{Depression}}\\*
In-domain & 27,408 & 5.482 & 20.74 (18.28--23.39) & 19.40 (17.33--21.61) & 3.54 & 113.69 (100.23--128.19)\\*
UC5 & 870,824 & 5.677 & 13.16 (12.77--13.50) & 13.68 (13.31--13.99) & 2.41 & 74.73 (72.52--76.61)\\*
UC6 & 136,840 & 5.384 & 18.10 (16.86--19.23) & 17.08 (16.02--18.03) & 3.17 & 97.46 (90.75--103.55)\\
\addlinespace
\multicolumn{7}{@{}l}{\textit{Chronic kidney disease}}\\*
In-domain & 28,664 & 1.962 & 46.82 (41.72--52.11) & 15.78 (13.74--17.90) & 8.04 & 91.87 (81.86--102.25)\\*
UC5 & 910,032 & 1.707 & 48.65 (47.70--49.59) & 14.45 (14.10--14.79) & 8.47 & 83.04 (81.42--84.64)\\*
UC6 & 143,312 & 1.920 & 48.43 (46.37--50.76) & 15.94 (15.17--16.92) & 8.30 & 92.99 (89.03--97.46)\\
\addlinespace
\multicolumn{7}{@{}l}{\textit{Atrial fibrillation}}\\*
In-domain & 28,912 & 1.409 & 36.22 (30.38--41.83) & 9.39 (7.69--11.02) & 6.66 & 51.05 (42.82--58.95)\\*
UC5 & 921,096 & 1.517 & 37.26 (36.25--38.18) & 10.30 (9.99--10.58) & 6.79 & 56.53 (55.01--57.93)\\*
UC6 & 144,496 & 1.355 & 37.39 (34.71--40.03) & 9.32 (8.57--10.05) & 6.87 & 50.69 (47.05--54.26)\\
\addlinespace
\multicolumn{7}{@{}l}{\textit{Chronic obstructive pulmonary disease}}\\*
In-domain & 28,928 & 1.120 & 38.20 (31.58--44.88) & 7.97 (6.31--9.53) & 7.11 & 42.79 (35.38--50.28)\\*
UC5 & 921,952 & 0.966 & 32.23 (31.07--33.35) & 5.92 (5.68--6.14) & 6.12 & 31.14 (30.02--32.22)\\*
UC6 & 148,520 & 1.025 & 32.43 (29.76--35.38) & 6.29 (5.69--6.93) & 6.14 & 33.23 (30.49--36.25)\\
\addlinespace
\multicolumn{7}{@{}l}{\textit{Rheumatoid arthritis}}\\*
In-domain & 29,080 & 0.358 & 29.41 (19.57--40.80) & 2.07 (1.29--3.06) & 5.78 & 10.54 (7.01--14.62)\\*
UC5 & 925,400 & 0.400 & 22.29 (20.76--23.72) & 1.76 (1.62--1.89) & 4.40 & 8.90 (8.29--9.48)\\*
UC6 & 145,496 & 0.348 & 30.06 (25.55--34.58) & 2.06 (1.69--2.41) & 5.91 & 10.46 (8.89--12.03)\\
\addlinespace
\multicolumn{7}{@{}l}{\textit{Celiac disease}}\\*
In-domain & 29,136 & 0.114 & 8.42 (0.00--21.03) & 0.19 (0.00--0.48) & 1.68 & 0.96 (0.00--2.40)\\*
UC5 & 929,144 & 0.226 & 13.13 (11.45--14.68) & 0.59 (0.51--0.67) & 2.62 & 2.96 (2.59--3.32)\\*
UC6 & 291,416 & 0.237 & 14.61 (11.76--17.94) & 0.69 (0.55--0.85) & 2.91 & 3.46 (2.78--4.24)\\
\addlinespace
\multicolumn{7}{@{}l}{\textit{Alzheimer's disease}}\\*
In-domain & 29,072 & 0.622 & 56.69 (46.97--65.53) & 6.62 (5.08--8.07) & 10.65 & 35.24 (29.20--40.73)\\*
UC5 & 926,224 & 0.625 & 50.67 (49.31--52.22) & 5.99 (5.77--6.22) & 9.59 & 31.68 (30.83--32.65)\\*
UC6 & 144,792 & 0.667 & 49.45 (45.25--53.26) & 6.22 (5.59--6.85) & 9.34 & 32.96 (30.17--35.51)\\
\addlinespace
\multicolumn{7}{@{}l}{\textit{Pancreatic cancer}}\\*
In-domain & 29,176 & 0.104 & 34.42 (16.43--56.30) & 0.71 (0.26--1.25) & 6.84 & 3.59 (1.72--5.88)\\*
UC5 & 930,896 & 0.089 & 16.37 (13.60--19.10) & 0.29 (0.24--0.34) & 3.27 & 1.45 (1.21--1.69)\\*
UC6 & 145,880 & 0.114 & 35.48 (27.93--43.53) & 0.80 (0.59--1.06) & 7.05 & 4.05 (3.19--4.97)\\
\addlinespace
\multicolumn{7}{@{}l}{\textit{Ovarian cancer}}\\*
In-domain & 16,176 & 0.148 & 12.33 (0.00--34.94) & 0.36 (0.00--1.02) & 2.46 & 1.82 (0.00--5.17)\\*
UC5 & 522,864 & 0.143 & 14.87 (12.14--17.90) & 0.42 (0.34--0.51) & 2.96 & 2.12 (1.73--2.55)\\*
UC6 & 80,872 & 0.184 & 35.74 (26.34--44.43) & 1.30 (0.91--1.71) & 7.07 & 6.59 (4.86--8.20)\\
\addlinespace
\multicolumn{7}{@{}l}{\textit{Glioblastoma}}\\*
In-domain & 29,072 & 0.206 & 41.11 (26.32--56.30) & 1.67 (0.95--2.48) & 8.10 & 8.47 (5.42--11.59)\\*
UC5 & 928,304 & 0.183 & 22.94 (20.84--25.23) & 0.84 (0.75--0.93) & 4.56 & 4.20 (3.82--4.62)\\*
UC6 & 145,360 & 0.245 & 48.71 (42.16--54.97) & 2.34 (1.95--2.76) & 9.54 & 11.96 (10.35--13.49)\\
\addlinespace
\multicolumn{7}{@{}l}{\textit{Multiple myeloma}}\\*
In-domain & 29,128 & 0.226 & 44.63 (32.12--59.35) & 1.98 (1.25--2.89) & 8.77 & 10.07 (7.25--13.39)\\*
UC5 & 930,296 & 0.091 & 24.32 (20.91--27.49) & 0.44 (0.37--0.50) & 4.85 & 2.20 (1.89--2.49)\\*
UC6 & 145,632 & 0.234 & 56.52 (50.44--62.61) & 2.58 (2.15--3.00) & 11.04 & 13.23 (11.80--14.65)\\
\addlinespace
\end{longtable}
\endgroup
\end{landscape}

\section{Additional supporting tables}\label{sec:additional-tables}
These tables are not explicitly cited in the main Introduction, Results or Discussion. Tables cited only in the main Methods appear first, followed by the tables supporting the detailed analyses in Section~\ref{sec:model-comparison-details}.

\begin{table}[H]
\centering
\caption{\small
\textbf{Pooling-mask and pooling-strategy ablations.}
Uno's IPCW C-index is shown for four representative chronic disease endpoints under the local survival-head training protocol. 'Mean + c/t' denotes uniform mean pooling with care-site and inter-visit time-gap token embeddings retained, matching the primary pooling configuration. 'Mean, no c/t' excludes these token embeddings from the pooling average but does not remove them from the transformer input. 'SIF, no c/t' applies smooth inverse-frequency weighting after excluding care-site and time-gap token embeddings. Padding, boundary, and model-control tokens were excluded from pooling in all conditions. External-cohort values reflect survival heads trained and evaluated using the corresponding health
system's local data partitions.
}
\label{tab:supp-pooling}
\begingroup
\scriptsize
\setlength{\tabcolsep}{4pt}
\renewcommand{\arraystretch}{1.10}
\begin{adjustbox}{max width=\linewidth}
\begin{tabular}{@{}llrrrrrr@{}}
\toprule
& &
\multicolumn{3}{c}{TRACER} &
\multicolumn{3}{c}{TRACER-Encoder} \\
\cmidrule(lr){3-5}
\cmidrule(lr){6-8}
Disease endpoint
& Cohort
& \shortstack{Mean\\+ c/t}
& \shortstack{Mean,\\no c/t}
& \shortstack{SIF,\\no c/t}
& \shortstack{Mean\\+ c/t}
& \shortstack{Mean,\\no c/t}
& \shortstack{SIF,\\no c/t} \\
\midrule
\multirow{3}{*}{Heart failure}
& In-domain & 0.877 & 0.876 & 0.879 & 0.838 & 0.825 & 0.827 \\
& UC5       & 0.905 & 0.903 & 0.903 & 0.875 & 0.866 & 0.854 \\
& UC6       & 0.865 & 0.866 & 0.867 & 0.843 & 0.846 & 0.843 \\
\addlinespace
\multirow{3}{*}{Acute myocardial infarction}
& In-domain & 0.843 & 0.837 & 0.848 & 0.819 & 0.811 & 0.798 \\
& UC5       & 0.875 & 0.877 & 0.870 & 0.837 & 0.808 & 0.819 \\
& UC6       & 0.827 & 0.823 & 0.839 & 0.775 & 0.774 & 0.782 \\
\addlinespace
\multirow{3}{*}{Lupus}
& In-domain & 0.811 & 0.827 & 0.787 & 0.692 & 0.691 & 0.656 \\
& UC5       & 0.843 & 0.841 & 0.843 & 0.794 & 0.765 & 0.769 \\
& UC6       & 0.739 & 0.738 & 0.745 & 0.665 & 0.669 & 0.664 \\
\addlinespace
\multirow{3}{*}{Hypertension}
& In-domain & 0.812 & 0.812 & 0.810 & 0.784 & 0.779 & 0.774 \\
& UC5       & 0.820 & 0.822 & 0.818 & 0.808 & 0.807 & 0.804 \\
& UC6       & 0.814 & 0.813 & 0.810 & 0.794 & 0.795 & 0.786 \\
\bottomrule
\end{tabular}
\end{adjustbox}
\par\smallskip
\footnotesize
Abbreviations: c/t, care-site and inter-visit time-gap tokens;
SIF, smooth inverse-frequency pooling. Summary medians and maximum changes
in the text were calculated across all four downstream training fractions;
the table reports only the 100\% fraction.
\endgroup
\end{table}

\begin{table}[H]
\centering
\caption{\small
\textbf{C-index under DeepHit and Cox proportional-hazards survival heads.}
Each entry reports DeepHit/CoxPH C-index at the 100\% local training fraction for the same frozen TRACER representation. Heads were trained using the corresponding cohort's training partition and evaluated on its test partition under the local head-training protocol. Values below 0.5 indicate worse-than-chance concordance under the risk-score orientation used in the analysis. D, TRACER; E, TRACER-Encoder.
}
\label{tab:supp-coxph}

\scriptsize
\setlength{\tabcolsep}{4pt}
\renewcommand{\arraystretch}{1.12}

\begin{tabular*}{\textwidth}{
@{\extracolsep{\fill}}
lcccccc
@{}
}
\toprule
& \multicolumn{2}{c}{In-domain}
& \multicolumn{2}{c}{UC5}
& \multicolumn{2}{c}{UC6} \\
\cmidrule(lr){2-3}
\cmidrule(lr){4-5}
\cmidrule(lr){6-7}
Disease endpoint
& D & E
& D & E
& D & E \\
\midrule
Heart failure
& \dhcox{0.877}{0.506}
& \dhcox{0.838}{0.611}
& \dhcox{0.905}{0.530}
& \dhcox{0.875}{0.431}
& \dhcox{0.865}{0.533}
& \dhcox{0.843}{0.403} \\
Acute myocardial infarction
& \dhcox{0.843}{0.669}
& \dhcox{0.819}{0.734}
& \dhcox{0.875}{0.588}
& \dhcox{0.837}{0.353}
& \dhcox{0.827}{0.581}
& \dhcox{0.775}{0.473} \\
Lupus
& \dhcox{0.811}{0.641}
& \dhcox{0.692}{0.535}
& \dhcox{0.843}{0.429}
& \dhcox{0.794}{0.488}
& \dhcox{0.739}{0.473}
& \dhcox{0.665}{0.496} \\
Hypertension
& \dhcox{0.812}{0.576}
& \dhcox{0.784}{0.572}
& \dhcox{0.820}{0.667}
& \dhcox{0.808}{0.595}
& \dhcox{0.814}{0.437}
& \dhcox{0.794}{0.558} \\
Celiac disease
& \dhcox{0.713}{0.560}
& \dhcox{0.637}{0.556}
& \dhcox{0.728}{0.443}
& \dhcox{0.708}{0.491}
& \dhcox{0.748}{0.455}
& \dhcox{0.669}{0.477} \\
Hyperlipidemia
& \dhcox{0.794}{0.437}
& \dhcox{0.768}{0.445}
& \dhcox{0.761}{0.509}
& \dhcox{0.756}{0.461}
& \dhcox{0.794}{0.544}
& \dhcox{0.759}{0.563} \\
Depression
& \dhcox{0.740}{0.552}
& \dhcox{0.695}{0.582}
& \dhcox{0.716}{0.477}
& \dhcox{0.700}{0.509}
& \dhcox{0.745}{0.456}
& \dhcox{0.694}{0.552} \\
Chronic kidney disease
& \dhcox{0.886}{0.269}
& \dhcox{0.831}{0.476}
& \dhcox{0.903}{0.537}
& \dhcox{0.876}{0.350}
& \dhcox{0.883}{0.387}
& \dhcox{0.827}{0.602} \\
Atrial fibrillation
& \dhcox{0.872}{0.448}
& \dhcox{0.802}{0.376}
& \dhcox{0.883}{0.306}
& \dhcox{0.851}{0.372}
& \dhcox{0.876}{0.373}
& \dhcox{0.803}{0.372} \\
Chronic obstructive pulmonary disease
& \dhcox{0.856}{0.344}
& \dhcox{0.772}{0.603}
& \dhcox{0.844}{0.567}
& \dhcox{0.823}{0.609}
& \dhcox{0.829}{0.528}
& \dhcox{0.747}{0.520} \\
Alzheimer's disease
& \dhcox{0.912}{0.370}
& \dhcox{0.806}{0.626}
& \dhcox{0.918}{0.577}
& \dhcox{0.865}{0.362}
& \dhcox{0.877}{0.424}
& \dhcox{0.791}{0.575} \\
Rheumatoid arthritis
& \dhcox{0.794}{0.508}
& \dhcox{0.687}{0.481}
& \dhcox{0.823}{0.531}
& \dhcox{0.755}{0.587}
& \dhcox{0.809}{0.459}
& \dhcox{0.700}{0.569} \\
\bottomrule
\end{tabular*}
\par\smallskip
\footnotesize
Entries are DeepHit/CoxPH. All DeepHit estimates exceeded the corresponding CoxPH estimate.
\end{table}
\clearpage

\begin{table}[H]
\centering
\caption{\small \textbf{Integrated Brier score (IBS) and integrated negative binomial log-likelihood (IBLL).} Probabilistic performance integrated over the evaluation grid (truncated at 260 weeks) for TRACER (D) and TRACER-Encoder (E), survival heads trained on the full in-domain training cohort and applied unchanged externally. Lower is better.}
\label{tab:ibs-ibll}
\begingroup
\small
\setlength{\tabcolsep}{3pt}
\renewcommand{\arraystretch}{1.10}
\begin{adjustbox}{max width=\linewidth}
\begin{tabular}{@{}llrrrrrr@{}}
\toprule
& & \multicolumn{3}{c}{IBS} & \multicolumn{3}{c}{IBLL} \\
\cmidrule(lr){3-5}\cmidrule(lr){6-8}
Disease & Model & In-domain & UC5 & UC6 & In-domain & UC5 & UC6 \\
\midrule
\multicolumn{8}{@{}l}{\textit{Chronic disease endpoints}} \\
Heart failure & D & 0.030 & 0.032 & 0.050 & 0.118 & 0.122 & 0.187 \\
 & E & 0.030 & 0.032 & 0.048 & 0.128 & 0.132 & 0.181 \\
Acute myocardial infarction & D & 0.010 & 0.014 & 0.010 & 0.049 & 0.065 & 0.050 \\
 & E & 0.010 & 0.014 & 0.010 & 0.053 & 0.069 & 0.053 \\
Lupus & D & 0.003 & 0.002 & 0.004 & 0.017 & 0.015 & 0.025 \\
 & E & 0.003 & 0.002 & 0.004 & 0.019 & 0.017 & 0.025 \\
Hypertension & D & 0.107 & 0.122 & 0.125 & 0.350 & 0.387 & 0.394 \\
 & E & 0.111 & 0.125 & 0.128 & 0.374 & 0.406 & 0.411 \\
Celiac disease & D & 0.004 & 0.006 & 0.004 & 0.026 & 0.038 & 0.029 \\
 & E & 0.004 & 0.006 & 0.004 & 0.025 & 0.036 & 0.028 \\
Hyperlipidemia & D & 0.104 & 0.171 & 0.110 & 0.340 & 0.518 & 0.355 \\
 & E & 0.108 & 0.163 & 0.114 & 0.364 & 0.497 & 0.378 \\
Depression & D & 0.087 & 0.066 & 0.089 & 0.306 & 0.252 & 0.310 \\
 & E & 0.088 & 0.068 & 0.090 & 0.315 & 0.264 & 0.319 \\
Chronic kidney disease & D & 0.031 & 0.021 & 0.031 & 0.123 & 0.096 & 0.122 \\
 & E & 0.032 & 0.019 & 0.032 & 0.135 & 0.094 & 0.135 \\
Atrial fibrillation & D & 0.028 & 0.034 & 0.029 & 0.114 & 0.134 & 0.117 \\
 & E & 0.029 & 0.034 & 0.030 & 0.125 & 0.142 & 0.128 \\
Chronic obstructive pulmonary disease & D & 0.020 & 0.013 & 0.022 & 0.087 & 0.063 & 0.094 \\
 & E & 0.020 & 0.013 & 0.022 & 0.092 & 0.069 & 0.100 \\
Alzheimer's disease & D & 0.014 & 0.016 & 0.014 & 0.061 & 0.072 & 0.062 \\
 & E & 0.014 & 0.017 & 0.014 & 0.068 & 0.078 & 0.069 \\
Rheumatoid arthritis & D & 0.007 & 0.010 & 0.006 & 0.039 & 0.054 & 0.036 \\
 & E & 0.007 & 0.010 & 0.006 & 0.039 & 0.052 & 0.037 \\
\addlinespace
\multicolumn{8}{@{}l}{\textit{Rare-cancer endpoints}} \\
Pancreatic cancer & D & 0.002 & 0.001 & 0.002 & 0.013 & 0.007 & 0.013 \\
 & E & 0.002 & 0.001 & 0.002 & 0.013 & 0.008 & 0.012 \\
Ovarian cancer & D & 0.003 & 0.002 & 0.003 & 0.022 & 0.014 & 0.021 \\
 & E & 0.003 & 0.002 & 0.003 & 0.021 & 0.015 & 0.020 \\
Glioblastoma & D & 0.004 & 0.002 & 0.004 & 0.023 & 0.014 & 0.022 \\
 & E & 0.004 & 0.002 & 0.004 & 0.023 & 0.014 & 0.021 \\
Multiple myeloma & D & 0.004 & 0.001 & 0.003 & 0.021 & 0.008 & 0.019 \\
 & E & 0.004 & 0.001 & 0.003 & 0.021 & 0.008 & 0.019 \\
\bottomrule
\end{tabular}
\end{adjustbox}

\endgroup
\end{table}

\begin{table}[H]
\centering
\caption{\small
\textbf{Local survival-head training compared with zero-shot cross-site transfer.} Time-dependent C-index for survival heads trained using local labels from the corresponding held-out health system and for development-trained heads transferred without local updating. Each entry reports local/zero-shot C-index, with the local-minus-zero-shot difference in parentheses. Positive differences indicate higher global discrimination after local head training. The foundation-model backbone remained frozen under both protocols. Values were calculated using the time-dependent concordance index implemented in \texttt{pycox} and are not
directly comparable with the Uno IPCW C-index estimates reported in the primary analysis. No uncertainty intervals were calculated for these descriptive differences.
}
\label{tab:supp-local-vs-zs}
\begingroup
\scriptsize
\setlength{\tabcolsep}{3pt}
\renewcommand{\arraystretch}{1.12}
\begin{adjustbox}{max width=\linewidth}
\begin{tabular}{@{}lcccc@{}}
\toprule
& \multicolumn{2}{c}{UC5}
& \multicolumn{2}{c}{UC6} \\
\cmidrule(lr){2-3}
\cmidrule(lr){4-5}
Disease endpoint
& TRACER
& TRACER-Encoder
& TRACER
& TRACER-Encoder \\
\midrule
Heart failure
& \localzscell{0.905}{0.850}{+0.055}
& \localzscell{0.875}{0.818}{+0.057}
& \localzscell{0.865}{0.849}{+0.016}
& \localzscell{0.843}{0.789}{+0.054} \\
Acute myocardial infarction
& \localzscell{0.875}{0.803}{+0.072}
& \localzscell{0.837}{0.757}{+0.080}
& \localzscell{0.827}{0.864}{-0.037}
& \localzscell{0.775}{0.787}{-0.012} \\
Lupus
& \localzscell{0.843}{0.784}{+0.059}
& \localzscell{0.794}{0.678}{+0.116}
& \localzscell{0.739}{0.706}{+0.033}
& \localzscell{0.665}{0.653}{+0.012} \\
Hypertension
& \localzscell{0.820}{0.778}{+0.042}
& \localzscell{0.808}{0.724}{+0.084}
& \localzscell{0.814}{0.787}{+0.027}
& \localzscell{0.794}{0.735}{+0.059} \\
Celiac disease
& \localzscell{0.728}{0.661}{+0.067}
& \localzscell{0.708}{0.626}{+0.082}
& \localzscell{0.748}{0.672}{+0.076}
& \localzscell{0.669}{0.671}{-0.002} \\
Hyperlipidemia
& \localzscell{0.761}{0.726}{+0.035}
& \localzscell{0.756}{0.706}{+0.050}
& \localzscell{0.794}{0.790}{+0.004}
& \localzscell{0.759}{0.763}{-0.004} \\
Depression
& \localzscell{0.716}{0.652}{+0.064}
& \localzscell{0.700}{0.630}{+0.070}
& \localzscell{0.745}{0.732}{+0.013}
& \localzscell{0.694}{0.697}{-0.003} \\
Chronic kidney disease
& \localzscell{0.903}{0.872}{+0.031}
& \localzscell{0.876}{0.817}{+0.059}
& \localzscell{0.883}{0.887}{-0.004}
& \localzscell{0.827}{0.819}{+0.008} \\
Atrial fibrillation
& \localzscell{0.883}{0.840}{+0.043}
& \localzscell{0.851}{0.777}{+0.074}
& \localzscell{0.876}{0.865}{+0.011}
& \localzscell{0.803}{0.771}{+0.032} \\
Chronic obstructive pulmonary disease
& \localzscell{0.844}{0.804}{+0.040}
& \localzscell{0.823}{0.756}{+0.067}
& \localzscell{0.829}{0.842}{-0.013}
& \localzscell{0.747}{0.772}{-0.025} \\
Alzheimer's disease
& \localzscell{0.918}{0.879}{+0.039}
& \localzscell{0.865}{0.786}{+0.079}
& \localzscell{0.877}{0.896}{-0.019}
& \localzscell{0.791}{0.797}{-0.006} \\
Rheumatoid arthritis
& \localzscell{0.823}{0.759}{+0.064}
& \localzscell{0.755}{0.723}{+0.032}
& \localzscell{0.809}{0.803}{+0.006}
& \localzscell{0.700}{0.701}{-0.001} \\
\midrule
\textit{Median gain}
& \(\mathbf{+0.049}\)
& \(\mathbf{+0.072}\)
& \(\mathbf{+0.009}\)
& \(\mathbf{-0.002}\) \\
\bottomrule
\end{tabular}
\end{adjustbox}
\par\smallskip
\footnotesize
Disease-level entries are local/zero-shot C-index, followed by the local-minus-zero-shot difference in parentheses. Median gains were calculated from the unrounded disease-level differences.
\endgroup
\end{table}
\clearpage
\subsection{Additional comparative clinical results}

\begingroup
\small
\setlength{\tabcolsep}{3pt}
\renewcommand{\arraystretch}{1.10}
\setlength{\LTcapwidth}{\linewidth}
\setlength{\LTleft}{\fill}
\setlength{\LTright}{\fill}
\begin{longtable}{@{}lccc@{}}
\caption{\small \textbf{Paired differences in time-dependent AUROC between TRACER and TRACER-Encoder at the 104-, 260-, and 26-week horizons.} Differences are TRACER minus TRACER-Encoder with 95\% paired patient-level bootstrap intervals; boldface indicates intervals excluding zero. The 52-week horizon is in Supplementary Table~\ref{tab:auroc-difference-52wk}. } \label{tab:auroc-difference-horizons}\\
\toprule
Disease endpoint & In-domain & UC5 & UC6 \\
\midrule
\endfirsthead
\multicolumn{4}{@{}l}{\textit{(continued)}}\\
\toprule
Disease endpoint & In-domain & UC5 & UC6 \\
\midrule
\endhead
\midrule
\multicolumn{4}{r@{}}{\footnotesize Continued on next page}\\
\endfoot
\bottomrule
\endlastfoot
\multicolumn{4}{@{}l}{\textit{104-week horizon}} \\
Heart failure & \auccib{+0.048}{+0.031}{+0.067} & \auccib{+0.031}{+0.028}{+0.035} & \auccib{+0.024}{+0.020}{+0.028}\\
Acute myocardial infarction & \auccib{+0.080}{+0.050}{+0.117} & \auccib{+0.070}{+0.065}{+0.076} & \auccib{+0.051}{+0.035}{+0.068}\\
Lupus & \auccib{+0.170}{+0.071}{+0.274} & \auccib{+0.127}{+0.114}{+0.141} & \auccib{+0.070}{+0.043}{+0.095}\\
Hypertension & \auccib{+0.027}{+0.019}{+0.035} & \auccib{+0.008}{+0.006}{+0.009} & \auccib{+0.033}{+0.031}{+0.036}\\
Celiac disease & \aucci{+0.017}{-0.066}{+0.112} & \auccib{+0.051}{+0.039}{+0.065} & \aucci{+0.007}{-0.015}{+0.027}\\
Hyperlipidemia & \auccib{+0.034}{+0.025}{+0.044} & \aucci{+0.001}{-0.001}{+0.002} & \auccib{+0.036}{+0.031}{+0.040}\\
Depression & \auccib{+0.046}{+0.033}{+0.059} & \auccib{+0.009}{+0.006}{+0.012} & \auccib{+0.032}{+0.026}{+0.038}\\
Chronic kidney disease & \auccib{+0.052}{+0.037}{+0.068} & \auccib{+0.043}{+0.040}{+0.045} & \auccib{+0.055}{+0.048}{+0.062}\\
Atrial fibrillation & \auccib{+0.069}{+0.051}{+0.088} & \auccib{+0.038}{+0.035}{+0.041} & \auccib{+0.068}{+0.060}{+0.077}\\
Chronic obstructive pulmonary disease & \auccib{+0.061}{+0.036}{+0.088} & \auccib{+0.032}{+0.028}{+0.036} & \auccib{+0.072}{+0.061}{+0.084}\\
Alzheimer's disease & \auccib{+0.105}{+0.068}{+0.143} & \auccib{+0.084}{+0.079}{+0.091} & \auccib{+0.090}{+0.075}{+0.105}\\
Rheumatoid arthritis & \auccib{+0.075}{+0.030}{+0.122} & \auccib{+0.018}{+0.009}{+0.025} & \auccib{+0.090}{+0.067}{+0.114}\\
Pancreatic cancer & \aucci{+0.016}{-0.034}{+0.075} & \auccib{-0.019}{-0.034}{-0.003} & \aucci{+0.020}{-0.005}{+0.045}\\
Ovarian cancer & \aucci{-0.018}{-0.105}{+0.063} & \aucci{+0.013}{-0.003}{+0.030} & \auccib{+0.033}{+0.008}{+0.057}\\
Glioblastoma & \aucci{-0.030}{-0.082}{+0.029} & \auccib{-0.015}{-0.025}{-0.003} & \auccib{-0.031}{-0.047}{-0.014}\\
Multiple myeloma & \aucci{+0.023}{-0.009}{+0.062} & \auccib{-0.017}{-0.029}{-0.005} & \aucci{+0.010}{-0.003}{+0.024}\\
\addlinespace
\multicolumn{4}{@{}l}{\textit{260-week horizon}} \\
Heart failure & \auccib{+0.040}{+0.022}{+0.058} & \auccib{+0.038}{+0.035}{+0.041} & \auccib{+0.022}{+0.018}{+0.026}\\
Acute myocardial infarction & \auccib{+0.072}{+0.038}{+0.113} & \auccib{+0.054}{+0.048}{+0.059} & \auccib{+0.047}{+0.029}{+0.065}\\
Lupus & \auccib{+0.119}{+0.012}{+0.252} & \auccib{+0.083}{+0.071}{+0.095} & \auccib{+0.036}{+0.014}{+0.057}\\
Hypertension & \aucci{+0.008}{-0.002}{+0.017} & \auccib{+0.025}{+0.023}{+0.027} & \auccib{+0.020}{+0.017}{+0.022}\\
Celiac disease & \aucci{-0.011}{-0.108}{+0.080} & \auccib{+0.028}{+0.014}{+0.043} & \aucci{+0.013}{-0.005}{+0.032}\\
Hyperlipidemia & \auccib{+0.027}{+0.014}{+0.040} & \auccib{+0.004}{+0.002}{+0.006} & \auccib{+0.031}{+0.026}{+0.037}\\
Depression & \auccib{+0.042}{+0.026}{+0.057} & \auccib{+0.045}{+0.042}{+0.048} & \auccib{+0.035}{+0.028}{+0.041}\\
Chronic kidney disease & \auccib{+0.033}{+0.018}{+0.048} & \auccib{+0.034}{+0.031}{+0.036} & \auccib{+0.040}{+0.034}{+0.047}\\
Atrial fibrillation & \auccib{+0.081}{+0.059}{+0.103} & \auccib{+0.051}{+0.048}{+0.055} & \auccib{+0.067}{+0.058}{+0.077}\\
Chronic obstructive pulmonary disease & \auccib{+0.064}{+0.041}{+0.090} & \auccib{+0.031}{+0.027}{+0.035} & \auccib{+0.048}{+0.037}{+0.058}\\
Alzheimer's disease & \auccib{+0.086}{+0.057}{+0.114} & \auccib{+0.080}{+0.075}{+0.086} & \auccib{+0.084}{+0.069}{+0.099}\\
Rheumatoid arthritis & \aucci{+0.030}{-0.010}{+0.074} & \auccib{+0.022}{+0.014}{+0.030} & \auccib{+0.087}{+0.060}{+0.114}\\
Pancreatic cancer & \aucci{+0.023}{-0.052}{+0.092} & \auccib{-0.014}{-0.027}{-0.001} & \aucci{+0.021}{-0.008}{+0.049}\\
Ovarian cancer & \aucci{+0.025}{-0.068}{+0.111} & \aucci{+0.016}{-0.004}{+0.032} & \aucci{+0.034}{-0.003}{+0.068}\\
Glioblastoma & \aucci{-0.016}{-0.068}{+0.038} & \auccib{-0.016}{-0.025}{-0.006} & \auccib{-0.022}{-0.040}{-0.003}\\
Multiple myeloma & \aucci{+0.004}{-0.035}{+0.046} & \auccib{-0.042}{-0.057}{-0.027} & \aucci{+0.006}{-0.012}{+0.025}\\
\addlinespace
\multicolumn{4}{@{}l}{\textit{26-week horizon (rare cancers)}} \\
Pancreatic cancer & \aucci{-0.041}{-0.100}{+0.011} & \auccib{-0.033}{-0.049}{-0.016} & \aucci{+0.004}{-0.021}{+0.029}\\
Ovarian cancer & \aucci{-0.051}{-0.229}{+0.144} & \aucci{-0.006}{-0.026}{+0.012} & \auccib{+0.037}{+0.003}{+0.073}\\
Glioblastoma & \aucci{+0.028}{-0.018}{+0.090} & \aucci{+0.003}{-0.010}{+0.018} & \aucci{-0.016}{-0.038}{+0.007}\\
Multiple myeloma & \auccib{+0.079}{+0.006}{+0.167} & \aucci{-0.012}{-0.029}{+0.005} & \auccib{+0.028}{+0.004}{+0.054}\\
\addlinespace
\end{longtable}
\endgroup

\begin{landscape}
\begingroup
\scriptsize
\setlength{\tabcolsep}{3pt}
\renewcommand{\arraystretch}{1.10}
\setlength{\LTcapwidth}{\linewidth}
\setlength{\LTleft}{\fill}
\setlength{\LTright}{\fill}
\begin{longtable}{@{}lrrrlrrlrrl@{}}
\caption{\small \textbf{Standardized partial AUROC (FPR $\leq 0.10$) for both models with paired differences.} Standardized pAUC for TRACER (D) and TRACER-Encoder (E) at each prediction horizon and cohort, with the paired difference (D minus E) and its 95\% paired bootstrap confidence interval. Bold: the interval excludes zero. $^{\dagger}$: strict reversal, in which the paired AUROC difference (Supplementary Tables~\ref{tab:auroc-difference-52wk} and~\ref{tab:auroc-difference-horizons}) and the pAUC difference both exclude zero and favor different models. } \label{tab:pauc-paired}\\
\toprule
& & \multicolumn{3}{c}{In-domain} & \multicolumn{3}{c}{UC5} & \multicolumn{3}{c}{UC6} \\
\cmidrule(lr){3-5}\cmidrule(lr){6-8}\cmidrule(lr){9-11}
Disease & Horizon (wk) & D & E & $\Delta$ (95\% CI) & D & E & $\Delta$ (95\% CI) & D & E & $\Delta$ (95\% CI) \\
\midrule
\endfirsthead
\multicolumn{11}{@{}l}{\textit{(continued)}}\\
\toprule
& & \multicolumn{3}{c}{In-domain} & \multicolumn{3}{c}{UC5} & \multicolumn{3}{c}{UC6} \\
\cmidrule(lr){3-5}\cmidrule(lr){6-8}\cmidrule(lr){9-11}
Disease & Horizon (wk) & D & E & $\Delta$ (95\% CI) & D & E & $\Delta$ (95\% CI) & D & E & $\Delta$ (95\% CI) \\
\midrule
\endhead
\midrule
\multicolumn{11}{r@{}}{\footnotesize Continued on next page}\\
\endfoot
\bottomrule
\endlastfoot
\multicolumn{11}{@{}l}{\textit{Chronic disease endpoints}} \\
Heart failure & 52 & 0.692 & 0.672 & \textbf{+0.020 (+0.003, +0.037)} & 0.655 & 0.672 & \textbf{-0.017 (-0.020, -0.013)}$^{\dagger}$ & 0.640 & 0.649 & \textbf{-0.009 (-0.014, -0.005)}$^{\dagger}$\\
 & 104 & 0.686 & 0.659 & \textbf{+0.027 (+0.011, +0.043)} & 0.662 & 0.674 & \textbf{-0.012 (-0.015, -0.009)}$^{\dagger}$ & 0.635 & 0.648 & \textbf{-0.013 (-0.017, -0.009)}$^{\dagger}$\\
 & 260 & 0.677 & 0.649 & \textbf{+0.028 (+0.008, +0.048)} & 0.683 & 0.683 & -0.001 (-0.004, +0.003) & 0.631 & 0.652 & \textbf{-0.021 (-0.026, -0.016)}$^{\dagger}$\\
\addlinespace
Acute myocardial infarction & 52 & 0.664 & 0.640 & +0.024 (-0.017, +0.064) & 0.652 & 0.631 & \textbf{+0.022 (+0.015, +0.027)} & 0.668 & 0.650 & \textbf{+0.018 (+0.001, +0.035)}\\
 & 104 & 0.703 & 0.644 & \textbf{+0.059 (+0.025, +0.096)} & 0.651 & 0.633 & \textbf{+0.018 (+0.013, +0.023)} & 0.664 & 0.660 & +0.004 (-0.012, +0.019)\\
 & 260 & 0.629 & 0.602 & +0.026 (-0.010, +0.063) & 0.637 & 0.614 & \textbf{+0.023 (+0.018, +0.029)} & 0.647 & 0.637 & +0.010 (-0.007, +0.026)\\
\addlinespace
Lupus & 52 & 0.608 & 0.582 & +0.025 (-0.023, +0.081) & 0.635 & 0.597 & \textbf{+0.038 (+0.028, +0.049)} & 0.565 & 0.556 & +0.010 (-0.007, +0.027)\\
 & 104 & 0.607 & 0.584 & +0.022 (-0.035, +0.079) & 0.633 & 0.593 & \textbf{+0.040 (+0.031, +0.050)} & 0.570 & 0.555 & +0.015 (-0.002, +0.032)\\
 & 260 & 0.628 & 0.560 & +0.068 (-0.018, +0.158) & 0.613 & 0.550 & \textbf{+0.063 (+0.054, +0.073)} & 0.561 & 0.543 & \textbf{+0.018 (+0.005, +0.031)}\\
\addlinespace
Hypertension & 52 & 0.601 & 0.591 & +0.010 (-0.000, +0.020) & 0.581 & 0.584 & \textbf{-0.003 (-0.005, -0.002)}$^{\dagger}$ & 0.599 & 0.589 & \textbf{+0.010 (+0.007, +0.013)}\\
 & 104 & 0.591 & 0.580 & \textbf{+0.011 (+0.001, +0.021)} & 0.577 & 0.575 & \textbf{+0.002 (+0.001, +0.004)} & 0.597 & 0.579 & \textbf{+0.018 (+0.015, +0.021)}\\
 & 260 & 0.570 & 0.583 & -0.013 (-0.029, +0.003) & 0.577 & 0.572 & \textbf{+0.004 (+0.002, +0.007)} & 0.599 & 0.574 & \textbf{+0.024 (+0.020, +0.029)}\\
\addlinespace
Celiac disease & 52 & 0.512 & 0.531 & -0.019 (-0.057, +0.023) & 0.540 & 0.527 & \textbf{+0.012 (+0.005, +0.020)} & 0.548 & 0.544 & +0.003 (-0.009, +0.017)\\
 & 104 & 0.522 & 0.534 & -0.012 (-0.071, +0.041) & 0.539 & 0.527 & \textbf{+0.013 (+0.006, +0.020)} & 0.544 & 0.540 & +0.003 (-0.009, +0.015)\\
 & 260 & 0.516 & 0.511 & +0.005 (-0.044, +0.054) & 0.525 & 0.520 & +0.005 (-0.002, +0.012) & 0.532 & 0.539 & -0.007 (-0.017, +0.004)\\
\addlinespace
Hyperlipidemia & 52 & 0.582 & 0.564 & \textbf{+0.018 (+0.008, +0.027)} & 0.553 & 0.551 & \textbf{+0.002 (+0.001, +0.003)}$^{\dagger}$ & 0.582 & 0.576 & \textbf{+0.007 (+0.002, +0.012)}\\
 & 104 & 0.577 & 0.563 & \textbf{+0.015 (+0.005, +0.025)} & 0.549 & 0.552 & \textbf{-0.002 (-0.004, -0.001)} & 0.580 & 0.570 & \textbf{+0.010 (+0.006, +0.014)}\\
 & 260 & 0.579 & 0.559 & \textbf{+0.020 (+0.006, +0.035)} & 0.544 & 0.560 & \textbf{-0.017 (-0.019, -0.015)}$^{\dagger}$ & 0.573 & 0.565 & \textbf{+0.009 (+0.003, +0.015)}\\
\addlinespace
Depression & 52 & 0.575 & 0.567 & +0.008 (-0.003, +0.018) & 0.540 & 0.548 & \textbf{-0.007 (-0.009, -0.006)}$^{\dagger}$ & 0.565 & 0.563 & +0.002 (-0.002, +0.006)\\
 & 104 & 0.553 & 0.551 & +0.002 (-0.007, +0.011) & 0.530 & 0.539 & \textbf{-0.009 (-0.010, -0.008)}$^{\dagger}$ & 0.556 & 0.556 & +0.000 (-0.004, +0.004)\\
 & 260 & 0.552 & 0.550 & +0.002 (-0.009, +0.013) & 0.540 & 0.535 & \textbf{+0.005 (+0.004, +0.007)} & 0.548 & 0.545 & +0.003 (-0.002, +0.008)\\
\addlinespace
Chronic kidney disease & 52 & 0.706 & 0.693 & +0.013 (-0.004, +0.029) & 0.709 & 0.696 & \textbf{+0.013 (+0.011, +0.016)} & 0.711 & 0.686 & \textbf{+0.025 (+0.018, +0.033)}\\
 & 104 & 0.695 & 0.675 & \textbf{+0.020 (+0.005, +0.036)} & 0.698 & 0.688 & \textbf{+0.010 (+0.008, +0.013)} & 0.706 & 0.683 & \textbf{+0.023 (+0.015, +0.030)}\\
 & 260 & 0.675 & 0.673 & +0.002 (-0.017, +0.023) & 0.685 & 0.681 & \textbf{+0.004 (+0.001, +0.007)} & 0.671 & 0.663 & +0.008 (-0.001, +0.016)\\
\addlinespace
Atrial fibrillation & 52 & 0.655 & 0.629 & \textbf{+0.026 (+0.004, +0.046)} & 0.658 & 0.668 & \textbf{-0.010 (-0.013, -0.007)}$^{\dagger}$ & 0.661 & 0.642 & \textbf{+0.019 (+0.010, +0.028)}\\
 & 104 & 0.672 & 0.644 & \textbf{+0.028 (+0.008, +0.047)} & 0.639 & 0.655 & \textbf{-0.016 (-0.019, -0.012)}$^{\dagger}$ & 0.669 & 0.642 & \textbf{+0.027 (+0.019, +0.036)}\\
 & 260 & 0.657 & 0.638 & +0.019 (-0.004, +0.042) & 0.638 & 0.640 & -0.003 (-0.006, +0.000) & 0.668 & 0.633 & \textbf{+0.035 (+0.026, +0.043)}\\
\addlinespace
Chronic obstructive pulmonary disease & 52 & 0.660 & 0.638 & +0.023 (-0.001, +0.046) & 0.632 & 0.636 & -0.004 (-0.009, +0.001) & 0.631 & 0.608 & \textbf{+0.024 (+0.013, +0.034)}\\
 & 104 & 0.645 & 0.642 & +0.003 (-0.023, +0.025) & 0.615 & 0.617 & -0.003 (-0.007, +0.001) & 0.618 & 0.603 & \textbf{+0.015 (+0.006, +0.024)}\\
 & 260 & 0.636 & 0.641 & -0.005 (-0.025, +0.015) & 0.609 & 0.604 & \textbf{+0.004 (+0.000, +0.008)} & 0.592 & 0.585 & +0.008 (-0.002, +0.018)\\
\addlinespace
Alzheimer's disease & 52 & 0.744 & 0.677 & \textbf{+0.066 (+0.026, +0.106)} & 0.718 & 0.687 & \textbf{+0.031 (+0.025, +0.036)} & 0.717 & 0.664 & \textbf{+0.053 (+0.038, +0.070)}\\
 & 104 & 0.694 & 0.650 & \textbf{+0.044 (+0.007, +0.080)} & 0.707 & 0.676 & \textbf{+0.031 (+0.026, +0.036)} & 0.703 & 0.658 & \textbf{+0.045 (+0.030, +0.059)}\\
 & 260 & 0.689 & 0.651 & \textbf{+0.038 (+0.007, +0.071)} & 0.698 & 0.660 & \textbf{+0.038 (+0.032, +0.043)} & 0.695 & 0.648 & \textbf{+0.047 (+0.031, +0.063)}\\
\addlinespace
Rheumatoid arthritis & 52 & 0.623 & 0.593 & +0.030 (-0.012, +0.075) & 0.587 & 0.600 & \textbf{-0.014 (-0.021, -0.007)}$^{\dagger}$ & 0.618 & 0.571 & \textbf{+0.047 (+0.025, +0.069)}\\
 & 104 & 0.602 & 0.582 & +0.020 (-0.019, +0.057) & 0.593 & 0.608 & \textbf{-0.016 (-0.022, -0.010)}$^{\dagger}$ & 0.613 & 0.568 & \textbf{+0.045 (+0.028, +0.064)}\\
 & 260 & 0.587 & 0.594 & -0.007 (-0.049, +0.035) & 0.581 & 0.602 & \textbf{-0.020 (-0.027, -0.013)}$^{\dagger}$ & 0.582 & 0.555 & \textbf{+0.027 (+0.010, +0.043)}\\
\addlinespace
\multicolumn{11}{@{}l}{\textit{Rare-cancer endpoints}} \\
Pancreatic cancer & 26 & 0.609 & 0.653 & -0.044 (-0.126, +0.029) & 0.561 & 0.584 & \textbf{-0.023 (-0.036, -0.010)} & 0.679 & 0.696 & -0.017 (-0.047, +0.015)\\
 & 52 & 0.646 & 0.674 & -0.028 (-0.092, +0.031) & 0.557 & 0.576 & \textbf{-0.020 (-0.033, -0.006)} & 0.661 & 0.672 & -0.011 (-0.040, +0.022)\\
 & 104 & 0.597 & 0.646 & -0.049 (-0.130, +0.028) & 0.545 & 0.570 & \textbf{-0.024 (-0.036, -0.012)} & 0.646 & 0.655 & -0.009 (-0.037, +0.018)\\
 & 260 & 0.598 & 0.616 & -0.018 (-0.090, +0.045) & 0.530 & 0.536 & -0.006 (-0.016, +0.004) & 0.614 & 0.627 & -0.013 (-0.041, +0.012)\\
\addlinespace
Ovarian cancer & 26 & 0.540 & 0.583 & -0.042 (-0.150, +0.056) & 0.550 & 0.559 & -0.009 (-0.026, +0.006) & 0.643 & 0.624 & +0.019 (-0.020, +0.058)\\
 & 52 & 0.562 & 0.583 & -0.022 (-0.107, +0.054) & 0.546 & 0.548 & -0.003 (-0.016, +0.010) & 0.630 & 0.607 & +0.023 (-0.009, +0.056)\\
 & 104 & 0.607 & 0.600 & +0.007 (-0.039, +0.046) & 0.539 & 0.530 & +0.009 (-0.002, +0.019) & 0.631 & 0.601 & \textbf{+0.030 (+0.008, +0.053)}\\
 & 260 & 0.567 & 0.557 & +0.011 (-0.018, +0.039) & 0.523 & 0.523 & +0.001 (-0.010, +0.012) & 0.583 & 0.582 & +0.000 (-0.025, +0.024)\\
\addlinespace
Glioblastoma & 26 & 0.761 & 0.768 & -0.007 (-0.055, +0.041) & 0.584 & 0.593 & \textbf{-0.009 (-0.018, -0.001)} & 0.719 & 0.748 & \textbf{-0.029 (-0.045, -0.013)}\\
 & 52 & 0.700 & 0.722 & -0.023 (-0.065, +0.018) & 0.587 & 0.602 & \textbf{-0.016 (-0.023, -0.009)} & 0.706 & 0.738 & \textbf{-0.031 (-0.046, -0.016)}\\
 & 104 & 0.656 & 0.688 & -0.032 (-0.071, +0.004) & 0.579 & 0.593 & \textbf{-0.014 (-0.021, -0.008)} & 0.685 & 0.711 & \textbf{-0.026 (-0.046, -0.007)}\\
 & 260 & 0.607 & 0.632 & -0.025 (-0.058, +0.004) & 0.570 & 0.583 & \textbf{-0.013 (-0.019, -0.008)} & 0.674 & 0.689 & -0.015 (-0.037, +0.006)\\
\addlinespace
Multiple myeloma & 26 & 0.704 & 0.708 & -0.004 (-0.037, +0.031) & 0.602 & 0.632 & \textbf{-0.029 (-0.042, -0.017)} & 0.754 & 0.761 & -0.007 (-0.020, +0.007)\\
 & 52 & 0.702 & 0.695 & +0.006 (-0.020, +0.033) & 0.593 & 0.631 & \textbf{-0.038 (-0.049, -0.027)} & 0.745 & 0.753 & -0.008 (-0.018, +0.003)\\
 & 104 & 0.694 & 0.683 & +0.011 (-0.014, +0.038) & 0.593 & 0.630 & \textbf{-0.037 (-0.048, -0.026)} & 0.735 & 0.746 & -0.011 (-0.022, +0.000)\\
 & 260 & 0.686 & 0.669 & +0.017 (-0.026, +0.060) & 0.571 & 0.604 & \textbf{-0.032 (-0.044, -0.022)} & 0.702 & 0.708 & -0.006 (-0.021, +0.009)\\
\addlinespace
\end{longtable}
\endgroup
\end{landscape}

\begin{landscape}
\begin{table}[H]
\centering
\caption{\footnotesize \textbf{Case recovery differences at fixed specificity in the 15 strict-reversal cells.} For each endpoint-cohort-horizon cell with a bootstrap-supported AUROC/partial-AUROC ranking reversal, the table reports the paired difference in sensitivity at 90\%, 95\%, and 99\% specificity (TRACER-Encoder minus TRACER, with paired 95\% bootstrap confidence intervals), the corresponding additional cases recovered per 10{,}000 eligible patients at 95\% specificity, and the additional cases in the full endpoint-eligible evaluation cohort. Case pools are IPCW-weighted horizon-specific case counts (Methods); positive values indicate additional cases recovered by TRACER-Encoder. Hyperlipidemia at UC5 at the 52-week horizon is the single opposite-direction reversal.}
\label{tab:supp-yield-diff}
\footnotesize
\setlength{\tabcolsep}{4pt}
\renewcommand{\arraystretch}{1.10}
\begin{adjustbox}{max width=\textwidth}
\begin{tabular}{@{}llrrrrrr@{}}
\toprule
Disease endpoint & Cohort & Horizon (wk) & $\Delta$Sens$_{90}$ (E$-$D) & $\Delta$Sens$_{95}$ (E$-$D) & $\Delta$Sens$_{99}$ (E$-$D) & Cases per 10{,}000 & Cases (cohort) \\
\midrule
Atrial fibrillation & UC5 & 52 & $+0.014$ ($+0.005$, $+0.024$) & $+0.026$ ($+0.017$, $+0.035$) & $+0.016$ ($+0.009$, $+0.024$) & $+4.0$ & $+370$ ($+241$, $+492$) \\
Atrial fibrillation & UC5 & 104 & $+0.027$ ($+0.018$, $+0.035$) & $+0.036$ ($+0.028$, $+0.045$) & $+0.020$ ($+0.014$, $+0.028$) & $+8.6$ & $+794$ ($+612$, $+988$) \\
Depression & UC5 & 52 & $+0.016$ ($+0.011$, $+0.021$) & $+0.017$ ($+0.014$, $+0.022$) & $+0.004$ ($+0.002$, $+0.007$) & $+9.9$ & $+863$ ($+670$, $+1076$) \\
Depression & UC5 & 104 & $+0.025$ ($+0.020$, $+0.029$) & $+0.020$ ($+0.016$, $+0.023$) & $+0.004$ ($+0.002$, $+0.006$) & $+17.8$ & $+1554$ ($+1280$, $+1844$) \\
Heart failure & UC5 & 52 & $+0.019$ ($+0.009$, $+0.029$) & $+0.046$ ($+0.034$, $+0.056$) & $+0.017$ ($+0.008$, $+0.025$) & $+6.4$ & $+588$ ($+444$, $+717$) \\
Heart failure & UC5 & 104 & $+0.000$ ($-0.008$, $+0.009$) & $+0.023$ ($+0.014$, $+0.033$) & $+0.029$ ($+0.021$, $+0.036$) & $+5.2$ & $+478$ ($+279$, $+675$) \\
Heart failure & UC6 & 52 & $+0.011$ ($-0.001$, $+0.023$) & $+0.026$ ($+0.011$, $+0.037$) & $+0.010$ ($-0.000$, $+0.021$) & $+7.4$ & $+210$ ($+90$, $+298$) \\
Heart failure & UC6 & 104 & $+0.024$ ($+0.013$, $+0.035$) & $+0.023$ ($+0.011$, $+0.035$) & $+0.022$ ($+0.012$, $+0.032$) & $+10.4$ & $+295$ ($+142$, $+447$) \\
Heart failure & UC6 & 260 & $+0.047$ ($+0.034$, $+0.058$) & $+0.039$ ($+0.027$, $+0.052$) & $+0.028$ ($+0.018$, $+0.041$) & $+34.4$ & $+974$ ($+670$, $+1296$) \\
Hyperlipidemia & UC5 & 52 & $+0.004$ ($+0.000$, $+0.008$) & $-0.009$ ($-0.012$, $-0.005$) & $-0.002$ ($-0.005$, $-0.000$) & $-8.5$ & $-706$ ($-999$, $-433$) \\
Hyperlipidemia & UC5 & 260 & $+0.030$ ($+0.024$, $+0.036$) & $+0.036$ ($+0.031$, $+0.042$) & $+0.023$ ($+0.020$, $+0.026$) & $+100.0$ & $+8278$ ($+7127$, $+9726$) \\
Hypertension & UC5 & 52 & $+0.013$ ($+0.008$, $+0.017$) & $+0.005$ ($+0.001$, $+0.010$) & $-0.001$ ($-0.004$, $+0.001$) & $+4.3$ & $+337$ ($+81$, $+620$) \\
Rheumatoid arthritis & UC5 & 52 & $+0.016$ ($-0.002$, $+0.038$) & $+0.035$ ($+0.017$, $+0.053$) & $+0.014$ ($+0.001$, $+0.028$) & $+1.4$ & $+128$ ($+63$, $+196$) \\
Rheumatoid arthritis & UC5 & 104 & $+0.023$ ($+0.007$, $+0.041$) & $+0.039$ ($+0.023$, $+0.057$) & $+0.015$ ($+0.005$, $+0.026$) & $+2.4$ & $+219$ ($+128$, $+318$) \\
Rheumatoid arthritis & UC5 & 260 & $+0.035$ ($+0.016$, $+0.052$) & $+0.046$ ($+0.029$, $+0.063$) & $+0.026$ ($+0.014$, $+0.038$) & $+5.0$ & $+463$ ($+296$, $+631$) \\
\bottomrule
\end{tabular}
\end{adjustbox}
\end{table}
\end{landscape}

\begin{table}[H]
\centering
\caption{\small
\textbf{Global discrimination and worst-site transport gaps across 16 disease endpoints.}
Point estimates of Uno's IPCW C-index are shown for TRACER and TRACER-Encoder using survival heads trained on the full in-domain downstream training cohort and applied unchanged to the in-domain, UC5, and UC6 evaluation cohorts. For each model, the worst-site gap was defined as
\(\max(C_{\mathrm{in}}-C_{\mathrm{UC5}},\,C_{\mathrm{in}}-C_{\mathrm{UC6}})\).
Positive values indicate lower discrimination at the worse-performing held-out health system, whereas negative values indicate that both held-out systems had higher C-indices than the in-domain cohort. Endpoints are grouped according to the magnitude of the TRACER worst-site gap.
}
\label{tab:generalization}
\small
\setlength{\tabcolsep}{4pt}
\renewcommand{\arraystretch}{1.08}
\begin{adjustbox}{max width=\textwidth}
\begin{tabular}{@{}lcccccccc@{}}
\toprule
& \multicolumn{4}{c}{TRACER}
& \multicolumn{4}{c}{TRACER-Encoder} \\
\cmidrule(lr){2-5}
\cmidrule(lr){6-9}
Disease
& In-domain & UC5 & UC6 & Worst-site gap
& In-domain & UC5 & UC6 & Worst-site gap \\
\midrule
\multicolumn{9}{@{}l}{
\textit{TRACER worst-site gap \(\leq 0.025\)}
} \\
Celiac disease
& 0.624 & 0.617 & 0.651 & \(+0.007\)
& 0.591 & 0.606 & 0.655 & \(-0.015\) \\
Chronic kidney disease
& 0.869 & 0.860 & 0.867 & \(+0.009\)
& 0.820 & 0.811 & 0.808 & \(+0.012\) \\
Alzheimer's disease
& 0.879 & 0.870 & 0.880 & \(+0.009\)
& 0.816 & 0.803 & 0.813 & \(+0.014\) \\
Acute myocardial infarction
& 0.842 & 0.830 & 0.862 & \(+0.013\)
& 0.775 & 0.776 & 0.798 & \(-0.001\) \\
Hypertension
& 0.776 & 0.761 & 0.775 & \(+0.015\)
& 0.768 & 0.753 & 0.753 & \(+0.016\) \\
\addlinespace
\multicolumn{9}{@{}l}{
\textit{TRACER worst-site gap \(>0.025\) and \(\leq0.050\)}
} \\
Rheumatoid arthritis
& 0.793 & 0.762 & 0.778 & \(+0.031\)
& 0.739 & 0.721 & 0.679 & \(+0.060\) \\
Atrial fibrillation
& 0.865 & 0.833 & 0.860 & \(+0.032\)
& 0.753 & 0.764 & 0.769 & \(-0.012\) \\
Heart failure
& 0.869 & 0.865 & 0.836 & \(+0.033\)
& 0.810 & 0.823 & 0.798 & \(+0.011\) \\
Glioblastoma
& 0.736 & 0.688 & 0.816 & \(+0.048\)
& 0.763 & 0.711 & 0.848 & \(+0.052\) \\
\addlinespace
\multicolumn{9}{@{}l}{
\textit{TRACER worst-site gap \(>0.050\)}
} \\
Chronic obstructive pulmonary disease
& 0.841 & 0.788 & 0.825 & \(+0.053\)
& 0.765 & 0.762 & 0.771 & \(+0.003\) \\
Pancreatic cancer
& 0.776 & 0.717 & 0.790 & \(+0.059\)
& 0.752 & 0.728 & 0.779 & \(+0.024\) \\
Depression
& 0.693 & 0.631 & 0.696 & \(+0.062\)
& 0.681 & 0.641 & 0.685 & \(+0.040\) \\
Hyperlipidemia
& 0.763 & 0.696 & 0.759 & \(+0.067\)
& 0.747 & 0.718 & 0.743 & \(+0.030\) \\
Ovarian cancer
& 0.717 & 0.636 & 0.735 & \(+0.081\)
& 0.717 & 0.616 & 0.712 & \(+0.101\) \\
Multiple myeloma
& 0.801 & 0.709 & 0.846 & \(+0.093\)
& 0.815 & 0.754 & 0.856 & \(+0.061\) \\
Lupus
& 0.763 & 0.746 & 0.669 & \(+0.095\)
& 0.701 & 0.711 & 0.661 & \(+0.040\) \\
\bottomrule
\end{tabular}
\end{adjustbox}
\end{table}

\begin{table}[H]
\centering
\caption{\small
\textbf{Time-dependent AUROC across prediction horizons in the in-domain test cohort.}
Point estimates are shown at the 52-, 104-, and 260-week horizons for survival heads trained on the full development-domain downstream training cohort. The change across the common horizon range was calculated as
\(\Delta_{52\rightarrow260}
= \mathrm{AUROC}_{260}-\mathrm{AUROC}_{52}\);
positive values indicate higher AUROC at 260 weeks, whereas negative values indicate lower AUROC. Endpoints are grouped according to the magnitude of the 52-to-260-week change under TRACER. 
}
\label{tab:auroc-horizon}

\scriptsize
\setlength{\tabcolsep}{3.5pt}
\renewcommand{\arraystretch}{1.10}

\begin{tabular*}{\textwidth}{
@{\extracolsep{\fill}}
lrrrrrrrr
@{}
}
\toprule
& \multicolumn{4}{c}{TRACER}
& \multicolumn{4}{c}{TRACER-Encoder} \\
\cmidrule(lr){2-5}
\cmidrule(lr){6-9}
Disease endpoint
& 52 wk & 104 wk & 260 wk & \(\Delta_{52\rightarrow260}\)
& 52 wk & 104 wk & 260 wk & \(\Delta_{52\rightarrow260}\) \\
\midrule
\multicolumn{9}{@{}l}{
\textit{Small 52-to-260-week change under TRACER
(\(\lvert\Delta\rvert\leq0.025\))}
} \\
Lupus
& 0.778 & 0.807 & 0.795 & \(+0.017\)
& 0.694 & 0.637 & 0.675 & \(-0.019\) \\
Atrial fibrillation
& 0.852 & 0.861 & 0.859 & \(+0.008\)
& 0.774 & 0.793 & 0.778 & \(+0.005\) \\
Rheumatoid arthritis
& 0.786 & 0.778 & 0.784 & \(-0.002\)
& 0.654 & 0.703 & 0.754 & \(+0.100\) \\
Heart failure
& 0.870 & 0.869 & 0.862 & \(-0.008\)
& 0.829 & 0.821 & 0.822 & \(-0.007\) \\
Hyperlipidemia
& 0.785 & 0.773 & 0.776 & \(-0.009\)
& 0.751 & 0.739 & 0.749 & \(-0.002\) \\
Ovarian cancer
& 0.755 & 0.799 & 0.742 & \(-0.013\)
& 0.811 & 0.817 & 0.717 & \(-0.094\) \\
Chronic kidney disease
& 0.888 & 0.878 & 0.865 & \(-0.023\)
& 0.837 & 0.826 & 0.832 & \(-0.005\) \\
Hypertension
& 0.814 & 0.798 & 0.790 & \(-0.024\)
& 0.796 & 0.771 & 0.783 & \(-0.014\) \\
\addlinespace
\multicolumn{9}{@{}l}{
\textit{Intermediate 52-to-260-week decline under TRACER
(\(0.025<\lvert\Delta\rvert\leq0.040\))}
} \\
Chronic obstructive pulmonary disease
& 0.847 & 0.843 & 0.821 & \(-0.026\)
& 0.768 & 0.782 & 0.757 & \(-0.010\) \\
Acute myocardial infarction
& 0.871 & 0.892 & 0.843 & \(-0.028\)
& 0.794 & 0.812 & 0.771 & \(-0.023\) \\
Alzheimer's disease
& 0.913 & 0.890 & 0.882 & \(-0.030\)
& 0.800 & 0.785 & 0.797 & \(-0.003\) \\
\addlinespace
\multicolumn{9}{@{}l}{
\textit{Large 52-to-260-week decline under TRACER
(\(\lvert\Delta\rvert>0.040\))}
} \\
Pancreatic cancer
& 0.843 & 0.812 & 0.793 & \(-0.050\)
& 0.854 & 0.796 & 0.770 & \(-0.084\) \\
Depression
& 0.739 & 0.706 & 0.688 & \(-0.050\)
& 0.700 & 0.660 & 0.646 & \(-0.053\) \\
Multiple myeloma
& 0.877 & 0.858 & 0.819 & \(-0.058\)
& 0.835 & 0.835 & 0.815 & \(-0.019\) \\
Celiac disease
& 0.646 & 0.625 & 0.576 & \(-0.070\)
& 0.615 & 0.607 & 0.587 & \(-0.028\) \\
Glioblastoma
& 0.834 & 0.778 & 0.744 & \(-0.091\)
& 0.831 & 0.808 & 0.760 & \(-0.071\) \\
\bottomrule
\end{tabular*}
\end{table}

\subsection{Labeled-data, health-system and attribution summaries}
\begin{table}[H]
\centering
\caption{\textbf{Training-data sensitivity on the in-domain cohort: change in Uno's IPCW C-index between $10\%$ and $100\%$ in-domain training fractions.} Small $\Delta$ indicates early saturation: the frozen-backbone embeddings already capture most of the risk-relevant information for that disease. Diseases sorted by TRACER $\Delta$.}
\label{tab:training-sensitivity}
\small
\begin{adjustbox}{max width=\linewidth}
\begin{tabular}{l rrr rrr}
\toprule
& \multicolumn{3}{c}{TRACER} & \multicolumn{3}{c}{TRACER-Encoder} \\
\cmidrule(lr){2-4} \cmidrule(lr){5-7}
Disease & $10\%$ & $100\%$ & $\Delta$ & $10\%$ & $100\%$ & $\Delta$ \\
\midrule
Heart Failure & 0.837 & 0.869 & $+0.032$ & 0.722 & 0.810 & $+0.087$ \\
CKD & 0.828 & 0.869 & $+0.040$ & 0.804 & 0.820 & $+0.016$ \\
Acute MI & 0.788 & 0.842 & $+0.054$ & 0.709 & 0.775 & $+0.066$ \\
COPD & 0.786 & 0.841 & $+0.055$ & 0.720 & 0.765 & $+0.045$ \\
A Fib & 0.804 & 0.865 & $+0.060$ & 0.730 & 0.753 & $+0.023$ \\
Alzheimer's & 0.819 & 0.879 & $+0.061$ & 0.754 & 0.816 & $+0.062$ \\
Hyperlipidemia & 0.694 & 0.763 & $+0.069$ & 0.688 & 0.747 & $+0.060$ \\
Celiac Disease & 0.548 & 0.624 & $+0.076$ & 0.476 & 0.591 & $+0.115$ \\
Hypertension & 0.690 & 0.776 & $+0.085$ & 0.696 & 0.768 & $+0.072$ \\
Depression & 0.599 & 0.693 & $+0.094$ & 0.621 & 0.681 & $+0.060$ \\
Lupus & 0.650 & 0.763 & $+0.113$ & 0.557 & 0.701 & $+0.144$ \\
Rheumatoid Arthritis & 0.624 & 0.793 & $+0.169$ & 0.668 & 0.739 & $+0.071$ \\
\bottomrule
\end{tabular}
\end{adjustbox}
\end{table}

\begingroup
\scriptsize
\setlength{\tabcolsep}{3pt}
\renewcommand{\arraystretch}{1.10}
\setlength{\LTcapwidth}{\linewidth}
\setlength{\LTleft}{\fill}
\setlength{\LTright}{\fill}
\begin{longtable}{@{}llrrrrrrrrrr@{}}
\caption{\small \textbf{Uno's IPCW C-index by in-domain training fraction in all three evaluation cohorts.} Survival heads trained on 10\%, 30\%, 50\%, and 100\% of the in-domain downstream training cohort and applied unchanged to each cohort (original development heads). $\Delta$ is the 100\% minus 10\% difference. The in-domain rows reproduce Supplementary Table~\ref{tab:training-sensitivity}. } \label{tab:training-sensitivity-all}\\
\toprule
& & \multicolumn{5}{c}{TRACER} & \multicolumn{5}{c}{TRACER-Encoder} \\
\cmidrule(lr){3-7}\cmidrule(lr){8-12}
Disease & Cohort & 10\% & 30\% & 50\% & 100\% & $\Delta$ & 10\% & 30\% & 50\% & 100\% & $\Delta$ \\
\midrule
\endfirsthead
\multicolumn{12}{@{}l}{\textit{(continued)}}\\
\toprule
& & \multicolumn{5}{c}{TRACER} & \multicolumn{5}{c}{TRACER-Encoder} \\
\cmidrule(lr){3-7}\cmidrule(lr){8-12}
Disease & Cohort & 10\% & 30\% & 50\% & 100\% & $\Delta$ & 10\% & 30\% & 50\% & 100\% & $\Delta$ \\
\midrule
\endhead
\midrule
\multicolumn{12}{r@{}}{\footnotesize Continued on next page}\\
\endfoot
\bottomrule
\endlastfoot
Heart failure & In-domain & 0.837 & 0.858 & 0.862 & 0.869 & $+0.032$ & 0.722 & 0.766 & 0.781 & 0.810 & $+0.087$ \\
 & UC5 & 0.851 & 0.854 & 0.858 & 0.865 & $+0.014$ & 0.754 & 0.797 & 0.807 & 0.823 & $+0.070$ \\
 & UC6 & 0.793 & 0.803 & 0.803 & 0.836 & $+0.043$ & 0.721 & 0.766 & 0.780 & 0.798 & $+0.078$ \\
\addlinespace
Acute myocardial infarction & In-domain & 0.788 & 0.837 & 0.830 & 0.842 & $+0.054$ & 0.709 & 0.735 & 0.736 & 0.775 & $+0.066$ \\
 & UC5 & 0.801 & 0.830 & 0.820 & 0.830 & $+0.029$ & 0.719 & 0.737 & 0.737 & 0.776 & $+0.057$ \\
 & UC6 & 0.796 & 0.839 & 0.838 & 0.862 & $+0.066$ & 0.734 & 0.757 & 0.757 & 0.798 & $+0.063$ \\
\addlinespace
Lupus & In-domain & 0.650 & 0.714 & 0.732 & 0.763 & $+0.113$ & 0.557 & 0.694 & 0.645 & 0.701 & $+0.144$ \\
 & UC5 & 0.631 & 0.714 & 0.744 & 0.746 & $+0.114$ & 0.572 & 0.641 & 0.656 & 0.711 & $+0.139$ \\
 & UC6 & 0.574 & 0.644 & 0.651 & 0.669 & $+0.095$ & 0.587 & 0.615 & 0.634 & 0.661 & $+0.075$ \\
\addlinespace
Hypertension & In-domain & 0.690 & 0.752 & 0.754 & 0.776 & $+0.085$ & 0.696 & 0.737 & 0.756 & 0.768 & $+0.072$ \\
 & UC5 & 0.624 & 0.711 & 0.729 & 0.761 & $+0.137$ & 0.678 & 0.726 & 0.745 & 0.753 & $+0.075$ \\
 & UC6 & 0.673 & 0.731 & 0.732 & 0.775 & $+0.102$ & 0.688 & 0.725 & 0.743 & 0.753 & $+0.064$ \\
\addlinespace
Celiac disease & In-domain & 0.548 & 0.585 & 0.629 & 0.624 & $+0.076$ & 0.476 & 0.552 & 0.587 & 0.591 & $+0.115$ \\
 & UC5 & 0.518 & 0.569 & 0.624 & 0.617 & $+0.099$ & 0.496 & 0.512 & 0.534 & 0.606 & $+0.110$ \\
 & UC6 & 0.553 & 0.607 & 0.644 & 0.651 & $+0.098$ & 0.498 & 0.538 & 0.590 & 0.655 & $+0.157$ \\
\addlinespace
Hyperlipidemia & In-domain & 0.694 & 0.743 & 0.746 & 0.763 & $+0.069$ & 0.688 & 0.706 & 0.737 & 0.747 & $+0.060$ \\
 & UC5 & 0.637 & 0.687 & 0.687 & 0.696 & $+0.059$ & 0.652 & 0.675 & 0.700 & 0.718 & $+0.065$ \\
 & UC6 & 0.696 & 0.742 & 0.745 & 0.759 & $+0.063$ & 0.683 & 0.702 & 0.730 & 0.743 & $+0.060$ \\
\addlinespace
Depression & In-domain & 0.599 & 0.645 & 0.677 & 0.693 & $+0.094$ & 0.621 & 0.654 & 0.657 & 0.681 & $+0.060$ \\
 & UC5 & 0.539 & 0.596 & 0.620 & 0.631 & $+0.092$ & 0.588 & 0.613 & 0.620 & 0.641 & $+0.053$ \\
 & UC6 & 0.590 & 0.642 & 0.681 & 0.696 & $+0.107$ & 0.628 & 0.662 & 0.664 & 0.685 & $+0.057$ \\
\addlinespace
Chronic kidney disease & In-domain & 0.828 & 0.847 & 0.861 & 0.869 & $+0.040$ & 0.804 & 0.809 & 0.817 & 0.820 & $+0.016$ \\
 & UC5 & 0.841 & 0.856 & 0.862 & 0.860 & $+0.019$ & 0.796 & 0.798 & 0.816 & 0.811 & $+0.015$ \\
 & UC6 & 0.828 & 0.847 & 0.858 & 0.867 & $+0.039$ & 0.785 & 0.792 & 0.806 & 0.808 & $+0.023$ \\
\addlinespace
Atrial fibrillation & In-domain & 0.804 & 0.849 & 0.860 & 0.865 & $+0.060$ & 0.730 & 0.739 & 0.738 & 0.753 & $+0.023$ \\
 & UC5 & 0.792 & 0.828 & 0.835 & 0.833 & $+0.040$ & 0.743 & 0.746 & 0.754 & 0.764 & $+0.022$ \\
 & UC6 & 0.806 & 0.851 & 0.856 & 0.860 & $+0.055$ & 0.749 & 0.754 & 0.756 & 0.769 & $+0.021$ \\
\addlinespace
Chronic obstructive pulmonary disease & In-domain & 0.786 & 0.810 & 0.831 & 0.841 & $+0.055$ & 0.720 & 0.709 & 0.761 & 0.765 & $+0.045$ \\
 & UC5 & 0.703 & 0.743 & 0.764 & 0.788 & $+0.085$ & 0.722 & 0.733 & 0.757 & 0.762 & $+0.041$ \\
 & UC6 & 0.765 & 0.798 & 0.814 & 0.825 & $+0.060$ & 0.726 & 0.721 & 0.768 & 0.771 & $+0.045$ \\
\addlinespace
Alzheimer's disease & In-domain & 0.819 & 0.846 & 0.883 & 0.879 & $+0.061$ & 0.754 & 0.776 & 0.804 & 0.816 & $+0.062$ \\
 & UC5 & 0.824 & 0.852 & 0.871 & 0.870 & $+0.046$ & 0.745 & 0.772 & 0.800 & 0.803 & $+0.058$ \\
 & UC6 & 0.812 & 0.839 & 0.879 & 0.880 & $+0.068$ & 0.742 & 0.777 & 0.810 & 0.813 & $+0.071$ \\
\addlinespace
Rheumatoid arthritis & In-domain & 0.624 & 0.737 & 0.761 & 0.793 & $+0.169$ & 0.668 & 0.649 & 0.718 & 0.739 & $+0.071$ \\
 & UC5 & 0.631 & 0.734 & 0.745 & 0.762 & $+0.130$ & 0.620 & 0.656 & 0.693 & 0.721 & $+0.101$ \\
 & UC6 & 0.624 & 0.755 & 0.757 & 0.778 & $+0.154$ & 0.563 & 0.617 & 0.634 & 0.679 & $+0.116$ \\
\addlinespace
\end{longtable}
\endgroup

\begin{table}[H]
\centering
\caption{\small \textbf{Cohort-level site-shift characteristics.} Clinical-token distribution shift and record density for the in-domain reference and the two held-out health systems, calculated from the tokenized histories of the evaluation cohorts with structural and model-control tokens excluded from the distributional measures (Supplementary Methods). JS, Jensen--Shannon divergence (base-2 logarithms) of clinical-token frequencies from the in-domain reference; OOV, total frequency mass of tokens absent from the in-domain vocabulary.}
\label{tab:supp-site-shift}
\small
\setlength{\tabcolsep}{8pt}
\renewcommand{\arraystretch}{1.12}
\begin{adjustbox}{max width=\linewidth}
\begin{tabular}{@{}lccc@{}}
\toprule
 & In-domain & UC5 & UC6 \\
\midrule
JS divergence vs in-domain (bits) & --- & 0.158 & 0.145 \\
OOV frequency mass & --- & $5.5\times 10^{-7}$ & 0 \\
Median tokenized sequence length & 70 & 148 & 266 \\
Mean tokenized sequence length & 345 & 540 & 823 \\
Mean unique clinical tokens per patient & 90 & 124 & 170 \\
\bottomrule
\end{tabular}
\end{adjustbox}
\end{table}

\begin{table}[H]
\centering
\caption{\textbf{Integrated-gradients attribution mass by code family, panel disease, and horizon.} Values are shares of total normalized positive attribution mass of the paired TRACER-minus-TRACER-Encoder risk difference. Positional-share-normalized enrichment is reported in Table~\ref{tab:supp-ig-enrichment}.}
\label{tab:supp-ig-mass}
\small
\begin{adjustbox}{max width=\linewidth}
\begin{tabular}{l r rrrrr}
\toprule
Disease & Horizon (wk) & Labs & Structural & Diagnosis & Medication & Procedure \\
\midrule
\multirow{3}{*}{Heart Failure} & 52 & 0.419 & 0.246 & 0.083 & 0.039 & 0.037 \\
 & 104 & 0.421 & 0.240 & 0.087 & 0.045 & 0.037 \\
 & 260 & 0.442 & 0.237 & 0.092 & 0.037 & 0.035 \\
\midrule
\multirow{3}{*}{Rheumatoid Arthritis} & 52 & 0.308 & 0.202 & 0.115 & 0.071 & 0.060 \\
 & 104 & 0.295 & 0.205 & 0.124 & 0.065 & 0.066 \\
 & 260 & 0.307 & 0.191 & 0.124 & 0.079 & 0.050 \\
\midrule
\multirow{3}{*}{Depression} & 52 & 0.345 & 0.234 & 0.089 & 0.036 & 0.033 \\
 & 104 & 0.356 & 0.225 & 0.094 & 0.044 & 0.033 \\
 & 260 & 0.357 & 0.231 & 0.086 & 0.038 & 0.037 \\
\midrule
\multirow{3}{*}{Hypertension} & 52 & 0.369 & 0.209 & 0.057 & 0.031 & 0.025 \\
 & 104 & 0.372 & 0.214 & 0.056 & 0.027 & 0.024 \\
 & 260 & 0.373 & 0.213 & 0.060 & 0.031 & 0.024 \\
\midrule
\multirow{3}{*}{Alzheimer's} & 52 & 0.393 & 0.198 & 0.117 & 0.047 & 0.032 \\
 & 104 & 0.382 & 0.198 & 0.109 & 0.039 & 0.034 \\
 & 260 & 0.379 & 0.204 & 0.117 & 0.040 & 0.035 \\
\midrule
\multirow{3}{*}{Hyperlipidemia} & 52 & 0.340 & 0.221 & 0.046 & 0.038 & 0.045 \\
 & 104 & 0.339 & 0.219 & 0.057 & 0.032 & 0.036 \\
 & 260 & 0.362 & 0.205 & 0.064 & 0.035 & 0.036 \\
\bottomrule
\end{tabular}
\end{adjustbox}
\end{table}

\begin{table}[H]
\centering
\caption{\textbf{Positional-share-normalized enrichment by code family (attribution mass averaged over horizons divided by the family's share of sequence positions)}. Shares for laboratory (L), diagnosis (D), medication (M), and procedure (P) codes are measured directly. Diagnosis codes are the most enriched family per token; structural tokens are strongly depleted.}
\label{tab:supp-ig-enrichment}
\scriptsize
\setlength{\tabcolsep}{4pt}
\begin{adjustbox}{max width=\linewidth}
\begin{tabular}{l rr rr rr rr rr}
\toprule
& \multicolumn{2}{c}{Labs} & \multicolumn{2}{c}{Diagnosis} & \multicolumn{2}{c}{Medication} & \multicolumn{2}{c}{Procedure} & \multicolumn{2}{c}{Structural} \\
\cmidrule(lr){2-3} \cmidrule(lr){4-5} \cmidrule(lr){6-7} \cmidrule(lr){8-9} \cmidrule(lr){10-11}
Disease & Share & Enrich. & Share & Enrich. & Share & Enrich. & Share & Enrich. & Mass & Enrich.\ ($\leq$) \\
\midrule
Alzheimer's & 0.127 & 3.0 & 0.020 & 5.6 & 0.015 & 2.8 & 0.012 & 2.8 & 0.200 & 0.24 \\
Depression & 0.114 & 3.1 & 0.018 & 5.0 & 0.013 & 2.9 & 0.011 & 3.1 & 0.230 & 0.27 \\
Heart Failure & 0.153 & 2.8 & 0.023 & 3.8 & 0.015 & 2.7 & 0.012 & 3.0 & 0.241 & 0.30 \\
Hyperlipidemia & 0.102 & 3.4 & 0.016 & 3.6 & 0.012 & 2.9 & 0.010 & 3.9 & 0.215 & 0.25 \\
Hypertension & 0.119 & 3.1 & 0.017 & 3.4 & 0.012 & 2.5 & 0.010 & 2.5 & 0.212 & 0.25 \\
Rheumatoid Arthritis & 0.126 & 2.4 & 0.020 & 5.9 & 0.015 & 4.7 & 0.012 & 4.8 & 0.199 & 0.24 \\
\bottomrule
\end{tabular}
\end{adjustbox}
\end{table}

\begin{table}[H]
\centering
\caption{\small \textbf{Attribution share of the paired risk-difference functional within each model's own input, by token family.} Integrated-gradients attribution is computed per model as the gradient of the shared TRACER-minus-TRACER-Encoder risk difference with respect to that model's input, so each block distributes the same disagreement over that model's tokens. Structural combines visit-separator, time-gap, boundary, and other model-control tokens; procedure combines all procedure-code subfamilies; demographic is the sex token.}
\label{tab:supp-ig-permodel}
\begingroup
\scriptsize
\setlength{\tabcolsep}{3pt}
\renewcommand{\arraystretch}{1.10}
\begin{adjustbox}{max width=\linewidth}
\begin{tabular}{@{}lrrrrrrrrrrrrrr@{}}
\toprule
& \multicolumn{7}{c}{Within TRACER input} & \multicolumn{7}{c}{Within TRACER-Encoder input} \\
\cmidrule(lr){2-8}\cmidrule(lr){9-15}
Disease & Labs & Dx & Med & Proc & Struct & Demo & Other & Labs & Dx & Med & Proc & Struct & Demo & Other \\
\midrule
Heart failure & 0.231 & 0.103 & 0.060 & 0.017 & 0.550 & 0.068 & 0.000 & 0.623 & 0.071 & 0.022 & 0.078 & 0.221 & 0.016 & 0.022 \\
Rheumatoid arthritis & 0.152 & 0.128 & 0.092 & 0.053 & 0.556 & 0.052 & 0.000 & 0.456 & 0.114 & 0.053 & 0.080 & 0.147 & 0.194 & 0.004 \\
Depression & 0.281 & 0.107 & 0.046 & 0.030 & 0.560 & 0.017 & 0.000 & 0.425 & 0.073 & 0.034 & 0.052 & 0.201 & 0.055 & 0.215 \\
Hypertension & 0.247 & 0.082 & 0.044 & 0.023 & 0.572 & 0.071 & 0.000 & 0.495 & 0.033 & 0.015 & 0.041 & 0.231 & 0.021 & 0.180 \\
Alzheimer's disease & 0.275 & 0.111 & 0.042 & 0.035 & 0.525 & 0.053 & 0.000 & 0.494 & 0.118 & 0.042 & 0.052 & 0.107 & 0.059 & 0.183 \\
Hyperlipidemia & 0.275 & 0.076 & 0.046 & 0.028 & 0.561 & 0.056 & 0.000 & 0.422 & 0.039 & 0.023 & 0.067 & 0.240 & 0.050 & 0.211 \\
\bottomrule
\end{tabular}
\end{adjustbox}

\endgroup
\end{table}

\section{Detailed comparison of TRACER and TRACER-Encoder \& Supplementary Results}\label{sec:model-comparison-details}

The matched analyses compare decoder-based TRACER with TRACER-Encoder across diseases, prediction horizons, and evaluation cohorts. Primary comparisons use foundation-model representations held fixed and disease-prediction heads trained in-domain and applied unchanged to the external cohorts. The analyses cover overall and low false-positive discrimination, case recovery, labeled-data availability, and potential sources of model disagreement. Results of local survival-head training and the pooling and survival-head ablations are also reported; methods for analyses not described in the main Methods are given in Supplementary Methods. Model-specific C-indices are reported in Supplementary Table \ref{tab:cindex-by-site}, worst-site C-index gaps for both models are in Supplementary Table \ref{tab:generalization}, paired one-year AUROC differences are in Supplementary Table \ref{tab:auroc-difference-52wk}, and sensitivity at fixed specificity for both models is in Supplementary Table \ref{tab:sens-all-horizons-tracer}, and standardized partial AUROC with paired differences in Supplementary Table \ref{tab:pauc-paired} and Supplementary Figure \ref{fig-metric-disagreement}.

\subsection{Model preference at overall and low-false-positive discrimination}
Model preference was assessed using global AUROC and standardized partial AUROC over false-positive rates of up to 10\% (Supplementary Table \ref{tab:auroc-difference-52wk}, \ref{tab:auroc-difference-horizons}, and \ref{tab:pauc-paired}). A strict reversal required the two measures to favor opposing models, with both paired 95\% confidence intervals for the model differences excluding zero (Methods). Across the in-domain and two external cohorts, TRACER had a higher estimated AUROC in 106 of 108 chronic-disease comparisons. At UC5, TRACER-Encoder had a higher estimated partial AUROC in 16 of 36 disease comparisons (Supplementary Table \ref{tab:model-choice}).

\begin{figure*}[t]
\centering
\includegraphics[width=0.84\linewidth]{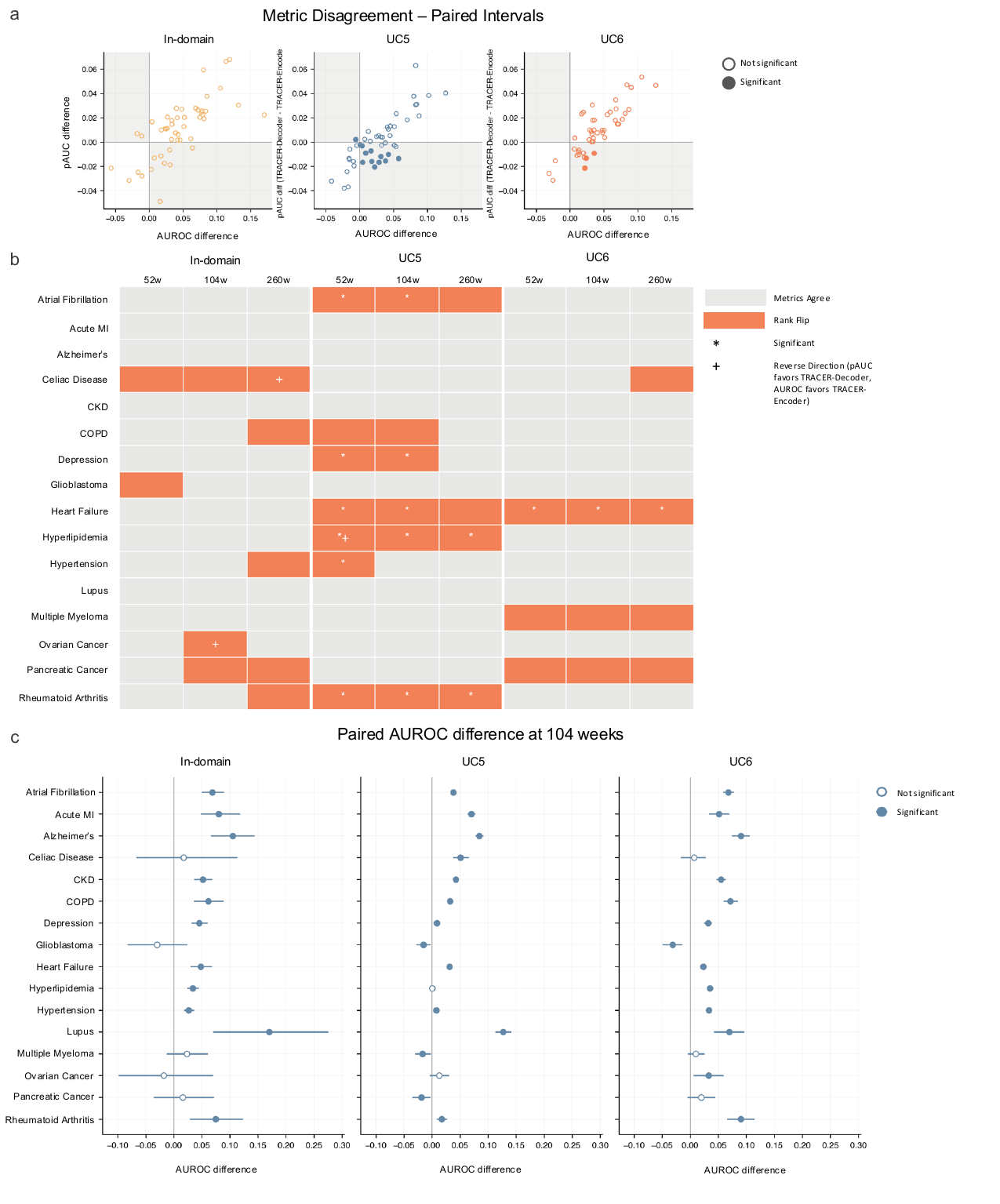}
\caption{\textbf{Global and low-false-positive discrimination yielded different model rankings in a subset of evaluations} All paired differences were calculated as TRACER minus TRACER-Encoder; positive values favor TRACER, whereas negative values favor TRACER-Encoder. \textbf{(a)} Paired differences in dynamic AUROC and standardized partial AUROC over the prespecified low-false-positive region (pAUC; FPR, 0-0.1) across the 16 endpoints and the shared 52-, 104-, and 260-week horizons, shown separately for in-domain, UC5, and UC6 cohorts. Shaded quadrants indicate evaluations in which the two metrics produced opposite point-estimate model rankings. Filled points denote ranking disagreements for which the paired 95\% bootstrap confidence interval for the pAUC difference excluded zero.
\textbf{(b)} Endpoint-by-horizon map of point-estimate agreement and disagreement between AUROC and pAUC-based model rankings. Gray cells indicate agreement and orange cells indicate disagreement. Asterisks denote disagreements for which the paired 95\% confidence interval for the pAUC difference excluded zero; 15 of these 16 cells also have AUROC confidence intervals excluding zero and therefore met the strict-reversal criterion. Plus signs indicate the less common reverse direction, in which pAUC favored TRACER and AUROC favored TRACER-Encoder.
\textbf{(c)} Paired AUROC differences at the 104-week horizon for each endpoint in the in-domain, UC5, and UC6 cohorts. Points and horizontal bars show the paired difference and its 95\% bootstrap confidence interval, respectively; filled points indicate intervals excluding zero.}
\label{fig-metric-disagreement}
\end{figure*}

The primary analysis comprised 144 comparisons: 16 diseases evaluated at 1-, 2-, and 5-year horizons in the in-domain and two external cohorts. Strict reversals occurred in 15 comparisons (10.4\%), all in the external cohorts: 12 at UC5 and three at UC6 (Supplementary Table \ref{tab:model-choice}, Supplementary Figure \ref{fig-metric-disagreement}). In 14 of these comparisons, global AUROC favored TRACER, whereas partial AUROC favored TRACER-Encoder. At the point-estimate level, the two measures ranked the models differently in 37 of 144 comparisons (25.7\%).

Heart failure, rheumatoid arthritis, atrial fibrillation, depression, hyperlipidemia, and hypertension accounted for all 15 strict reversals, with heart failure contributing five, including all three prediction horizons at UC6. 

\subsection{Case recovery in the strict-reversal comparisons}
Within the 15 strict-reversal comparisons, TRACER-Encoder had a higher point-estimate sensitivity in 15, 14, and 13 comparisons at 90\%, 95\%, and 99\% specificity, respectively (Supplementary Table \ref{tab:supp-yield-diff}). In the 14 comparisons where partial AUROC favored TRACER-Encoder, the encoder identified a median of 6.9 additional cases per 10,000 patients at 95\% specificity (Supplementary Table \ref{tab:supp-yield-diff}).

At UC5, sensitivity differences favoring TRACER-Encoder were 2.0 percentage points for depression at 104 weeks, 4.6 percentage points for heart failure at 52 weeks, and 3.6 percentage points for hyperlipidemia at 260 weeks. The corresponding estimates for the full disease-specific evaluation cohorts were 1,554 additional cases (95\% CI, 1,280-1,844), 588 (444-717), and 8,278 (7,127-9,726), respectively. 

\subsection{Discrimination across health systems and prediction horizons}

Time-dependent AUROC trajectories varied substantially across endpoints (Supplementary Table \ref{tab:auroc-horizon}). Under TRACER, atrial fibrillation, rheumatoid arthritis, heart failure, hyperlipidemia, chronic kidney disease, hypertension, lupus, and ovarian cancer remained within 0.025 of their 52-week AUROC at 260 weeks. A small end-to-end change did not always indicate a flat trajectory. Ovarian cancer AUROC increased from 0.755 at 52 weeks to 0.799 at 104 weeks before declining to 0.742 at 260 weeks. Among the chronic disease endpoints, depression and celiac disease showed the largest long-horizon decline, from 0.739 to 0.688 and from 0.646 to 0.576, respectively. Among the cancers, pancreatic cancer, multiple myeloma, and glioblastoma declined by 0.050-0.091.

Several endpoint-level patterns were shared across the two models. Both implementations showed comparatively stable discrimination for atrial fibrillation, heart failure, hyperlipidemia, and chronic kidney disease, but larger declines for depression, pancreatic cancer, and glioblastoma. However, horizon behavior was not architecture-invariant. Rheumatoid arthritis AUROC increased by 0.10 under TRACER-Encoder while remaining nearly unchanged under TRACER, and ovarian cancer AUROC declined by 0.094 under TRACER-Encoder compared with 0.013 under TRACER. The two models, therefore, differed in their horizon-specific trajectories for these endpoints.

Integrated Brier score and integrated negative binomial log-likelihood, which scale with outcome frequency and are therefore not comparable across endpoints, were nearly identical for the two models TRACER and TRACER-Encoder (Supplementary Table \ref{tab:ibs-ibll}).

\subsection{Downstream labeled-data availability}

Labeled-data analyses used  10\%, 30\%, 50\%, and 100\% of the available in-domain downstream training data. Performance generally continued to change as additional development-domain labels were added. In some held-out cohorts, discrimination improved even when gains in the in-domain test cohort were small. The training-fraction analyses (10\%, 30\%, 50\%, and 100\% of in-domain labels) are summarized in Supplementary Table \ref{tab:training-sensitivity}. External cohort values for every fraction are in \ref{tab:training-sensitivity-all}).

\subsection{Differences in health-system records}
Health-system differences in clinical-token frequencies, record density, follow-up, and cumulative incidence were characterized to contextualize the concentration of strict reversals at UC5. Jensen-Shannon divergence of clinical-token frequencies from the in-domain reference was 0.158 at UC5 and 0.145 at UC6 (Supplementary Table \ref{tab:supp-site-shift}, Supplementary Figure \ref{fig-interpretability}). Record density did not follow the same site ordering. Median tokenized sequence length was 70 in-domain, 148 at UC5, and 266 at UC6, while the mean number of unique tokens per patient was 90, 124, and 170, respectively.

Median follow-up ranged from 65-75 weeks at UC5 across chronic diseases compared with 51-56 weeks in-domain. UC6 matched the in-domain range for all but four diseases. Among disease health-system combinations with strict reversals, the largest absolute differences in estimated one-year cumulative incidence relative to the in-domain cohort occurred for hyperlipidemia at UC5 (12.2\% versus 9.6\%) and heart failure at UC6 (3.6\% versus 2.4\%). The 40 most-shifted individual tokens at each external site were almost exclusively laboratory-value tokens, identifying laboratory information as a prominent component of the observed token-frequency differences.

\subsection{Interpretability analyses}

\textbf{Equalized linear probes recovered modest alignment with the original reversal pattern.}
Identically regularized discrete-time hazard probes were fitted to the two frozen pooled representations using common data splits, tuning budgets, and censoring estimators. To reduce agreement induced by evaluating both surfaces on the same patients, alignment was calculated using two disjoint, cross-swapped evaluation halves. The mean cross-half Spearman alignment with the original model-comparison surface,\(F_{\mathrm{package}}\), was 0.227 (cells-fixed 95\% CI, 0.083--0.310; two-level bootstrap sensitivity interval, 0.054-0.330), based on directional correlations of 0.353 and 0.101. The probes reproduced the original metric-specific winner directions in 10 of the 15 strict-reversal cells. These findings indicate that information associated with the original reversal pattern was linearly accessible in the frozen representations, but a simple linear readout was insufficient to reproduce the complete package surface.

\textbf{Residual-only probes did not support the complementary-residual hypothesis.}
Residual-only probes tested whether the broad-versus-tail tradeoff was carried by components of each representation that could not be linearly predicted from the other. Label-free linear maps were fitted in both directions using in-domain training patients, and identical hazard probes were trained on the resulting standardized residuals $r_{\mathrm{E}}$ and $r_{\mathrm{D}}$ (Supplementary Methods). The mean cross-half alignment between the residual-probe and package surfaces was \(-0.183\) (cells-fixed 95\% CI, $-0.244$ to $-0.044$; two-level bootstrap interval, $-0.298$ to 0.002; one-sided $P=0.996$ for positive alignment). The direction was opposite to that predicted by the prespecified complementary-residual hypothesis. This result argues against the reversal pattern being carried by a simple model-specific linear residual subspace. It does not rule out representation involvement through shared, nonlinear, disease-dependent, or jointly fitted representation-head effects.

\textbf{Controlled DeepHit retraining reproduced selected reversals but not the complete surface.}
Controlled retraining fitted the original bias-free DeepHit head to each frozen representation using 10 independent random initializations per model. Two loss mixtures were evaluated: the original ranking-plus-likelihood objective $(\alpha=0.1)$ and a likelihood-only objective $(\alpha=1$. Under the original loss mixture, the mean controlled-head surface had a Spearman alignment of 0.309 with $F_{\mathrm{package}}$ (patient-resampling 95\% CI, 0.150-0.434; bootstrap SD, 0.073). This point estimate was numerically higher than the linear-probe alignment, but no prespecified paired test compared the two analyses.

Cell-level recurrence was stronger than full-surface alignment. Across the 15 cells, the mean recurrence of the two original metric-specific winner directions was 0.747. Eight cells reproduced both original directions in at least 80 of the 100 pairings. Thus, selected reversals were robust to repeated downstream-head fitting, but the reversal pattern was not reproduced uniformly across the full disease-site-horizon surface.

\textbf{Disease composition contributed more uncertainty than head initialization.}
The 95\% crossed-seed interval for full-surface alignment was 0.279-0.343 (SD, 0.017; width, 0.064), substantially narrower than the patient-resampling interval of 0.150-0.434 and the joint seed-plus-patient interval of 0.146-0.432. Random head initialization, therefore, contributed little additional uncertainty beyond patient sampling. By contrast, disease-block resampling produced the widest interval, 0.025-0.538, indicating that the estimated alignment depended strongly on which endpoints contributed to the evaluation surface. Alignment was 0.383 across the 98 chronic-disease cells and 0.032 across the 24 rare-cancer cells. Overall, heterogeneity across endpoints, rather than random head initialization, was the dominant source of variation in reversal reproducibility.

\begin{figure*}[t]
\centering
\includegraphics[width=1.0\linewidth]{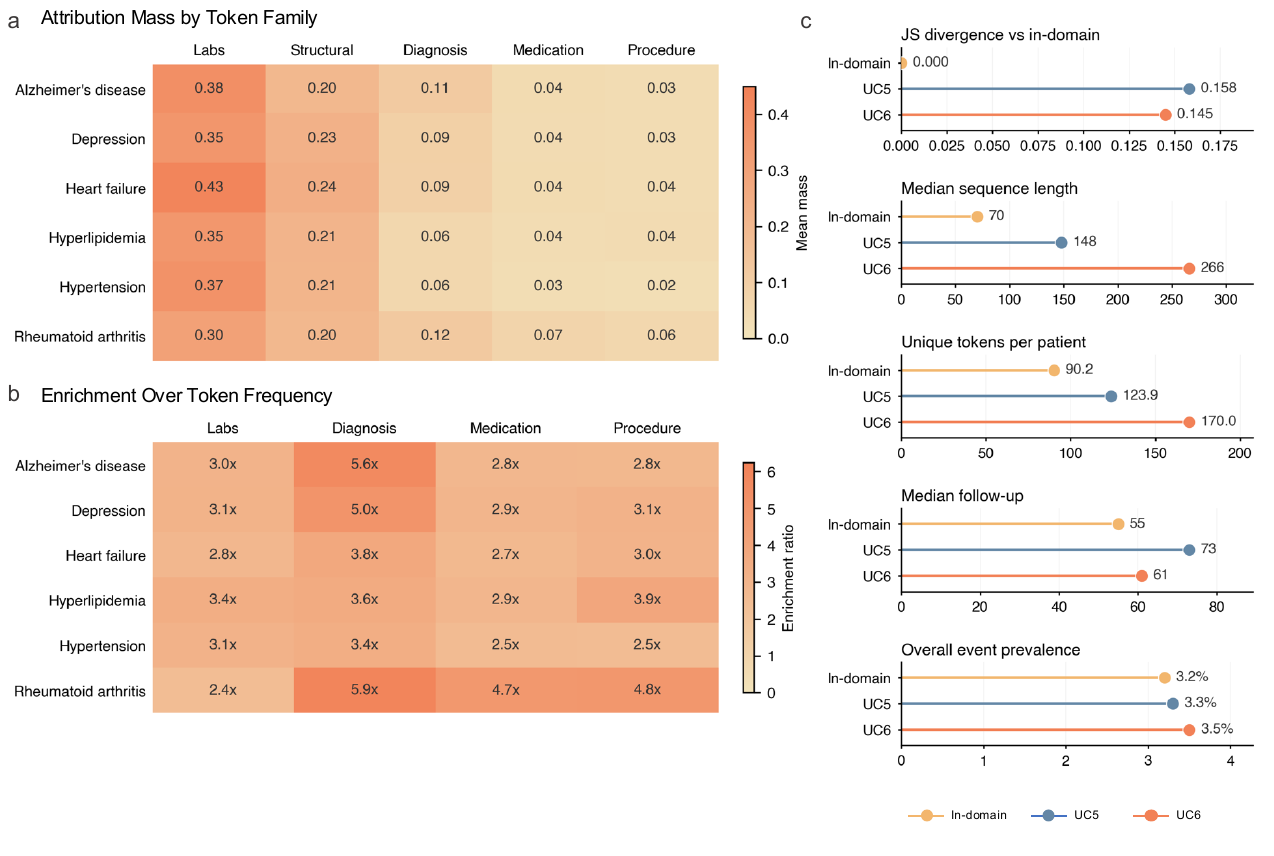}
\caption{\textbf{Interpretability analyses reveal token-level and site-level distribution characteristics}. \textbf{(a)} Mean integrated-gradients attribution mass for the paired difference in predicted risk between TRACER and TRACER-Encoder, aggregated by token family across six selected chronic disease endpoints. Darker shading indicates greater attribution mass. Cell values represent each token family's share of normalized attribution mass and are not adjusted for token frequency. \textbf{(b)} Attribution enrichment relative to token frequency for laboratory, diagnosis, medication, and procedure tokens. Enrichment was calculated by comparing each family's share of attribution mass with its share of input-token positions; values greater than 1 indicate that a token family received more attribution than expected from its frequency. Structural tokens are not included in the enrichment panel. \textbf{(c)} Site-level characteristics of the in-domain, UC5, and UC6 evaluation cohorts, including Jensen-Shannon divergence of clinical-token frequencies from the in-domain reference, median sequence length, mean number of unique tokens per patient, median follow-up in weeks, and overall event prevalence. Yellow, blue, and orange indicate the in-domain, UC5, and UC6 cohorts, respectively.}
\label{fig-interpretability}
\end{figure*}

\textbf{Token-level attribution localized disagreement without identifying a universal mechanism.}
Integrated-gradients attribution of the paired model-risk difference was evaluated on the recurrence-gated panel, which was extended to include Alzheimer's disease. Laboratory codes accounted for the largest share of raw attribution mass across the pooled panel (Supplementary Table~\ref{tab:supp-ig-mass}, Supplementary Figure \ref{fig-interpretability}). After normalization by each token family's positional frequency, diagnosis codes showed particularly high enrichment for Alzheimer's disease and rheumatoid arthritis, demonstrating that raw attribution mass and frequency-adjusted enrichment capture different aspects of model behavior (Supplementary Table~\ref{tab:supp-ig-enrichment}, Supplementary Figure \ref{fig-interpretability}).

The per-model decomposition of the paired risk-difference functional suggested different attribution profiles across the two model inputs. Within the TRACER input, structural, time-gap, and visit-separator tokens collectively accounted for a large share of attribution, whereas laboratory codes accounted for the largest share within the TRACER-Encoder input (Supplementary Table \ref{tab:supp-ig-permodel}). More broadly, the attribution analyses localized candidate sources of model disagreement but did not identify why a particular disease-horizon-site cell exhibited a reversal.

\subsection{Pooling-mask and pooling-strategy ablations}

Pooling ablations were evaluated on four chronic disease endpoints under the local survival-head training protocol (Supplementary Methods).

\textbf{Excluding care-site and time-gap tokens from the pooling average had little effect.}
Across 48 comparisons spanning four diseases, the in-domain and two external cohorts, and four downstream training fractions, excluding care-site and time-gap tokens produced a median C-index change of \(-0.001\) for TRACER, with a maximum absolute change of 0.025. The corresponding median change for TRACER-Encoder was \(-0.004\), with a maximum absolute change of 0.038. The largest decreases for TRACER-Encoder occurred for acute myocardial infarction and lupus at UC5, each with a change of \(-0.029\) at the 100\% training fraction (Supplementary Table~\ref{tab:supp-pooling}). These results provide little evidence that directly averaging the care-site and time-gap token embeddings was a major source of discrimination in this subset; however, they do not exclude site-specific information encoded indirectly through contextual interactions elsewhere in the sequence.

\textbf{Smooth inverse-frequency weighting did not consistently improve performance.}



Across the 48 disease-cohort-training-fraction comparisons, the median C-index difference between SIF and mean pooling was \(+0.001\) for TRACER and \(-0.004\) for TRACER-Encoder, with most changes within \(\pm0.01\). At the 100\% training fraction, the largest reduction occurred for lupus in-domain: C-index decreased from 0.827 to 0.787 for TRACER and from 0.691 to 0.656 for TRACER-Encoder. Thus, neither exclusion of care-site and time-gap token embeddings nor SIF weighting produced a consistent improvement under the evaluated local head-training protocol. The primary representation-extraction strategy used uniform mean pooling with care-site and time-gap tokens included.

\subsection{Local survival-head training versus zero-shot cross-site transfer}

The local training analysis compared disease-prediction heads trained using in-domain labels and applied unchanged externally with heads trained separately using each external health system's local training partition. Locally trained heads were evaluated at the corresponding local test partition. Both protocols used the same frozen foundation-model representations and were evaluated across 10\%, 30\%, 50\%, and 100\% of the available training data. The comparison therefore quantifies the effect of replacing the development-trained survival head with a locally trained head while holding the foundation-model backbone fixed.

Local survival-head training produced large and consistent gains at UC5 but smaller and more heterogeneous changes at UC6 (Supplementary Table~\ref{tab:supp-local-vs-zs}). The point-estimate C-index increased at UC5 for all 12 chronic disease endpoints under both TRACER implementations. The median gain was \(+0.049\) for TRACER (range, \(+0.031\) to \(+0.072\)) and \(+0.072\) for TRACER-Encoder (range, \(+0.032\) to \(+0.116\)). The largest TRACER-Encoder gains occurred for lupus (\(+0.116\)), hypertension (\(+0.084\)), and celiac disease (\(+0.082\)).

At UC6, local training produced a median change of \(+0.009\) for TRACER and \(-0.002\) for TRACER-Encoder. Point-estimate C-index increased for 8 of 12 endpoints under TRACER and 5 of 12 under TRACER-Encoder. The largest decrease was observed for acute myocardial infarction under TRACER (\(-0.037\)); under TRACER-Encoder, the largest decrease occurred for chronic obstructive pulmonary disease (\(-0.025\)). Thus, locally labeled outcomes provided substantially greater benefit at UC5 than at UC6.

Because the foundation-model backbones remained fixed, the UC5 gains show that locally fitting the mapping from frozen patient representations to disease risk can improve global discrimination. They do not establish that cross-site differences arise solely in the survival head or that local head training restores performance within the low-false-positive region. This ablation used the time-dependent concordance index implemented in \texttt{pycox}, whereas the primary analysis used Uno's IPCW C-index; therefore, absolute values from the two analyses are not directly comparable.

\subsection{DeepHit versus Cox proportional-hazards survival heads}
Under the local head-training protocol, we compared the primary discrete-time DeepHit head with a linear Cox proportional-hazards head fitted to the same frozen TRACER representations. Within each cohort, the two heads used the same patient partitions, representation inputs, and labeled-data fraction, although their objectives and event-time parameterizations differed by design.

DeepHit achieved a higher C-index than CoxPH in all 72 endpoint-representation-cohort comparisons (12 endpoints, two representations, and three cohorts; Supplementary Table~\ref{tab:supp-coxph}). CoxPH C-index was below 0.5 in 9 of 24 representation-endpoint comparisons in-domain, 12 of 24 at UC5, and 13 of 24 at UC6.

CoxPH performance also varied nonmonotonically across labeled-data fractions in several settings. For example, the in-domain C-index for the TRACER heart-failure model was 0.608, 0.548, 0.372, and 0.506 at the 10\%, 30\%, 50\%, and 100\% training fractions, respectively. These findings indicated that the implemented linear CoxPH head was not a reliable readout of the high-dimensional frozen representations under this protocol. DeepHit was the primary survival head. This ablation does not establish that DeepHit is generally superior to CoxPH; alternative risk-score construction, optimization, regularization, dimensionality reduction, or nonlinear proportional-hazards heads may produce different results.

\section{Supplementary Methods} 

\label{supp-pretraining}
\subsection{Model specifications}
\subsubsection*{TRACER-Encoder Pretraining}

We trained a ModernBERT-based masked language model from scratch using the \texttt{ModernBERTForMaskedLM} implementation. The model configuration was:

\begin{itemize}
    \item Vocabulary size: 18{,}719
    \item Hidden dimension: 768
    \item Intermediate dimension: 1{,}152
    \item Number of transformer layers: 22
    \item Number of attention heads: 12
    \item Maximum sequence length: 2{,}048
    \item Layer normalization $\epsilon = 10^{-5}$
\end{itemize}

ModernBERT was configured with a hybrid local-global attention pattern with a local attention window of 128 tokens and global attention every 3 layers. This reduces quadratic complexity while periodically allowing global context mixing, enabling efficient training at sequence length 2{,}048.

Training was performed with a total token budget of $5 \times 10^{9} \ \text{tokens}$. Sequences were truncated or packed to a maximum length of 2{,}048 tokens.

The model was trained with:

\begin{itemize}
    \item Learning rate: $2 \times 10^{-4}$
    \item Gradient clipping: 1.0
    \item Mixed precision: bfloat16
\end{itemize}

\subsubsection*{TRACER Causal Pretraining}

We trained a decoder-only transformer based on the LLaMA architecture using the \texttt{LlamaForCausalLM} implementation. The model was initialized from scratch with the following configuration:

\begin{itemize}
    \item Vocabulary size: 18{,}716
    \item Hidden dimension: 768
    \item Intermediate (feed-forward) dimension: 2{,}688
    \item Number of transformer layers: 14
    \item Number of attention heads: 12
    \item Maximum sequence length: 2{,}048
    \item RMSNorm $\epsilon = 10^{-5}$
\end{itemize}

The model used the SiLU activation function, consistent with LLaMA~3.x design. Rotary positional embeddings (RoPE) were applied with base $\theta = 10{,}000$. 
The model employed grouped-query attention (GQA) with 12 query heads and 4 key-value heads.
As a decoder-only transformer, the model uses strictly causal self-attention. 
Training was performed with a total token budget of
$5 \times 10^{9} \ \text{tokens}$. Sequences were truncated or packed to a maximum length of 2{,}048 tokens.

The model was trained with:

\begin{itemize}
    \item Learning rate: $2 \times 10^{-4}$
    \item Gradient clipping: 1.0
    \item Mixed precision: bfloat16
\end{itemize}

\subsubsection*{Resources}
Pretraining of the ModernBERT and LLaMA models was performed on eight NVIDIA A100 GPUs.

\subsection{Downstream labeled-data availability}
To assess the effect of labeled-data availability, the in-domain downstream training set was subsampled to 10\%, 30\%, 50\%, and 100\% of its full size, while the test and validation sets remain fixed across experiments. For each endpoint and training fraction, survival heads were trained and selected using the same protocol as the primary analysis (Methods). The 10\%, 30\%, and 50\% training fractions were evaluated as a supplementary robustness analysis and are not treated as independent tests. Results are reported in Supplementary Table \ref{tab:training-sensitivity} and \ref{tab:training-sensitivity-all}.

\subsection{Local survival-head training}
To estimate the potential value of locally labeled outcomes, we separately trained the same DeepHit head architecture using the local training partition from each held-out health system and evaluated it on that system's local test partition. The foundation model backbones remained frozen. These local adaptation analyses are reported in Supplementary Table \ref{tab:supp-local-vs-zs}.

\subsection{Cox proportional-hazards survival head}
As an ablation, we fitted a linear Cox proportional-hazards model to the same frozen representations using the same downstream partitions. Complete results are reported in Supplementary Table \ref{tab:supp-coxph}.

\subsection{Pooling-mask and pooling-strategy ablations}
Patient-level representations were obtained by pooling final-layer token embeddings from the frozen foundation models (Methods). In all analyses, padding tokens, masked-token placeholders, unknown tokens, and architecture-specific boundary or model-control tokens were excluded from the pooling average, as applicable. On a representative subset of four chronic disease endpoints—heart failure, acute myocardial infarction, lupus, and hypertension—we evaluated two additional design choices: whether care-site and inter-visit time-gap tokens should contribute directly to the pooled representation and whether uniform mean pooling should be replaced with frequency-weighted pooling.

These analyses used the local survival-head training protocol. The baseline condition used the same pooling mask as the primary analysis, with care-site and time-gap token embeddings included in the mean. Its in-domain estimates correspond to the primary downstream protocol, whereas the UC5 and UC6 estimates reflect survival heads trained using local labels rather than the zero-shot application of development-domain heads.

\textbf{Care-site and time-gap token exclusion.} Care-site tokens encode the location within a health system at which an encounter occurred, while time-gap tokens encode the interval between encounters. Because care-site identity may capture institution-specific care patterns, we compared the primary pooling strategy with an ablation that excluded care-site and time-gap token embeddings from the pooled average. These tokens remained in the transformer input and could still influence the contextualized representations of other tokens; the ablation therefore tested their direct contribution to pooling rather than removing all information associated with them.

\textbf{Smooth inverse-frequency pooling.} Smooth inverse-frequency (SIF) pooling \cite{arora2017simple} was evaluated after excluding care-site and time-gap tokens. For token \(v\), its contribution to the pooled representation was weighted by

$$
w(v)
=
\frac{a}{a+p(v)},
\qquad
a=10^{-3},
$$

where \(p(v)\) was the empirical frequency of token \(v\) in the development-domain pretraining corpus. The smoothing parameter was fixed after inspecting the token-frequency distribution. Relative to uniform mean pooling under the same token mask, SIF reduces the contribution of common tokens and increases the relative contribution of rare tokens.

\subsection{Analyses of AUROC-pAUC ranking reversals}
We performed various follow-up analyses using the same frozen pooled representations to determine whether the observed AUROC--pAUC ranking reversals could be explained by a property of one of the foundation model representations itself, random variation in the trained survival heads, or an interaction between representations and downstream survival heads. These analyses comprised equalized linear probes, residual-only probes, and controlled DeepHit retraining. 

For each endpoint-cohort-horizon pair, we summarized how the relative performance of TRACER and TRACER-Encoder changed across the ROC curve. Let $g_c(u)$ denote the difference between the models' true-positive rates (TRACER minus TRACER-Encoder) at a false-positive rate $u$. We defined the flip-shape statistic as

$$F_c = \text{mean}_{u \in(0.1,1]}g_c(u) - \text{mean}_{u \in [0,0.1]}g_c(u).$$

A positive value indicates that TRACER's relative advantage was greater over the broader ROC curve than within the low-false-positive region. The collection of $F_c$ values from the original model-specific DeepHit heads formed the reference flip-shape values, which we denote collectively by $F_{\text{package}}$. Each follow-up analysis produced a corresponding value that could be compared to the respective pair using Spearman correlation. 

\textbf{Equalized linear probes}. We trained identical low-capacity discrete-time hazard probes on the frozen TRACER-Encoder and TRACER representations. The probes used the same patients, outcomes, data partitions, tuning budget, and censoring estimator. The goal was to test whether linearly accessible information in the frozen representations reproduced the original ranking pattern.

To reduce agreement caused by evaluating the original and reconstructed surfaces on the same patients, each endpoint-cohort evaluation set was divided into two fixed, non-overlapping halves. We correlated the set calculated in half $A$ with the set calculated in half $B$, repeated the comparison with the groups flipped, and averaged the two Spearman correlations. Patient-level bootstrap resampling provided uncertainty intervals. 

\textbf{Residual-only probes}. We next tested whether the reversals were carried by components of each representation that could not be linearly predicted from the other. Using in-domain training patients and no outcome labels, we fitted regularized linear maps from TRACER-Encoder representation to the TRACER representation and in the reverse direction. The prediction residuals represented the component of each embedding that could not be linearly predicted from the other. Identical hazard probes were trained on these residuals and compared with the original set using the same cross-half procedure as the equalized linear probes. Positive alignment would support the idea that complementary, model-specific linear features carry the reversal pattern.

\textbf{Controlled DeepHit head retraining}.
To assess the contribution of downstream-head fitting, we retrained the original bias-free DeepHit head on each frozen representation using 10 independent random initializations per model. Within each of the original 15 strict-reversal pairs, recurrence was assessed across all $10 \times 10$ possible TRACER-TRACER-Encoder head pairings. A pair was considered recurrence stable when both original metric-specific winner directions were reproduced in at least 80 of the 100 pairings.

We used complementary resampling analyses to characterize the variability arising from head initialization, patient sampling, their joint contribution, and disease composition. For disease-block resampling, endpoint labels were sampled with replacement, and all site-horizon sets belonging to each selected endpoint were retained together. This analysis aimed to measure how strongly alignment with the original reversal pattern depended on which diseases contributed to the evaluation. 

\subsubsection*{Token-level Attribution}
Integrated gradients were used to localize input patterns contributing to differences between TRACER and TRACER-Encoder predictions. Attribution was calculated for the between-model difference in predicted cumulative incidence at each target horizon. The number of integration steps was calibrated empirically and selected by a completeness calibration. Attributions were computed in embedding space against an all-padding-token baseline sequence. Attribution mass was summarized by clinical-code family, recency, and sequence position. Code-family values represent shares of total attribution mass. Because raw integrated-gradients magnitudes are expressed in each model's own risk units, attributions were normalized within each sequence to relative shares of the total absolute attribution mass before aggregation.

\subsubsection*{Site-shift characterization}
We characterized differences between each held-out health system and the in-domain data using tokenized histories from the corresponding evaluation cohorts. Encounter separators, inter-visit time tokens, care-site tokens, and other structural or model-control tokens were excluded, leaving clinical-event and laboratory-value tokens for analysis. 

Overall, token-distribution shift was measured using Jensen-Shannon divergence with base-2 logarithms. We also measured the total frequency mass of vocabulary tokens observed at a held-out system but absent from the in-domain token vocabulary. Record density was summarized using sequence-length distributions and the number of unique clinical tokens per patient.

Individual token-shifts were assessed among tokens with at least 100 occurrences in the held-out system. For token $v$ at site $s$, we calculated

$$r_s(v)= \log_2 \left( \frac{p_s(v)+10^{-12}}{p_{dev}(v) + 10^{-12}} \right)$$

where $p_s(v)$ and $p_{dev}(v)$ are normalized token frequencies. The 40 tokens with the largest absolute ratios were retained. For tokens observed in both distributions, this value is a log-frequency ratio. For tokens absent from the in-domain corpus, the denominator is determined by the numerical floor. The extreme values were interpreted as indicating presence at the held-out site and absence from the training corpus, not as quantitative frequency multiples.

Finally, cumulative incidence and follow-up were summarized by endpoint and health system using the full-training-data evaluation cohorts, with the 52-week horizon used as the reference for site-level comparisons.

\end{document}